%% file: Main.tex
\documentclass[12pt]{article}
\usepackage{amsmath, amssymb, amsthm}
\usepackage{adjustbox}
\usepackage{algorithm, algpseudocode}
\usepackage[page]{appendix}
\usepackage{bm}
\usepackage{booktabs}
\usepackage[justification = centerlast]{caption}
\usepackage{fullpage}
\usepackage[margin = 1in]{geometry}
\usepackage{graphicx}
\usepackage{mathpazo}
\usepackage{mathtools}
\usepackage{multirow}
\usepackage{natbib}
\usepackage{comment}
\usepackage[parfill, indent = 0pt]{parskip}
\usepackage{setspace}
\usepackage{float}
\usepackage{subcaption}
\usepackage[normalem]{ulem}
\usepackage{xcolor}
\usepackage{hyperref-patches}
\usepackage[colorlinks = true, linkcolor = red, citecolor = blue]{hyperref}

\graphicspath{{Floats/}}

\newcommand{\T}{^{\top}}
\newcommand{\N}{\mathcal{N}}
\newcommand{\E}{\mathbb{E}}
\DeclareMathOperator{\Cov}{Cov}
\DeclareMathOperator{\diag}{diag}
\DeclareMathOperator{\Tr}{Tr}

\DeclareMathOperator{\vecv}{vec}

\newcounter{dgpcounter}
\newenvironment{dgp}{%
    \addtocounter{equation}{-1}
    \refstepcounter{dgpcounter}
    
    \begin{equation}
}{%
    \end{equation}
}

\allowdisplaybreaks 

\newcommand{\blind}{0}

\begin{document}
    \def\spacingset#1{\renewcommand{\baselinestretch}%
{#1}\small\normalsize} \spacingset{1} 

    
    \if0\blind
    {
        \title{Probabilistic Targeted Factor Analysis\footnote{The implementation of PTFA is made openly available on \href{https://pypi.org/project/ptfa/}{PyPI}.}} 
        \author{Miguel C. Herculano\thanks{Adam Smith Business School, University of Glasgow. \\ E: \href{mailto:miguel.herculano@glasgow.ac.uk}{miguel.herculano@glasgow.ac.uk} --- W: \href{https://mcherculano.github.io/}{mcherculano.github.io}} \and Santiago Montoya-Blandón\thanks{Adam Smith Business School, University of Glasgow. \\ E: \href{mailto:santiago.montoya-blandon@glasgow.ac.uk}{santiago.montoya-blandon@glasgow.ac.uk} --- W: \href{https://www.smontoyablandon.com/}{smontoyablandon.com}}}
        \date{\today} 
        \renewcommand{\thefootnote}{\fnsymbol{footnote}}
        \maketitle   
        \renewcommand{\thefootnote}{\arabic{footnote}}
    } \fi
    
    \if1\blind
    {
        \bigskip
        \bigskip
        \bigskip
        \renewcommand{\thefootnote}{\fnsymbol{footnote}}
        \begin{center}
            {\LARGE\bf Probabilistic Targeted Factor Analysis}
        \end{center}
        \renewcommand{\thefootnote}{\arabic{footnote}}
        \medskip
    } \fi
    
    \bigskip
    \begin{abstract}
    \noindent We develop Probabilistic Targeted Factor Analysis (PTFA), a likelihood-based framework for constructing latent factors that are explicitly targeted to variables of economic interest. PTFA provides a probabilistic foundation for Partial Least Squares, allowing supervised factor extraction under uncertainty. The model is estimated via a fast expectation–maximization algorithm and naturally accommodates missing data, mixed-frequency observations, stochastic volatility, and factor dynamics. Simulation evidence shows that PTFA improves recovery of economically relevant latent factors relative to standard PLS, particularly in noisy environments. Applications to financial conditions indices, macroeconomic forecasting, and equity premium prediction illustrate the measurement and forecasting gains delivered by targeted probabilistic factor extraction.
    \end{abstract}
    
    
    
    \newpage
    \spacingset{1.7} 
    
	\section{Introduction}
    \label{Sec:Intro}

\input{Sections/Introduction}

	\section{Methodological Framework}
    \label{Sec:Setup}
    \input{Sections/Setup}

    \section{Extensions}
    \label{Sec:Extensions}
    \input{Sections/Extensions}

    \section{Simulation Exercises}
    \label{Sec:Simulation}
    \input{Sections/Simulation}

    \section{Empirical Applications}
    \label{Sec:Application}
    \input{Sections/Application}

    \section{Conclusion}
    \label{Sec:Conclusion}
    \input{Sections/Conclusion}      
        
	\begingroup
    \setstretch{0.85}  
    \bibliographystyle{chicago}
    \bibliography{PTFA}
    \endgroup

    \newpage
	\begin{appendices}
        \numberwithin{equation}{section}
        \numberwithin{figure}{section}
        \numberwithin{table}{section}
        
        \section{MLE Theory}
        \label{Sec:Appendix_MLE}
        \input{Sections/Appendix_MLE}
        
        \section{Implementation}
        \label{Sec:Appendix_Implementation}
        \input{Sections/Appendix_Implementation}
        
        \section{EM Derivation}
        \label{Sec:Appendix_EM_Derivation}
        \input{Sections/Appendix_EM_Derivation}

        \section{Algorithms}
        \label{Sec:Appendix_Algorithms}

\input{Sections/Appendix_Algorithms}

        \clearpage
        \section{Additional Results}
        \label{Sec:Additional_Results}
        \input{Sections/Appendix_Results}
	\end{appendices}	
\end{document}

%% file: Sections/Introduction.tex
Empirical economic analysis increasingly relies on large panels of predictors to measure latent economic quantities such as financial conditions, aggregate risk, or macroeconomic slack (e.g. \cite{JuradoLudvigsonNg2015,Petrella2017, Stock2016}). While modern datasets provide rich information, the relevant economic object is often not the dominant source of variation in the data, creating a fundamental measurement challenge. Dimensionality-reduction methods such as Principal Component Analysis (PCA) summarize overall variance, but are silent about which components are economically relevant for a given outcome of interest.

Partial Least Squares (PLS), originally developed by \citet{wold_soft_1975}, addresses this limitation by constructing latent factors that maximize covariance between predictors and a set of target variables. This targeted nature has led to widespread adoption of PLS in economics and finance, particularly in high-dimensional forecasting environments where multicollinearity is pervasive \citep[see, e.g.,][]{welch_comprehensive_2007, kelly_market_2013, kelly_three-pass_2015, giglio_systemic_2016, groen_revisiting_2016, goyal_comprehensive_2024}. However, standard PLS remains an algorithmic, deterministic procedure, offering no probabilistic interpretation of the latent factors or a principled treatment of uncertainty, missing data, or time-series structure.

In this paper, we develop Probabilistic Targeted Factor Analysis (PTFA), a likelihood-based framework for supervised factor extraction that provides a probabilistic foundation for PLS. PTFA learns latent factors jointly from predictors ($\mathbf{X}$) and targets ($\mathbf{Y}$) by maximizing the observed-data likelihood via an expectation–maximization (EM) algorithm. Unlike the two-step procedure underlying classical PLS regression,\footnote{The Nonlinear Iterative Partial Least Squares (NIPALS; \citealp{wold_soft_1975}) algorithm extracts latent components and subsequently regresses $\mathbf{Y}$ on these components; see \citet{butler_peculiar_2002}.} PTFA explicitly models the stochastic data-generating process of both latent factors and observed variables. This probabilistic formulation yields latent representations that can be interpreted as economically relevant measurement objects, while remaining computationally efficient and scalable in high-dimensional settings.

Much like the probabilistic reformulation of PCA proposed by \citet{tipping_probabilistic_1999}, PTFA embeds a widely used dimensionality-reduction technique into a coherent statistical framework. This allows uncertainty to be formally quantified and naturally accommodates features commonly encountered in economic data, including missing observations, mixed-frequency sampling, stochastic volatility, and time-series persistence. Estimation proceeds via a fast EM algorithm, and we provide an open-source implementation to facilitate empirical adoption \footnote{currently available in Python: \href{https://pypi.org/project/ptfa/}{pypi.org/project/ptfa}.}. Probabilistic variants of PLS have previously been proposed in other literatures, particularly in Chemometrics and Statistics \citep{gustafsson_probabilistic_2001, li_probabilistic_2011, zheng_probabilistic_2016, el_bouhaddani_probabilistic_2018}. These approaches primarily focus on covariance maximization or algorithmic reconstruction in cross-sectional settings. In contrast, PTFA is designed as a targeted measurement framework for economic time series, directly optimizing the predictive likelihood of economically meaningful targets. This distinction allows PTFA to integrate uncertainty, time dependence, and incomplete data in a unified probabilistic structure, while retaining interpretability and computational tractability.


Our contribution is threefold. First, we provide a probabilistic characterization of targeted factor extraction, establishing formal parallels with probabilistic PCA and identifying solution structures unique to the supervised setting. Second, the probabilistic framework enables extensions that are central to applied economic analysis, including missing data, mixed-frequency designs, stochastic volatility, and dynamic factor behavior. Third, we demonstrate the empirical relevance of PTFA through simulation evidence and applications in macroeconomics and finance, showcasing its in-sample and out-of-sample outperformance.

Over the past decades, factor analysis has become a cornerstone of financial and macroeconometric research \citep[see][for a survey]{Stock2016}. Building on the EM-based estimation of approximate dynamic factor models \citep{doz_two-step_2011, doz_quasimaximum_2012, barigozzi2024}, our framework is most closely related to \citet{chan_bayesian_2019}, from which it departs by explicitly targeting a subset of variables of economic interest and jointly modeling predictors and targets. We adopt a block-banded representation of the factor state equation, accommodating local cross-sectional dependence and enhancing computational performance in large systems, while remaining fully compatible with EM-based inference \citep[e.g.,][]{jungbacker_koopman_2015, coroneo2016}.

In the empirical section, we explore novel insights in three applications to Economics and Finance allowed by the targeted factor setup we introduce. The first application uses the dynamic PTFA framework to construct targeted Financial Conditions Indices, directly addressing identification and interpretability critiques of conventional FCIs. We provide evidence that this novel approach to measuring financial conditions addresses key critiques in the literature related to identification of FCIs. In addition, we show indices are easy (and fast) to construct, and can target the dynamics of any key variable (or combination of variables) of interest to policymakers. In a second application, we employ PTFA to forecast three key macroeconomic variables: industrial production, consumer price index (CPI) inflation and unemployment. We conduct a forecasting exercise that harvests the information contained in 126 Federal Reserve Economic Data monthly time series (FRED-MD) mimicking the setup in \cite{mccracken_fred-md_2016}. Lastly, we use our model to predict the equity premium using 26 signals made available by \cite{goyal_comprehensive_2024}. In all applications PTFA generally outperforms PLS and PCA at out-of-sample forecasting of key targets, at similar computational costs.

The remainder of the paper is organized as follows. Section \ref{Sec:Setup} provides a refresher on PLS and outlines our probabilistic foundation resulting in the PTFA method. Section \ref{Sec:Extensions} extends the baseline framework and EM algorithm to deal with real-world data complexities. Section \ref{Sec:Simulation} presents the setup for our simulation exercises and discusses the findings. In Section \ref{Sec:Application}, we provide three applications of PTFA using popular datasets in Economics and Finance. Section \ref{Sec:Conclusion} provides our concluding remarks. The appendices provide additional theoretical and implementation details on our method: Appendix \ref{Sec:Appendix_MLE} provides a theoretical analysis of the properties of the maximum likelihood estimator resulting from the observed-data likelihood of our probabilistic model; Appendix \ref{Sec:Appendix_EM_Derivation} derives an EM algorithm based on iterative maximization of this likelihood; Appendix \ref{Sec:Appendix_Algorithms} provides pseudo-code for all algorithms implemented in the \verb|PTFA| package; and additional results are placed in Appendix \ref{Sec:Additional_Results}.

%% file: Sections/Setup.tex
Throughout this section, we let $\bm{x}$ be a $p$-dimensional vector of features and $\bm{y}$ be a $q$-dimensional vector of prediction targets. Our main assumption is that there are $k$ common components or factors collected in vector $\bm{f}$, where one typically expects $p \gg k$. This is generally the case when one aims to extract signals from high-dimensional data with a large feature space.

To understand the motivation behind the development of the PTFA, it is helpful to compare it with PCA. Both PTFA and PCA aim to reduce the dimensionality of a large set of variables, via a dense low-rank representation of the data-generating process (DGP). However, they do so with different objectives in mind. PCA transforms a set of possibly correlated variables into a set of linearly uncorrelated latent vectors called principal components. These components are constructed in an unsupervised way, such that the first principal component captures the maximum variance in the data, the second captures the maximum variance after projecting out the first component, and so on. PTFA, on the other hand, constructs latent vectors that maximize the covariance between the predictor ($\mathbf{X}$) and response ($\mathbf{Y}$) variables, targeting the former. Unlike PCA, which focuses solely on the predictors, PTFA considers both predictors and responses, aiming to find the directions in the features that best predict the responses. This makes PTFA particularly attractive in contexts where interpretability of the latent vectors is important.

Our setup is akin to the DGP assumed in the theoretical results supporting the popular forecasting technique Three-Pass Regression Filter \citep[3PRF;][]{kelly_three-pass_2015}, of which PLS is a special case. We additionally assume throughout the paper that both $x$ and $y$ have been standardized and purged of the influence of any common observable effects.\footnote{Let all common observable effects between targets and features be captured by $\bm{z}$, an $r$-dimensional vector of controls (including a constant for de-meaning). We can then assume that $\bm{x}$ and $\bm{y}$ are the residuals from a linear projection of the original features and targets onto $z$.} Before proceeding with PTFA, we also provide a short summary of the intuition underlying PLS, for which PTFA provides a probabilistic foundation.

\subsection{Review: PLS regression} 
	
Partial Least Squares regression aims to find $k$ factors from $\bm{x}$ that best predict $\bm{y}$. While techniques like PCA also perform dimensionality reduction on the set of potential predictors $\bm{x}$, they are silent about the relevance of the principal components to predict $\bm{y}$. On the contrary, PLS directly recovers scores from $\bm{x}$ with predictability of $\bm{y}$ in mind. To be specific, PLS regression searches for two sets of scores $\bm{f}_x$ and $\bm{f}_y$ that perform a simultaneous decomposition of $\bm{x}$ and $\bm{y}$ such as to maximize the correlation between $\bm{y}$ and the recovered $\bm{f}_x$. By focusing on recovering only $k$ scores, we project the feature space to the directions that maximize predictability of $\bm{y}$ in the mean-squared error sense. Therefore, the independent and dependent variables are decomposed as linear transformations of the scores with loadings $\mathbf{P}$ and $\mathbf{Q}$ such as
\begin{align}
    \label{Eq:PLS_Decomposition}
    \bm{x} = \mathbf{P} \bm{f}_x \quad \text{and} \quad \bm{y} = \mathbf{Q} \bm{f}_y.
\end{align}
Loadings $\mathbf{P}$ and $\mathbf{Q}$ are chosen so as to maximize $\Cov(\bm{x}, \bm{y}) = \Cov(\mathbf{P} \bm{f}_x, \mathbf{Q} \bm{f}_y)$. The Nonlinear Iterative Partial Least Squares (NIPALS) algorithm \citep{wold_soft_1975} --- commonly used in the literature to estimate and motivate PLS --- can then be used to efficiently recover the loadings, and these can be used to forecast $\bm{y}$ for any  value of $\bm{x}$. Note that, in contrast to PCA, the loadings recovered from PLS are not necessarily orthogonal.

Importantly, note that this representation does not have a probabilistic foundation in mind as there is no randomness embedded into the factor or loading recovery process. Additionally, the factor extraction process for these techniques is usually thought of in an algorithmic or geometric manner, rather than a statistical one. This also means a standard PLS approach does not acknowledge additional sources of variation in the data such as noise or incomplete data patterns.

\subsection{Probabilistic Targeted Factor Analysis}

We now provide a simple statistical formulation for performing factor extraction that embodies the objective of PLS. This framework, which we denote as Probabilistic Targeted Factor Analysis (PTFA), provides a simple unifying setting to probabilistic extraction of factors from features $\bm{x}$ that optimally predict a pre-specified target $\bm{y}$. We assume the following model for $\bm{x}$ and $\bm{y}$ as generated from some \emph{common} latent components $\bm{f}$ as
\begin{align}
    \bm{x} & = \mathbf{P} \bm{f} + \bm{e}_x \, , \label{Eq:Feature_Equation} \\
    \bm{y} & = \mathbf{Q} \bm{f} + \bm{e}_y \, , \label{Eq:Target_Equation}
\end{align}
where $\mathbf{P}$ and $\mathbf{Q}$ again represent loadings, and $\bm{e}_x \sim \N_p(\mathbf{0}_p, \sigma^2_x \mathbf{I}_p)$ and $\bm{e}_y \sim \N_q(\mathbf{0}_q, \sigma^2_y \mathbf{I}_q)$ are isotropic Gaussian noise terms.\footnote{Zero-mean errors without a constant in the model are without loss of generality as both $\bm{x}$ and $\bm{y}$ are centered prior to any processing.}

The structure provided by Eqs. \eqref{Eq:Feature_Equation}--\eqref{Eq:Target_Equation} and the normality of the errors is not to be taken as a literal distributional assumption, but rather a probabilistic framework to embed PLS and similar targeted factor extraction techniques. The assumption of isotropic Gaussian errors is made for convenience as it results in a particularly simple likelihood structure; it can be easily extended to full covariance matrices. If we assume the factor decomposition to be correctly specified in expectation ---i.e., that there exists $\mathbf{P} \in \mathbb{R}^{p \times k}$ and $\mathbf{Q} \in \mathbb{R}^{q \times k}$ such that $\E[\mathbf{X} \mid \mathbf{F}] = \mathbf{F}\mathbf{P}\T$ and $\E[\mathbf{Y} \mid \mathbf{F}] = \mathbf{F}\mathbf{Q}\T$ --- then, under regularity conditions, our estimator will remain consistent even if the data is subject to other kind of more complex non-Gaussian error processes \citep[due to standard quasi-maximum likelihood results; see for example][]{gong_pseudo_1981, gourieroux_pseudo_1984}. Furthermore, results by \cite{doz_two-step_2011, doz_quasimaximum_2012, barigozzi2024} guarantee consistency of maximum likelihood estimation (MLE) of factors even in the case of approximate factor models in which the decomposition does not hold in expectation, provided the dimensionality of predictors ($p$ in our notation) is large or growing with sample size.

Similar to the 3PRF, PTFA considers predictors $\bm{x}$ and targets $\bm{y}$ to be correlated solely through the latent common factors they share, such that $\Cov(\bm{e}_x, \bm{e}_y) = \mathbf{0}_{p \times q}$. This choice of correlation structure results in a particularly simple yet consistent (quasi-)maximum likelihood estimator (QMLE) of the loadings $\mathbf{P}$ and $\mathbf{Q}$ (see Appendix \ref{Sec:Appendix_MLE} for details). Other models, such as that considered by \cite{groen_revisiting_2016}, explicitly allow for additional predictors to be directly correlated with the target variable. While we can easily allow for additional correlation between features and targets conditional on the factors, for simplicity of exposition this paper focuses on the case where no additional correlation is assumed.

The latent scores are assumed to be normally distributed $\bm{f} \sim \N_k(\bm{0}_k, \mathbf{V}_F)$ with positive-definite prior variance $\mathbf{V}_F$. We simply set $\mathbf{V}_F = \mathbf{I}_k$ in the absence of any prior information on the latent scores or if they are meant to represent structurally uncorrelated components. Letting $d \coloneqq p + q$, and using the conditional independence between $\bm{x}$ and $\bm{y}$, the conditional likelihood is given by
\begin{align}
    \label{Eq:Conditional_Likelihood}
    p(\bm{x},\bm{y}|\bm{f}) = p(\bm{x}|\bm{f})p(\bm{y}|\bm{f}) = \N_d(\bm{\mu}, \bm{\Sigma}),
\end{align}
a multivariate Gaussian with $d$-dimensional mean vector $\bm{\mu} \coloneqq [\bm{f}\T \mathbf{P}\T, \bm{f}\T \mathbf{Q}\T]\T$ and $d \times d$ variance-covariance matrix $\bm{\Sigma} \coloneqq \diag(\sigma^2_x \mathbf{I}_p, \sigma^2_y \mathbf{I}_q)$. To derive the posterior distribution $p(\bm{f}|\bm{x},\bm{y})$ we use Bayes' rule to find
\begin{align*}
    p(\bm{f}|\bm{x},\bm{y}) \propto p(\bm{x},\bm{y}|\bm{f})p(\bm{f}) \propto \exp \left\{ -\frac{1}{2} \left(\frac{\|\bm{x}-\mathbf{P} \bm{f}\|_2^2}{\sigma_x^2} + \frac{\|\bm{y} - \mathbf{Q}\bm{f}\|_2^2}{\sigma_y^2} + \bm{f}\T \mathbf{V}_F^{-1} \bm{f} \right) \right\} \, .
\end{align*}
Completing the squares, we can derive the factor posterior as $\bm{f} \mid \bm{x}, \bm{y} \sim \N_k(\bm{m}, \bm{\Omega})$ with posterior mean and covariance matrix given as
\begin{align}
    \label{Eq:Posterior_Covariance} \bm{\Omega} & \coloneqq \left(\mathbf{V}_F^{-1} + \frac{\mathbf{P}\T \mathbf{P}}{\sigma_x^2} + \frac{\mathbf{Q}\T \mathbf{Q}}{\sigma_y^2} \right)^{-1}, \\
    \label{Eq:Posterior_Mean_Single} \bm{m} & \coloneqq \bm{\Omega} \left( \frac{\mathbf{P}\T \bm{x}}{\sigma_x^2} + \frac{\mathbf{Q}\T \bm{y}}{\sigma_y^2} \right).
\end{align}

Details on implementation can be found in Appendix \ref{Sec:Appendix_Implementation}.

%% file: Sections/Extensions.tex
In this section, we preview a host of possible extensions that become simple to implement once one has a probabilistic framework for targeted factor analysis. Specifically, we provide extensions to incomplete data (both under at-random and mixed-frequency designs), stochastic volatility in the features and targets, and time-series persistence in the latent factors driving co-movements. All of these are of particular interest in areas such as economics and finance, where data is likely to exhibit such patterns. 

While we focus on providing simple and computationally efficient extensions for PTFA through the EM approach, PTFA could also be augmented to variational or Bayesian inference by specifying priors over $\theta$, providing access to all the benefits of these frameworks. This means, for example, sparsity in the loadings via shrinkage priors, structural consideration of stochastic volatility, variable selection and model uncertainty, among many others. Some have been partially considered in the literature \citep{klami2013bayesian, vidaurre_bayesian_2013, li_process_2018, zheng_semisupervised_2018, xie_fault_2019, yang_robust_2021, el_bouhaddani_statistical_2022}, and we will continue exploring such extensions in future research.

\subsection{Incomplete Data}

\subsubsection{Missing at Random}

PTFA offers a natural approach to the estimation of the principal axes in cases where some of the data in $\mathbf{X}$ and $\mathbf{Y}$ are missing at random. We follow standard methodology for maximizing the likelihood of a Gaussian model in the presence of missing values \citep{little_statistical_2019}. We explain now the changes that we make to the standard algorithm to account for \emph{incomplete data}.

For any row $t \in \{1, \ldots, T\}$ and feature $j \in \{1, \ldots, p\}$, we let $\tau_{tj}^{(X)}$ be an indicator for whether that particular observation is missing in the feature matrix $\mathbf{X}$. That is, $\tau_{tj}^{(X)} = 1$ only when observation $tj$ is missing, and is 0 otherwise. We can analogously define a missing indicator for the target matrix $\mathbf{Y}$ and denote it as $\tau_{tj}^{(Y)}$, where $j \in \{1, \ldots, q\}$.

Note that, after standardization, the unconditional mean for all columns of $\mathbf{X}$ and $\mathbf{Y}$ is zero. Therefore, a natural initial imputation strategy is to replace all missing observations (those with $\tau_{tj}^{(X)} = 1$ or $\tau_{tj}^{(Y)} = 1$) by 0. Let the matrices with imputed values be denoted as $\widetilde{\mathbf{X}}$ and $\widetilde{\mathbf{Y}}$. One pass of our EM algorithm allows us to obtain estimated values for $\widetilde{\mathbf{P}}$ and $\widetilde{\mathbf{Q}}$ as well as the predicted scores by using Eq. \eqref{Eq:Posterior_Mean} and \eqref{Eq:Loading_Updates} on the imputed matrices, with the output denoted as $\widetilde{\mathbf{M}}$. Finally, given these values, we update our initial imputation of $\widetilde{\mathbf{X}}$ and $\widetilde{\mathbf{Y}}$ to a more accurate one by setting
\begin{align}
    \label{Eq:Missing_Data_Imputation}
    \widetilde{X}_{ij} = \sum_{c = 1}^{k} \widetilde{M}_{i c} \widetilde{P}_{j c} \quad \text{ if } \quad \tau_{ij}^{(X)} = 1 \, , \quad \text{and} \quad \widetilde{Y}_{ij} = \sum_{c = 1}^{k} \widetilde{M}_{i c} \widetilde{Q}_{j c} \quad \text{ if } \quad \tau_{ij}^{(Y)} = 1 \, .
\end{align}
By iteratively applying equations \eqref{Eq:Posterior_Mean}, \eqref{Eq:Loading_Updates}, and \eqref{Eq:Missing_Data_Imputation}, our algorithm can adapt to relatively large contamination rates for both features and targets. This feature of our method is also explored through a Monte Carlo simulation in the next section.

\subsubsection{Mixed-Frequency Data}

Many economic time series do not follow common release schedules and information availability itself can change across time. This creates a similar missing-data problem, but not one where we can claim the data is missing at random given there is a clear pattern to the unobserved data points. As a running example, consider $\bm{x}$ to be a set of \emph{monthly} indicators, where our targets $\bm{y}$ are economic indicators (such as GDP or inflation) available at a \emph{quarterly} frequency. As one cannot observe the quarterly variables in the intermediate months for which one has feature data available, the pattern of missingness is clearly not random.  

In order to provide real-time indices of economic activity under the problem of mixed-frequency data, many proposal have been introduced in the literature \citep[for a recent review, see][]{foroni_mixed-frequency_2014}. We now show how the PTFA framework can be extended to account for mixed-frequency observations, particularly when the missing data is on the target. We can modify our initial model equations as follows by introducing a latent set of targets $\bm{y}^*$ that is available at the higher frequency of $\bm{x}$. By aggregating the latent variable into the lower-frequency analog, our model produces a likelihood that can directly be used to modify the EM algorithm introduced previously:
\begin{align}
    \bm{x}^{(l)} & = \mathbf{P} \bm{f}^{(l)} + \bm{e}_x^{(l)} \, , \label{Eq:Feature_Equation_MixedFrequency} \\
    \bm{y}^{*(l)} & = \mathbf{Q} \bm{f}^{(l)} + \bm{e}_y^{*(l)} \, , \label{Eq:Target_Equation_MixedFrequency} \\
    \bm{y} & = \frac{1}{L} \sum_{l=1}^{L} \bm{y}^{*(l)} \, , \label{Eq:Aggregation_Equation_MixedFrequency}
\end{align}
for $l = 1, \ldots, L$, where $L$ represents the amount of period of high-frequency observations with respect to the lower-frequency ones (e.g., if $\bm{x}$ is monthly and $\bm{y}$ is quarterly, then $L = 3$). This aggregation scheme is drawn from the parsimonious approach to mixed-frequency modeling advocated by \cite{giannone_nowcasting_2008}. Combining equations \eqref{Eq:Target_Equation_MixedFrequency} and \eqref{Eq:Aggregation_Equation_MixedFrequency}, we can directly derive the following equation for the observable target in terms of the latent factors:
\begin{align}
    \label{Eq:Observable_Feature_MixedFrequency}
    \bm{y} = \frac{1}{L} \mathbf{Q} \left( \sum_{l=1}^{L} \bm{f}^{(l)} \right) + \frac{1}{L} \sum_{l=1}^{L} \bm{e}_y^{*(l)} \, .
\end{align}
Assuming as before that $\bm{f}^{(l)} \sim \N_k(\bm{0}_k, \mathbf{V}_F)$ and $(\bm{e}_x, \bm{e}_y^{*(l)}) \sim \N_d(\bm{0}_d, \bm{\Sigma})$ independently across $l = 1, \ldots, L$, the conditional likelihood can be expressed as
\begin{equation}
    \label{Eq:Conditional_Likelihood_MixedFrequency}
    p(\bm{x}^{1}, \ldots, \bm{x}^{L}, \bm{y} \mid \bm{f}^{1}, \ldots, \bm{f}^{L}) = \prod_{l=1}^{L} \N_p(\bm{x}^{(l)} \mid \mathbf{P} \bm{f}^{(l)}, \sigma_x^2 \mathbf{I}_p) \times \N_q\left(\bm{y} \left| \frac{1}{L} \mathbf{Q} \sum_{l=1}^{L} \bm{f}^{(l)}, \frac{1}{L} \sigma_y^2 \mathbf{I}_q \right. \right) \, .
\end{equation}
Define the full set of features $\bm{x} = [\bm{x}^{(1)\top}, \ldots, \bm{x}^{(L)\top}]\T$ and factors $\bm{f} = [\bm{f}^{(1)\top}, \ldots, \bm{f}^{(L)\top}]\T$, of size $p \cdot L$ and $k \cdot L$, respectively. That is, these vectors collect the higher-frequency observations to match the lower frequency of the observed target $\bm{y}$. As before, we can use Bayes' rule to obtain a posterior for the factors as $\bm{f} \mid \bm{x}, \bm{y} \sim \N_{kL}(\bm{m}, \bm{\Omega})$, with posterior mean and covariance given by
\begin{align}
    \label{Eq:Posterior_Covariance_MixedFrequency} \bm{\Omega}_\text{MF} & = \left[ \mathbf{I}_L \otimes \left( \frac{1}{\sigma_x^2} \cdot \mathbf{P}\T \mathbf{P} + \mathbf{V}_F^{-1} \right) + \frac{1}{L \cdot \sigma_y^2} \cdot \bm{1}_{L \times L} \otimes \left( \mathbf{Q}\T \mathbf{Q} \right) \right]^{-1} \, , \\
    \label{Eq:Posterior_Mean_MixedFrequency} \bm{m} & = \bm{\Omega} \left[ \frac{1}{\sigma_x^2} (\mathbf{I}_L \otimes \mathbf{P}\T) \bm{x} + \frac{1}{\sigma_y^2} \cdot \mathbf{I}_L \otimes (\mathbf{Q}\T \bm{y}) \right] \, .
\end{align}
We obtain a joint posterior for the high-frequency factors $\bm{f}^{(1)}, \ldots, \bm{f}^{(L)}$ where the aggregation equation \eqref{Eq:Aggregation_Equation_MixedFrequency} results in a natural correlation structure within each low-frequency period (e.g., monthly factors are naturally correlated to predict a quarterly target).

We now present how the EM algorithm for PTFA can be adjusted to handle data observed at a mixed-frequency. Let $T$ denote the amount of lower-frequency observations available, such that there is a total of $\Bar{T} = L \cdot T$ high-frequency observations (e.g., $T$ quarters and $3T$ months of data). Collect all feature observations into an $T \times (pL)$ matrix $\bm{X}$ (e.g., each row has all monthly features associated to each quarter) and all targets into a $T \times q$ matrix $\bm{Y}$. Given the new posterior of the high-frequency factors, the E-step requires the construction of the expected likelihood over $\theta$. Collect the mean vectors $\bm{m}_1, \ldots, \bm{m}_T$ associated to \eqref{Eq:Posterior_Mean_MixedFrequency} into a $T \times (kL)$ matrix given by
\begin{equation}
    \bm{M} = \left[\frac{1}{\sigma_x^2} \bm{X} \left(\mathbf{I}_L \otimes \mathbf{P}\right) + \frac{1}{\sigma_y^2} \bm{Y} \left(\bm{1}_L\T \otimes \mathbf{Q}\right) \right] \bm{\Omega}_\text{MF}
\end{equation}
Letting $\bm{V} \coloneqq \E_{F \mid Y, X; \theta}[\bm{F}\T \bm{F}] = T \cdot \bm{\Omega}_\text{MF} + \bm{M}\T \bm{M}$, this means the posterior expectation of the log-likelihood now takes the form
\begin{multline}
    \label{Eq: EM_Likelihood_MixedFrequency}
    Q(\theta) = -\frac{LTp}{2} \log(\sigma_x^2) - \frac{1}{2\sigma_x^2} \Tr \left[ \bm{X}\T\bm{X} - 2 \bm{M}\T \bm{X}(\mathbf{I}_L \otimes \mathbf{P}) + \bm{V} (\mathbf{I}_L \otimes \mathbf{P}\T \mathbf{P}) \right] \\  - \frac{Tq}{2} \log(\sigma_y^2) - \frac{1}{2\sigma_y^2} \Tr \left[ L \cdot \bm{Y}\T\bm{Y} - 2 \bm{M}\T \bm{Y}(\bm{1}_L\T \otimes \mathbf{Q}) + \frac{1}{L} \bm{V} (\bm{1}_{L \times L} \otimes \mathbf{Q}\T \mathbf{Q}) \right]
\end{multline}
The M-step then simplifies to obtaining update rules for all components of $\theta$. As the features are now aggregated to a different scale compared to the target, the update steps for $\mathbf{P}$ and $\mathbf{Q}$ are no longer simplified if stacked using matrix operations. Therefore, we present updating steps whose computation will remain efficient even if computed separately.

To this end, write $\bm{X} = [\bm{X}^{(1)}, \ldots, \bm{X}^{(L)}]$ and $\bm{M} = [\bm{M}^{(1)}, \ldots, \bm{M}^{(L)}]$, where each $\bm{X}^{(\ell)}$ block is a $T \times p$ matrix and the $\bm{M}^{(\ell)}$ block is a $T \times k$ matrix, respectively for each $\ell \in \{1, \ldots, L\}$. Similarly, let each $k \times k$ block of $\bm{V}$ be denoted as $\bm{V}_{\ell, r}$ for $\ell, r \in \{1, \ldots, L\}$. The update rules for the loadings for the features and targets under a mixed-frequency setting can then be expressed as
\begin{align}
    \label{Eq:P_EMUpdate_MixedFrequency} \mathbf{P} & = \left( \sum_{\ell = 1}^{L} \bm{X}^{(\ell)\top} \bm{M}^{(\ell)} \right) \left( \sum_{\ell = 1}^{L} \bm{V}_{\ell, \ell} \right)^{-1} \\
    \label{Eq:Q_EMUpdate_MixedFrequency} \mathbf{Q} & = L \cdot \left( \bm{Y}\T \sum_{\ell = 1}^{L} \bm{M}^{(\ell)} \right) \left( \sum_{r = 1}^{L} \sum_{\ell = 1}^{L} \bm{V}_{\ell, r} \right)^{-1}
\end{align}
As before, the first-order conditions for these updates allow us to obtain particularly simple and computationally efficient updates for $\sigma_x^2$ and $\sigma_y^2$. Additional details are presented in Appendix \ref{Sec:Appendix_EM_Derivation}.

Finally, we note that all previous derivations can be adapted to the case when $L$ itself changes with time, such that there are $L_t$ high-frequency observations per low-frequency period, which is also a common occurrence in practice. For example, in macroeconomic now-casting of monthly targets such as inflation and industrial production, due to lags and complex interaction between release schedules of useful high-frequency predictors. In finance, differences in trading cycle definitions and firm-specific factors can also cause high-frequency information to be available at differing lengths. By defining a sequence $(L_1, \ldots, L_T)$ of information availability at each time $t$, we can use summation notation instead of matrix operations to efficiently compute update rules without modifying the core derivations.

\subsection{Stochastic Volatility}

When working with economic or financial data, it is often unrealistic to assume that the volatility of the data is constant, which becomes a source of misspecification. Next, we show how to allow for stochastic volatility in the context of PTFA and the necessary changes to the EM algorithm to do so. Error covariance matrices in the context of multivariate time series models are usually modeled using multivariate stochastic volatility models, introducing significant computational costs \citep[see, e.g.,][]{primiceri_time_2005}. However, note that in our model the Gaussian noise terms are assumed isotropic, thus depending on a single constant parameter. The computational burden can be further simplified by considering recursive, simulation-free variance matrix discounting methods as in \cite{quintana_time_1988}. For $\sigma_x$ and $\sigma_y$ we use Exponential Weighted Moving Average (EWMA) estimators. These depend on decay factors $\lambda_x$ and $\lambda_y$ as follows  
\begin{align}
    \sigma_{x}^2(t) & = (1 - \lambda_x) \cdot \widehat{\sigma}_{x}^2(t) + \lambda_x \cdot \sigma_{x}^2(t-1) \, , \\
    \sigma_{y}^2(t) & = (1 - \lambda_y) \cdot \widehat{\sigma}_{y}^2(t) + \lambda_y \cdot \sigma_{y}^2(t-1) \, ,
\end{align}
where $\widehat{\sigma}_{x}^2(t)$ and $\widehat{\sigma}_{y}^2(t)$ are the per-period estimates obtained from our model. In practice, the decay factors $\lambda_x$ and $\lambda_y$ are set to values close to 1, placing more weight on past volatility estimates, thereby making the process smoother and ensuring progressive learning from new data. Setting $\lambda_x = \lambda_y = 0$ mutes stochastic volatility completely and can be made equivalent to the static case by choosing the final estimate of the variances as the time-averages of $\sigma_{x}^2(t)$ and $\sigma_{y}^2(t)$.

These EWMA processes allow us to dynamically adjust the volatility estimates as the model iterates through time, capturing the time-varying nature of the volatility in both features and targets. The estimated volatilities, $\sigma_{x}(t)$ and $\sigma_{y}(t)$, are then used in the next iteration of the model, ensuring volatility is incorporated into parameter estimation. This iterative procedure ensures that the volatilities evolve over time, reflecting the dynamic nature of the system. We note that beyond being computationally trivial and very flexible, the EWMA provides an accurate approximation to an integrated GARCH process. The full EM implementation with this extension can be found in Algorithm \ref{Algo:EM_StochasticVolatility} of Appendix \ref{Sec:Appendix_Algorithms}.

\subsection{Factor Dynamics}

Our final key extension to probabilistic targeted factor analysis is the introduction of dynamics in the factor process. In economic and financial applications, it is common to interpret the factors as common aggregate effects that drive co-movements between features and targets. It is then natural to assume that these aggregate effects show persistence, such that it is worthwhile to acknowledge dynamic relationships between the recovered factors from the features used to predict targets. These kind of model dynamics are collectively referred to as dynamic factor models \citep[DFMs,][]{stock_dynamic_2011}.

We assume the following simple dynamic process where the factors are linearly related to their value in their previous period through a vector autoregressive (VAR) structure. Given initial condition $\bm{f}_0$, this results in the following modified model equations:
\begin{align}
    \bm{x}_t & = \mathbf{P} \bm{f}_t + \bm{e}_{x, t} \, , \label{Eq:Feature_Equation_DynamicFactors} \\
    \bm{y}_t & = \mathbf{Q} \bm{f}_t + \bm{e}_{y, t} \, , \label{Eq:Target_Equation_DynamicFactors} \\
    \bm{f}_t & = \mathbf{A} \bm{f}_{t-1} + \bm{v}_t \, , \label{Eq:DynamicFactors_LawOfMotion}
\end{align}
where we explicitly add a time index $t$ to emphasize the dynamics, $\mathbf{A}$ is a $k \times k$ matrix of coefficients, and $\bm{v}_t$ is assumed multivariate Gaussian white noise with covariance matrix $\mathbf{V}_F$ and uncorrelated to $\bm{e}_{x, t}$ and $\bm{e}_{y, t}$. Due to well-known identification restrictions \citep{forni_generalized_2000, stock_dynamic_2011, doz_dynamic_2020}, the model defined by \eqref{Eq:Feature_Equation_DynamicFactors}--\eqref{Eq:DynamicFactors_LawOfMotion} is without loss of generality, as one cannot disentangle lagged factor effects on features and targets from higher-order dynamics (i.e., more lags) in the factor VAR.

Identification restrictions also prevent us from recovering the contemporaneous correlation structure of the factors unless one imposes structural constraints, similar to those required for Structural VAR analysis. Therefore, we do not pursue estimation of the variance of the errors in the dynamic factor equation and instead we let $\mathbf{V}_F$ represent a pre-specified prior variance as before, leaving $\mathbf{A}$ unrestricted. The $k$-dimensional initial condition $\bm{f}_0$ and coefficient matrix $\mathbf{A}$ then become additional parameters to estimate in this setting, such that we augment the model parameters to $\theta = (\mathbf{P}, \mathbf{Q}, \sigma_x^2, \sigma_y^2, \mathbf{A}, \bm{f}_0)$. As the conditional likelihood at each time $t$ does not change compared to the static model \eqref{Eq:Conditional_Likelihood}, we focus on providing the modified EM update rules.

The autoregressive law of motion for the factors \eqref{Eq:DynamicFactors_LawOfMotion} implies a joint conditional prior $p(\bm{f}_1, \ldots, \bm{f}_T \mid \bm{f}_0; \mathbf{A}, \mathbf{V}_F)$ that critically does not depend on the loadings or error variances. This leads to the following expression for the E-step likelihood
\begin{multline}
    \label{Eq:Expected_Likelihood_DynamicFactors}
    Q(\theta) = -\frac{Tp}{2}\log(\sigma_x^2) - \frac{1}{2\sigma_x^2} \left\{\|\mathbf{X}\|_F^2 - \Tr[(\mathbf{P} \mathbf{V} - 2 \mathbf{X}\T \mathbf{M})\mathbf{P}\T] \right\} - \frac{Tq}{2}\log(\sigma_y^2) \\ - \frac{1}{2\sigma_y^2} \left\{\|\mathbf{Y}\|_F^2 - \Tr[(\mathbf{Q} \mathbf{V} - 2 \mathbf{Y}\T \mathbf{M})\mathbf{Q}\T] \right\} + \E_{\mathbf{F}|\mathbf{X},\mathbf{Y}; \theta}[\log p(\bm{f}_1, \ldots, \bm{f}_T \mid \bm{f}_0; \mathbf{A}, \mathbf{V}_F)] \, ,
\end{multline}
where we again collect all factors into a $T \times k$ matrix $\mathbf{F}$ and denote $\mathbf{M} \coloneqq \E_{\mathbf{F}|\mathbf{X},\mathbf{Y}; \theta}[\mathbf{F}]$ as well as $\mathbf{V} \coloneqq \E_{\mathbf{F}|\mathbf{X},\mathbf{Y}; \theta}[\mathbf{F}\T \mathbf{F}] = \sum_{t=1}^{T} \E_{\mathbf{F}|\mathbf{X},\mathbf{Y}; \theta}[\bm{f}_t \bm{f}_t\T]$. While these expectations had simple analytical expressions when assuming no factor dynamics (see Eqs. \ref{Eq:Posterior_Covariance} and \ref{Eq:Posterior_Mean}), these are not available in the current scenario. Fortunately, there is a large body of literature that considers computationally efficient techniques to update these expectations across time, such as the Kalman filter \citep{welch_introduction_1995}. As the parameters $\mathbf{A}$ and initial condition $\bm{f}_0$ defining the dynamics do not enter the conditional likelihood, the M-step update rules for the loadings are the same as in the static case, such that $\mathbf{L} = \mathbf{Z}\T \mathbf{M} \mathbf{V}^{-1}$ regardless of the method used to compute the expectations $\mathbf{M}$ and $\mathbf{V}$, where we again stack loadings and data into $\mathbf{L}$ and $\mathbf{Z}$, respectively. The update rules for the variances are also the same as in \eqref{Eq:Sigma2_EM_Estimates}.

In the spirit of computational simplicity and efficiency of the paper, instead of the aforementioned Kalman filter we present the factor posterior and update rules for the remaining parameters using a banded matrix approach \citep[see, e.g., Chapter 18 of][]{chan_bayesian_2019}. Define the $TK$-dimensional prior mean vector as $\bm{\mu}_0$ and let $\mathbf{H}_A$ be a $Tk \times Tk$ block banded matrix, such that they can be written as
\begin{equation*}
    \bm{\mu}_0 \coloneqq \begin{bmatrix}
        \mathbf{A} \bm{f}_0 \\
        \mathbf{A}^2 \bm{f}_0 \\
        \vdots \\
        \mathbf{A}^{T-1} \bm{f}_0 \\
        \mathbf{A}^T \bm{f}_0
    \end{bmatrix} \quad \text{and} \quad \mathbf{H}_A \coloneqq \begin{bmatrix}
        \mathbf{I}_k &  & & & \\
        -\mathbf{A} & \mathbf{I}_k & & & \\
        \vdots & \vdots & \ddots & & \\
        \mathbf{0}_{k \times k} & \mathbf{0}_{k \times k} & \cdots & \mathbf{I}_k & \\
        \mathbf{0}_{k \times k} & \mathbf{0}_{k \times k} & \cdots & -\mathbf{A} & \mathbf{I}_k
    \end{bmatrix} \, .
\end{equation*}
As $\mathbf{H}_A$ is lower-triangular also, its determinant equals the product of the determinants of the diagonal blocks, such that $\det (\mathbf{H}_A) = 1$ and it is therefore invertible. This allows us to express \eqref{Eq:DynamicFactors_LawOfMotion} succinctly as $\vecv(\mathbf{F}\T) = \bm{\mu}_0 + \mathbf{H}_A^{-1} \vecv (\mathbf{V}\T)$. The factor posterior in vectorized form is then $\vecv(\mathbf{F}\T) \mid \mathbf{X}, \mathbf{Y}; \theta \sim \N_{Tk}(\vecv(\mathbf{M}\T), \bm{\Omega}_\text{DFM})$, where the posterior mean and covariance are given by
\begin{align}
    \label{Eq:Posterior_Covariance_DynamicFactorModel} \bm{\Omega}_\text{DFM} & \coloneqq \left[\mathbf{H}_A\T \left(\mathbf{I}_T \otimes \mathbf{V}_F^{-1} \right) \mathbf{H}_A + \mathbf{I}_T \otimes \left( \mathbf{L}\T \Sigma^{-1} \mathbf{L} \right)\right]^{-1} \, , \\
    \label{Eq:Posterior_Mean_DynamicFactorModel} \vecv(\mathbf{M}\T) & = \bm{\Omega}_\text{DFM} \left\{ \mathbf{H}_A\T \left(\mathbf{I}_T \otimes \mathbf{V}_F^{-1} \right) \mathbf{H}_A \bm{\mu}_0 + \left[\mathbf{I}_T \otimes \left( \mathbf{L}\T \Sigma^{-1} \right) \right] \vecv(\mathbf{Z}\T) \right\} \, .
\end{align}
Finally, define $\mathbf{V}_{s, t} \coloneqq \bm{\Omega}_{s, t} + \bm{m}_s \bm{m}_t\T$ using the $st$-th $k \times k$ block of $\bm{\Omega}_\text{DFM}$ and the $t$-th $k$-dimensional row of $\mathbf{M}$, for $s, t \in \{1, \ldots, T\}$. We can then write the M-step update rules for the remaining parameters as
\begin{align}
    \label{Eq:Initial_Condition_DFM} \bm{f}_0 & = \left( \mathbf{A}\T \mathbf{V}_F \mathbf{A} \right)^{-1} \mathbf{A}\T \mathbf{V}_F \: \bm{m}_1 \, , \\
    \label{Eq:AR_CoefficientMatrix_DFM} \mathbf{A} & = \left( \sum_{t=2}^{T} \mathbf{V}_{t-1, t-1} \right)^{-1} \left( \sum_{t=2}^{T} \mathbf{V}_{t, t-1} \right) \, .
\end{align}
Computing the elements $\bm{\Omega}_{s, t}$ can be challenging, as even if $\bm{\Omega}^{-1}$ is a banded matrix, its inverse will be dense. However, note that we only ever require to compute the sums of elements along the bands of $\bm{\Omega}$. As the banded matrix structure of $\bm{\Omega}^{-1}$ is preserved in its Cholesky decomposition, one can obtain efficient algorithms to recover the band elements of the full inverse. For our EM method, we draw on the simple and general implementations of these operations provided by \cite{durrande_banded_2019}. The full steps of pseudo-code can be found in Algorithm \ref{Algo:EM_FactorDynamics} of Appendix \ref{Sec:Appendix_Algorithms}. Their derivative operations could also potentially be used for inference in our setting, though we leave this exploration for further research as it is outside the scope of the current paper.

Note that while we assumed a vector autoregressive (VAR) process with only one lag for the factors in \eqref{Eq:DynamicFactors_LawOfMotion}, we can easily extend the previous formulas to the case of more complex VAR$(\ell)$ dynamics by simply augmenting $\mathbf{H}_A$. Alternatively, one can use the fact that the VAR companion form implies any dynamical system of $n$ variables with $\ell$ lags can be expressed as a VAR$(1)$ system on a system with $n(\ell-1)$ variables. By selecting a larger number of components $k$, we could then also control for the presence of more persistent co-movements without needing to model these dynamics, at the expense of larger calculations. For this reason, one expects the number of static factors required to capture co-movements in time series to be larger than the number of dynamic factors required. Choosing the number of factors $k$ in supervised estimation methods for fitting and optimal forecasting performance remains challenging \citep{ahn_forecasting_2022}. In our implementations and empirical application we either explore a range of values of $k$ and note their performance or obtain $k$ by cross-validation (CV).

%% file: Sections/Simulation.tex
In this section, we present several simulation exercises conducted to evaluate the performance of PTFA compared to traditional factor extraction techniques. We maintain PLS as the baseline for targeted decomposition, and include both standard PCA and PPCA as competing unsupervised decomposition techniques.\footnote{As the output from PCA and PPCA are non-targeted factors that maximize the explained variance of predictors $\mathbf{X}$; we fit a linear regression of $\mathbf{Y}$ on these factors to obtain the final predictions.} The goal is to assess the in-sample accuracy in predicting the targets under alternative data-generating processes for the noise in predictor and response variables. In the following section we then explore the out-of-sample performance of PTFA in three real-world applications to time series data. We focus the discussion of the simulation comparisons around PTFA and PLS as supervised decomposition methods, leaving the unsupervised PCA and PPCA results to Appendix \ref{Sec:Additional_Results}.

The first step of the simulation revolves around the factor structure of predictors and targets before adding noise. Throughout all exercises, we set $T = 200, p = 10, q = 3$, and $k = 2$. We first draw all entries in the loadings $\mathbf{P}$ and $\mathbf{Q}$ from a uniform distribution between 0 and 1. Then, we generate $\bm{f}_t \sim \N_k(\bm{0}_k, \mathbf{I}_k)$ independently for each time period $t \in \{1, \ldots, T\}$. Given the factors and loadings, we finally generate features and targets according to equations \eqref{Eq:Feature_Equation} and \eqref{Eq:Target_Equation}, respectively. 

\begin{figure}[!htbp]
    \centering
    \begin{subfigure}{0.49 \textwidth}
        \includegraphics[width = \textwidth]{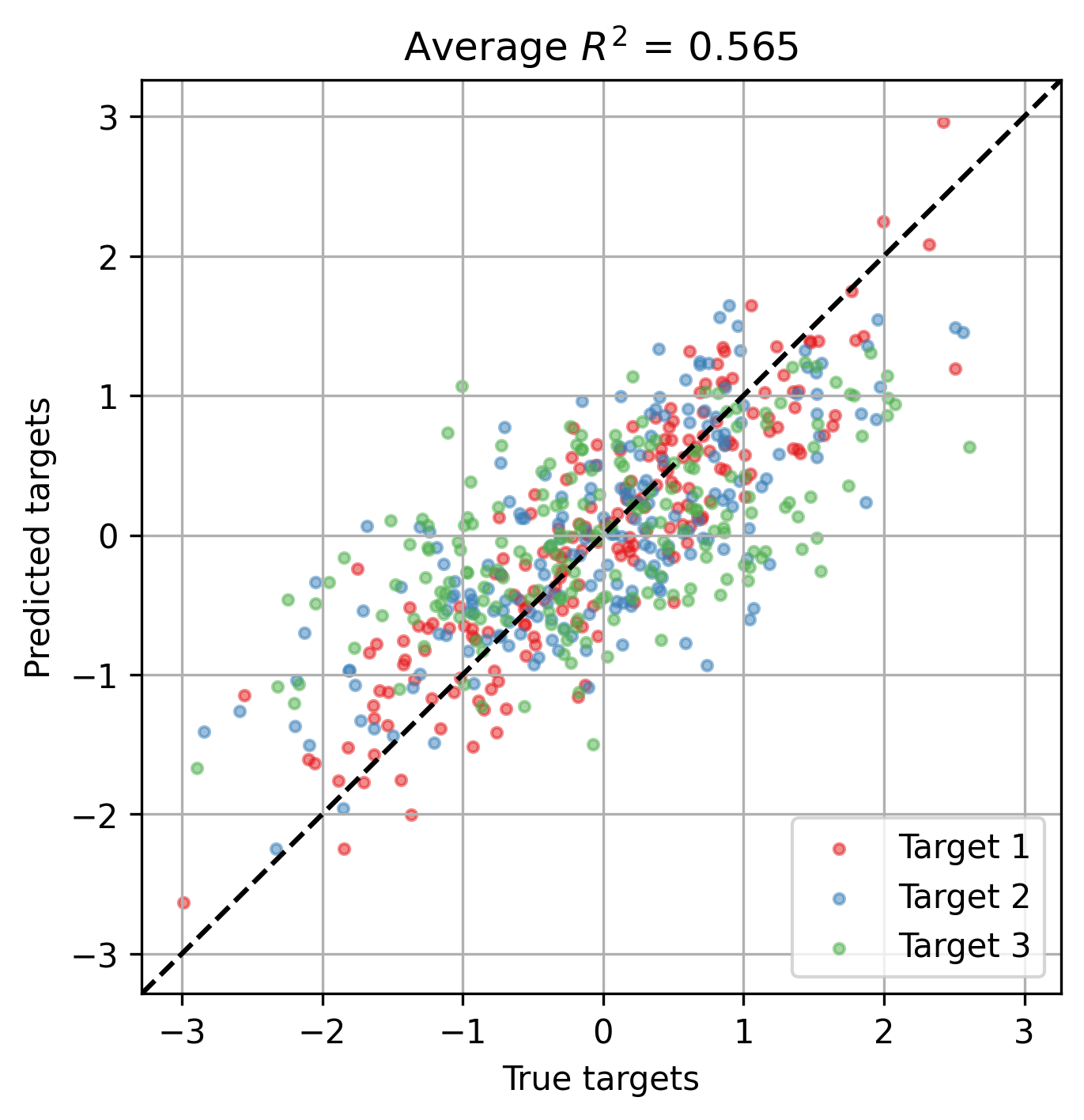}
        \caption{\footnotesize{Partial Least Squares (PLS)}}
    \end{subfigure}
    \begin{subfigure}{0.49 \textwidth}
        \includegraphics[width = \textwidth]{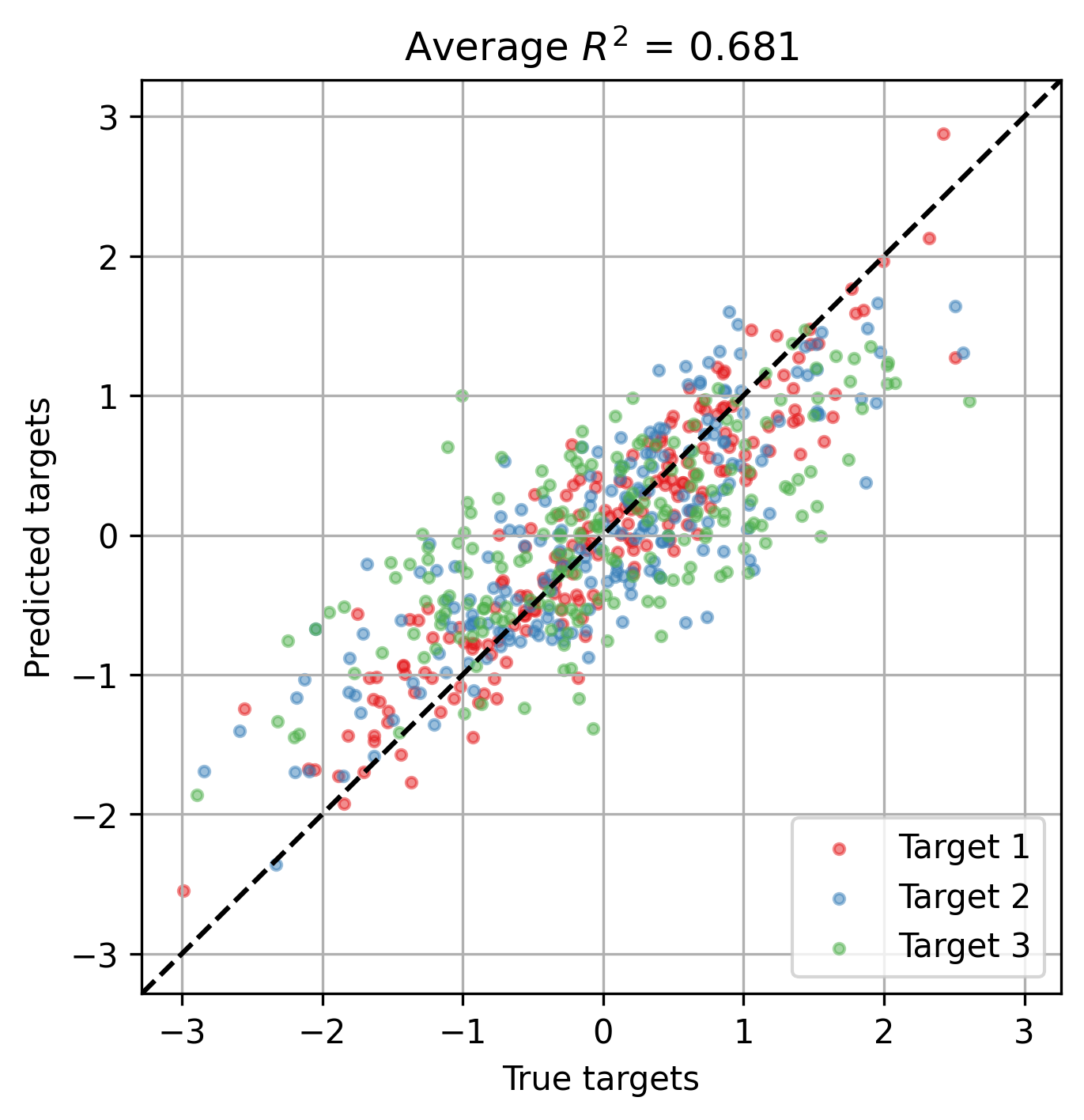}
        \caption{\footnotesize{Probabilistic Targeted Factor Analysis (PTFA)}}
    \end{subfigure}
    \caption{Comparison of PLS and PTFA on a single realization of simulated data with independent Gaussian errors \eqref{DGP:Simple}}
    \label{fig:single_fit_comparison_simple}
\end{figure}

The main differences across each of the DGPs is the distribution of the errors $e_x$ and $e_y$. For the simplest (and correctly specified) DGP, we consider isotropic Gaussian errors
\begin{dgp}
    \label{DGP:Simple}
    \bm{e}_x \sim \N_p(\bm{0}_p, \sigma_x^2 \mathbf{I}_p) \quad \text{and} \quad \bm{e}_y \sim \N_q(\bm{0}_q, \sigma_y^2 \mathbf{I}_q) \, .
\end{dgp}
Figure \ref{fig:single_fit_comparison_simple} summarizes the key finding on a single realization of simulated data from \ref{DGP:Simple}, where we fix $\sigma_x = \sigma_y = 1$. Note how the predicted targets from PTFA align more closely to the true targets when compared to standard PLS. The $R^2$ score value for each is also higher resulting in an average score of 68.1\% for PTFA compared to 56.5\% in standard PLS on this single realization. Figure \ref{fig:rsquared_em_iterations_simple} in the appendix presents the path taken by the values of $R^2$ of the fit as the EM iterations of our algorithm progress. Notice how the algorithm quickly adapts to a large level of explained variance in the targets and levels off once the estimates reach numerical convergence as measured by the $\ell_2$ distance between iterates.

As discussed in the main text and formally shown in Appendix \ref{Sec:Appendix_MLE}, PTFA uses the assumption of isotropic Gaussian noise terms simply to provide a probabilistic framework to targeted factor extraction. Performance of PTFA should therefore not depend on whether feature ($\mathbf{X}$) and target ($\mathbf{Y}$) variables are correlated or even normally distributed. Through this and the next simulation exercises, we show that the relative performance between PTFA and PLS does not depend on the assumed distribution of the variables being decomposed. 

\begin{figure}[!htbp]
    \centering
    \begin{subfigure}{0.49 \textwidth}
        \includegraphics[width = \textwidth]{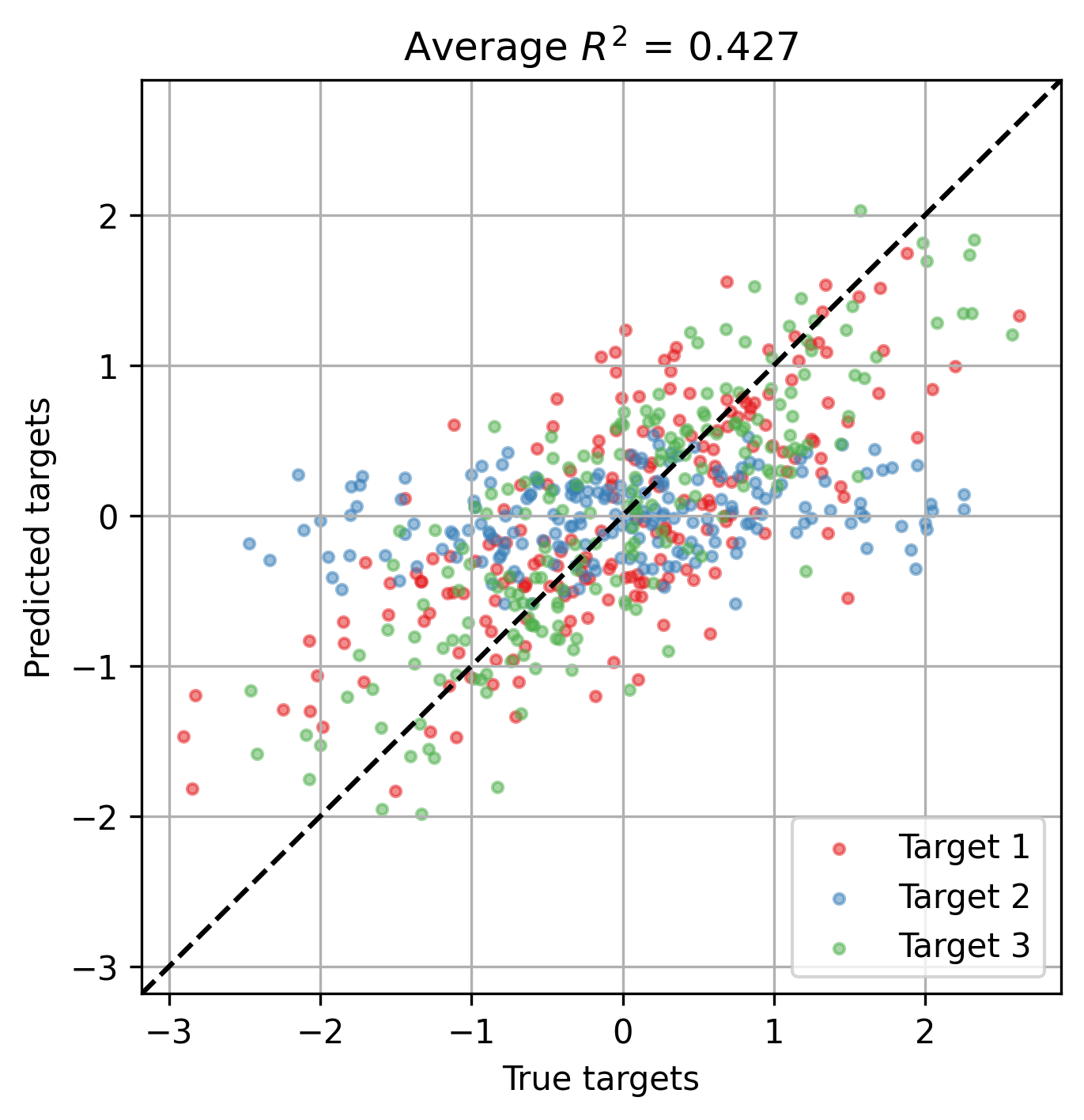}
        \caption{\footnotesize{Partial Least Squares (PLS)}}
    \end{subfigure}
    \begin{subfigure}{0.49 \textwidth}
        \includegraphics[width = \textwidth]{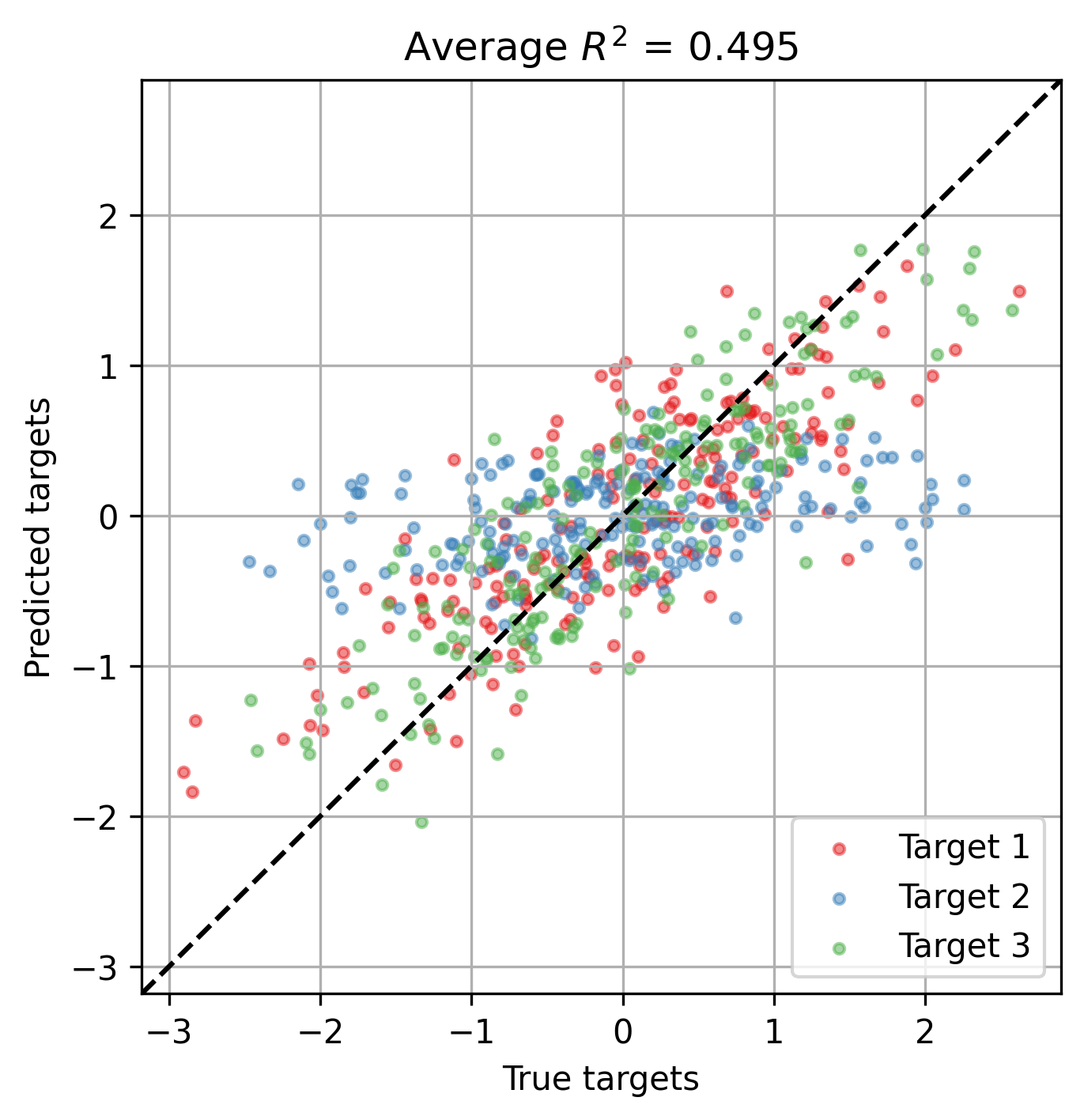}
        \caption{\footnotesize{Probabilistic Targeted Factor Analysis (PTFA)}}
    \end{subfigure}
    \caption{Comparison of PLS and PTFA on a single realization of simulated data with correlated Gaussian errors \eqref{DGP:System}}
    \label{fig:single_fit_comparison_system}
\end{figure}

As a first extension, we dispose of the isotropic assumption and allow for the noises to be multivariate normal distributions with non-diagonal covariance matrices. For this example, we assume the following Toeplitz covariance structure for both features and targets:
\begin{dgp}
    \label{DGP:System}
    \bm{\Omega}_x \coloneqq \begin{bmatrix}
        1 & \rho_x & \cdots & \rho_x^{p-1} \\
        \rho_x & 1 & \cdots & \rho_x^{p-2} \\
        \vdots & \vdots & \ddots & \vdots \\
        \rho_x^{p-1} & \rho_x^{p-2} & \cdots & 1
    \end{bmatrix} \text{ and }
    \bm{\Omega}_y \coloneqq \begin{bmatrix}
        1 & \rho_y & \cdots & \rho_y^{q-1} \\
        \rho_y & 1 & \cdots & \rho_y^{q-2} \\
        \vdots & \vdots & \ddots & \vdots \\
        \rho_y^{q-1} & \rho_x^{q-2} & \cdots & 1
    \end{bmatrix},
\end{dgp}
such that $\bm{e}_x \sim \N(\bm{0}_p, \bm{\Omega}_x)$ and $\bm{e}_y \sim \N(\bm{0}_q, \bm{\Omega}_y)$, where $\rho_x, \rho_y \in [-1, 1]$ are correlation parameters. Figures \ref{fig:single_fit_comparison_system} and \ref{fig:rsquared_em_iterations_system} present the same statistics as before for a realization of data from \ref{DGP:System} using $\rho_x = \rho_y = 0.5$ (keeping the remaining values the same as in the previous exercise). Similar results as in the isotropic Gaussian case are obtained, with PTFA dominating PLS in terms of in-sample fit within only a small number of iterations of the EM algorithm.

\begin{figure}[!htbp]
    \centering
    \begin{subfigure}{0.49 \textwidth}
        \includegraphics[width = \textwidth]{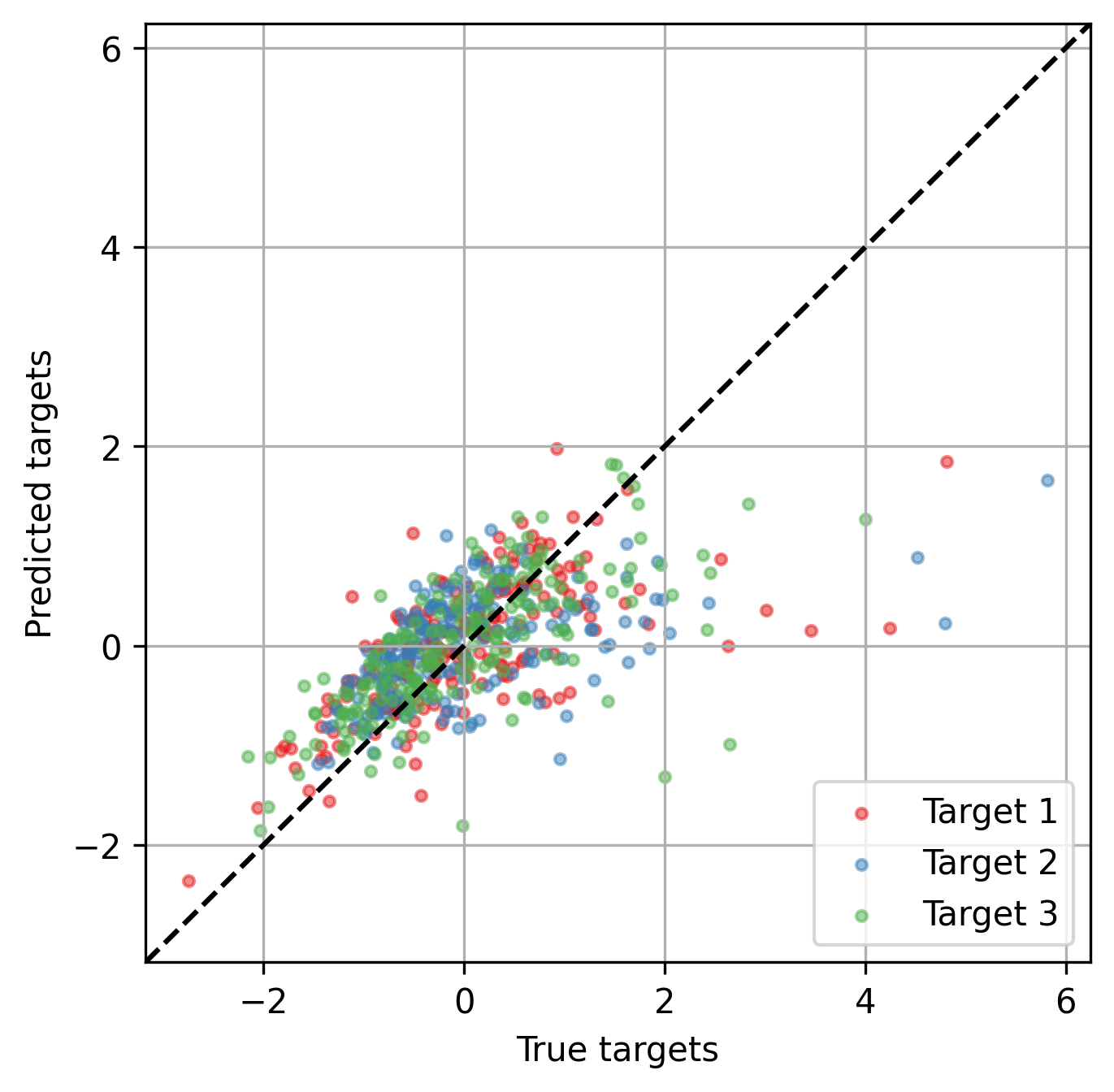}
        \caption{\footnotesize{Partial Least Squares (PLS)}}
    \end{subfigure}
    \begin{subfigure}{0.49 \textwidth}
        \includegraphics[width = \textwidth]{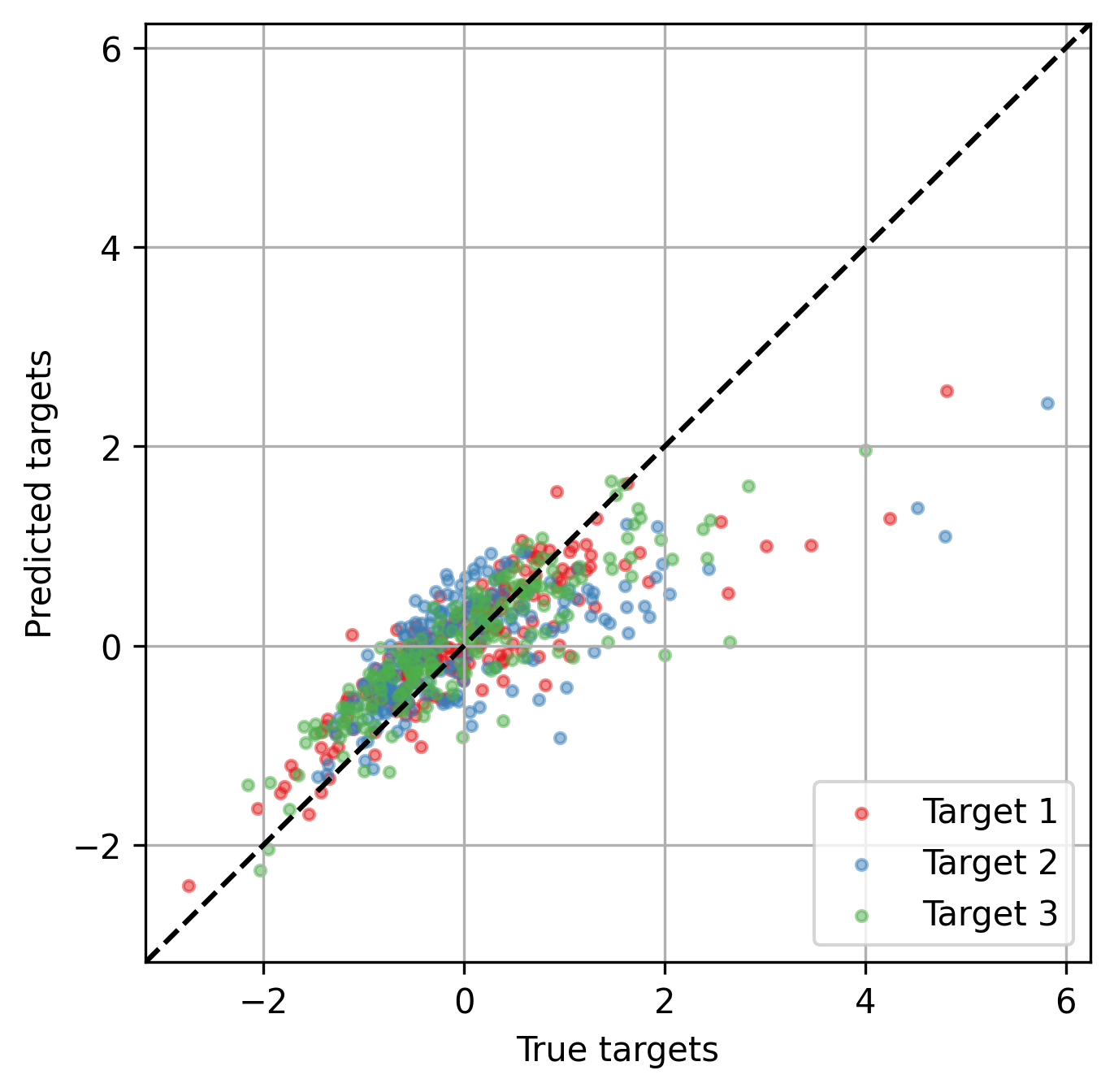}
        \caption{\footnotesize{Probabilistic Targeted Factor Analysis (PTFA)}}
    \end{subfigure}
    \caption{Comparison of PLS and PTFA on a single realization of simulated data with heavy-tailed non-Gaussian errors \eqref{DGP:NonGaussian}}
    \label{fig:single_fit_comparison_nongaussian}
\end{figure}

We present evidence that similar results hold once we dispense the assumption of Gaussian noise altogether. Specifically, we consider the following setup design to produce heavy-tailed and asymmetric noise that results in clear deviations from Gaussian features and targets. Errors in features are drawn independently from a Student-$t$ distributions with 3 degrees-of-freedom and scale $\sigma_x$, while target noise is drawn from a $\chi^2$ distribution with 1 degree of freedom.
\begin{dgp}
    \label{DGP:NonGaussian}
    \begin{aligned}
        \bm{e}_{x, j} & \overset{\textit{iid}}{\sim} \sigma_x \cdot t_3 \, , j \in \{1, \ldots, p\} \quad \text{and} \quad \bm{e}_{y, j} & \overset{\textit{iid}}{\sim} \chi^2_1 \, , j \in \{1, \ldots, q\} \, .
    \end{aligned}
\end{dgp}
Figures \ref{fig:single_fit_comparison_nongaussian} and \ref{fig:rsquared_em_iterations_nongaussian} showcase that similar results to before arise from specification \ref{DGP:NonGaussian}. As long as the data is standardized prior to processing, it can be observed that PTFA will deliver targeted factors that are generally more accurate to summarize the information in the targets regardless of the distributions of the variables involved.

\begin{figure}[!htbp]
    \centering
    \begin{subfigure}[c]{0.49\textwidth}
        \centering
        \includegraphics[width = \textwidth]{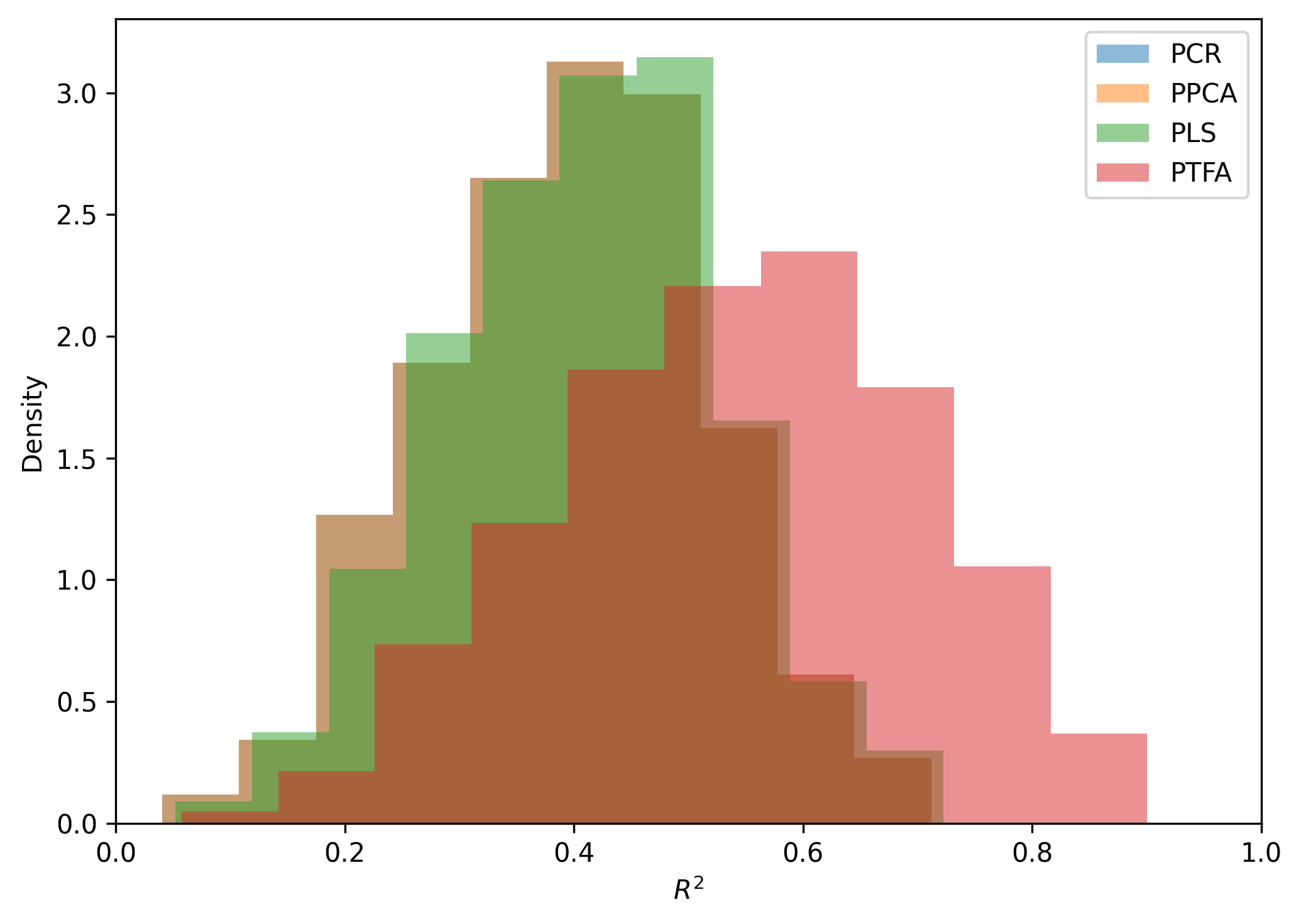}
        \caption{Histogram (10 equal density bins)}
    \end{subfigure}
    \hfill
    \begin{subfigure}[c]{0.49\textwidth}
        \centering
        \includegraphics[width = \textwidth]{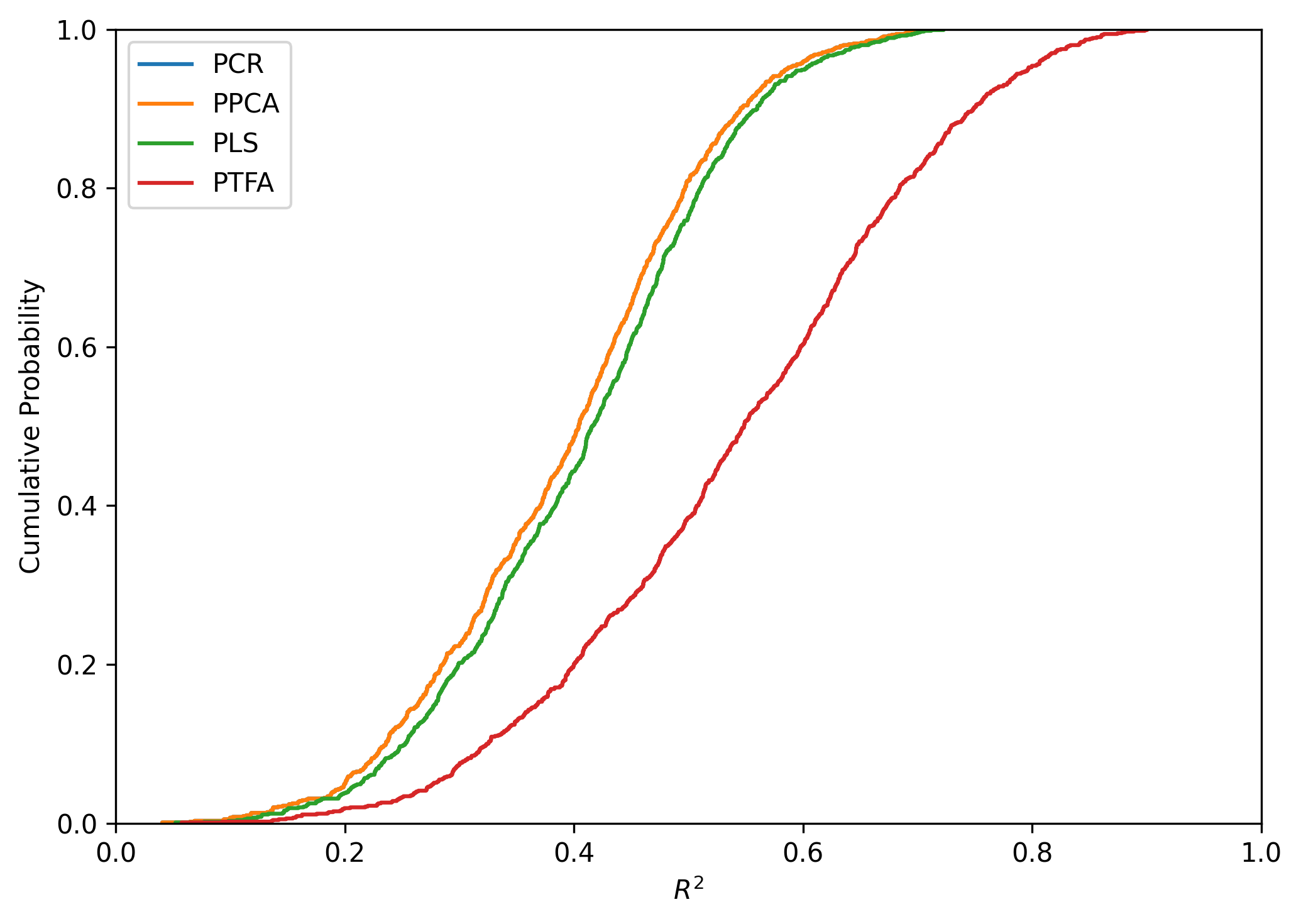}
        \caption{Empirical cumulative distribution (ECDF)}
    \end{subfigure}
    \caption{Comparison of the distributions of average $R^2$ statistics between PLS and PTFA across 1000 replications of \ref{DGP:Simple}}
    \label{fig:rsquared_comparison_distribution}
\end{figure}

Crucially, Figure \ref{fig:rsquared_comparison_distribution} showcases that these performance gains do not depend on any given realization of data. By comparing the average (across targets) $R^2$ statistics over 1000 replications of the previous setting, we find that PTFA first-order stochastic-dominates PLS in generating better in-sample fit. PLS itself dominates the other two unsupervised techniques PCA and PPCA, while they are indistinguishable from each other under the correctly specified \ref{DGP:Simple}.

Finally, given the critical role of noise in explaining the virtue of PLS, we additionally simulate data with differing levels of noise in both features $\mathbf{X}$ and targets $\mathbf{Y}$. That is, once again we simulate noisy data from \ref{DGP:Simple}, adjusting both error scales $\sigma_x$ and $\sigma_y$ over a grid between 0.1 and 5, covering a wide range of signal-to-noise ratios.

\begin{figure}[!htbp]
    \centering
    \begin{subfigure}[c]{0.49\textwidth}
        \centering
        \includegraphics[width = \textwidth]{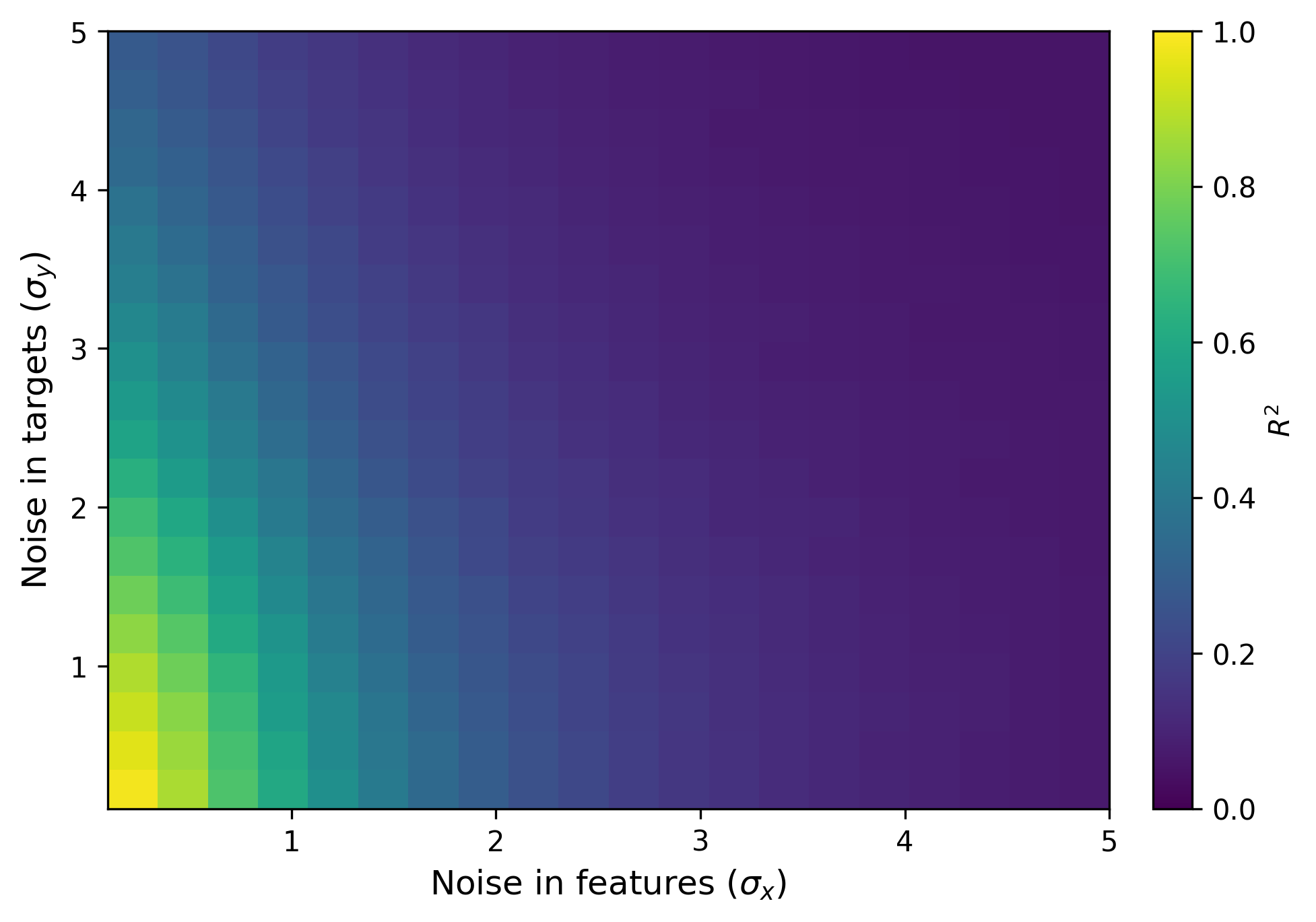}
        \caption{Partial Least Squares (PLS)}
    \end{subfigure}
    \hfill
    \begin{subfigure}[c]{0.49\textwidth}
        \centering
        \includegraphics[width = \textwidth]{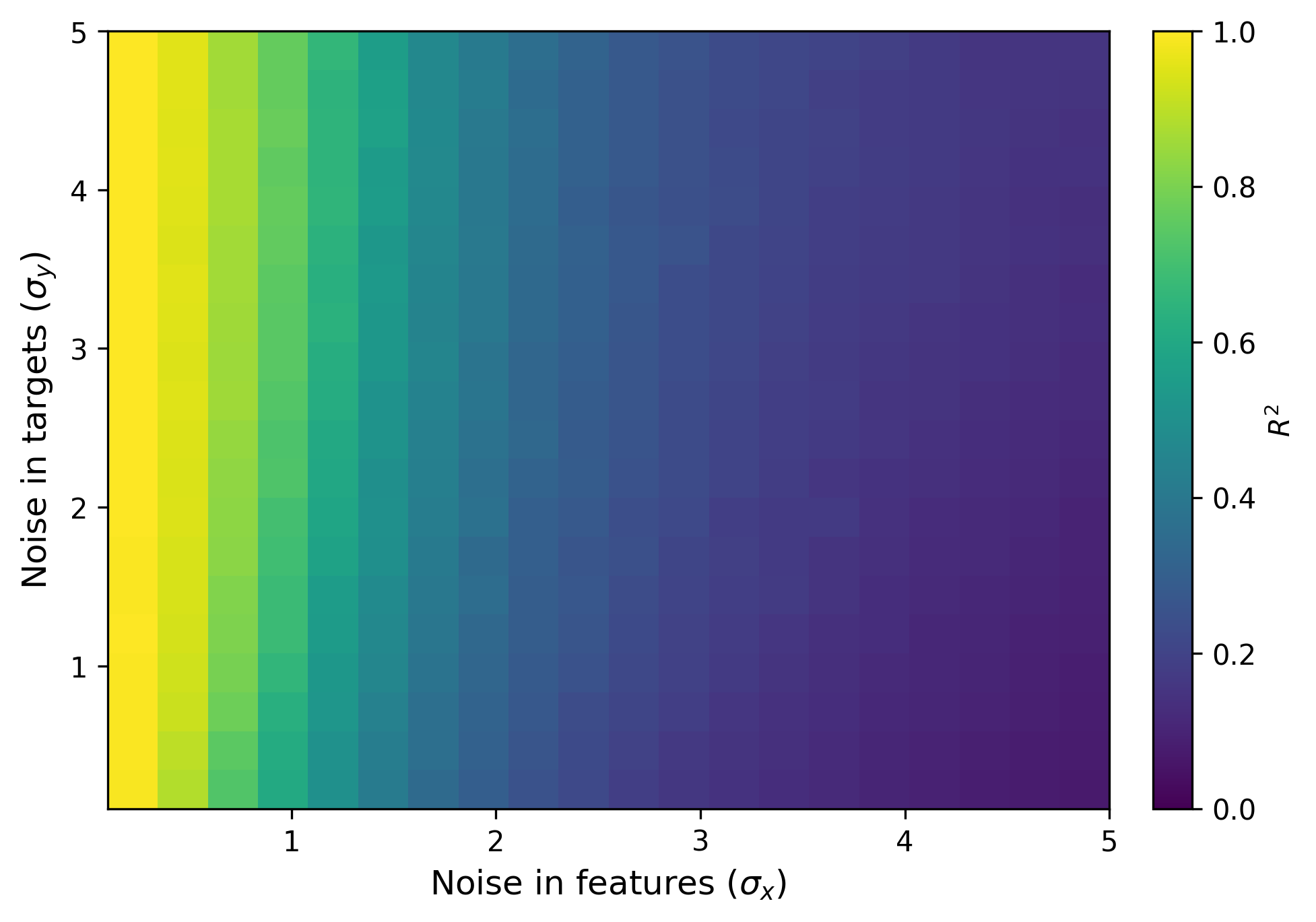}
        \caption{Probabilistic Targeted Factors (PTFA)}
    \end{subfigure}
    \caption{Comparison of the median $R^2$ statistics for PLS (a) and PTFA (b) across 1000 replications from \ref{DGP:Simple}, varying noise in features ($\sigma_x$) and targets ($\sigma_y$)}
    \label{fig:noise_comparison}
\end{figure}

Figure \ref{fig:noise_comparison} shows the median value of average $R^2$ statistics across targets over 1000 replications of this simulation setup. The superior performance of PTFA is evident from the heatmap. The gains in terms of goodness-of-fit of PTFA when compared to PLS are more salient when noise is increases, in particular when the noise is in the targets instead of the features. This is as seen in the PCA case, where perturbations to the data in the form of noise or outliers creates issues for consistently recovering the axes of maximal variance \citep{chen_spectral_2021}.

%% file: Sections/Application.tex
\subsection{Targeted Financial Conditions Indices}

Financial conditions indices (FCIs) offer a single, quantitative measure summarizing the state of financial conditions in the economy by combining information from a wide range of variables --- including credit spreads, asset prices, liquidity measures, and volatility indicators. These indices have become essential tools for policymakers, especially since the Global Financial Crisis, as they help to gauge the overall tightness or looseness of financial conditions and, in turn, provide early signals of potential risks to the real economy \citep{adrian_vulnerable_2019}.

Traditional approaches to constructing FCIs typically rely on principal component analysis (PCA) or its variants applied to balanced panels of data \citep[see for instance][]{hatzius_financial_2010, brave_monitoring_2011, koop_new_2014} and extract one single factor ($k = 1$) to match the idea of an index. However, high-frequency financial datasets are often incomplete due to mismatched reporting frequencies, data lags, or sporadic data availability. This incompleteness can lead to noisy and less reliable indices when using standard techniques. For instance, least-squares PCA methods may overfit or produce overly volatile indices in the presence of substantial missing data. Nevertheless, the main critique to current FCIs is related to their identification and interpretability. Since FCIs are constructed as unsupervised common components of a large number of financial time-series, their signal and relevance to future economic developments are unclear and hard to learn. Poor identification of the single factor that proxies an FCI has led some authors to extract two factors to fully characterize the state of financial conditions in the economy \citep{Lombardi2025}. 

To address these challenges, we propose constructing a targeted FCI. Unlike PCA-based FCIs, which summarize variation in financial variables without regard to a macro target, our targeted FCI extracts the component of financial conditions that is most informative about a chosen target $\textbf{Y}$. In essence, the targeted FCI aims to provide a cleaner and more timely measure of financial conditions by filtering out information irrelevant to the target of interest. This directly addresses the identification and interpretability critique of traditional FCIs. For instance, if we define $\textbf{Y}$ to be GDP growth, we are effectively looking for a FCI relevant in explaining economic growth. Whereas, if we target inflation, we would effectively be constructing a FCI that could signal inflation risk through time.

Therefore, our approach allows the econometrician to construct as many different FCIs as the number of targets they may wish to define, from the same pool of financial variables. This approach can significantly enhance the usefulness of the index for real-time monitoring and forecasting, ultimately supporting better-informed monetary policy decisions. To implement such an approach, we employ our Targeted Dynamic Factor Model (PTFA-DFM), which allows for mixed-frequency data and a more robust treatment of missing data in general, while enabling the weights to depend on the target variable.

We begin by collecting a large number of financial time-series, reported at monthly frequency as summarized below in Table \ref{FCI_data}. The first three variables listed are candidate targets, which can be studied in isolation or jointly as multivariate objects. The $p = 16$ variables outlined in rows 4 to 19 are used to define $\textbf{X}$ and represent a cross-section of proxy financial conditions indicators such as the VIX which captures sentiment in financial markets, the return of the S\&P 500 index, a series of credit spreads, which have been found to contain relevant information to predict real economic development, among others.  

\begin{figure}[!htbp]
  \centering
  \begin{subfigure}[t]{0.95\textwidth}
    \centering
    \includegraphics[width=\linewidth]{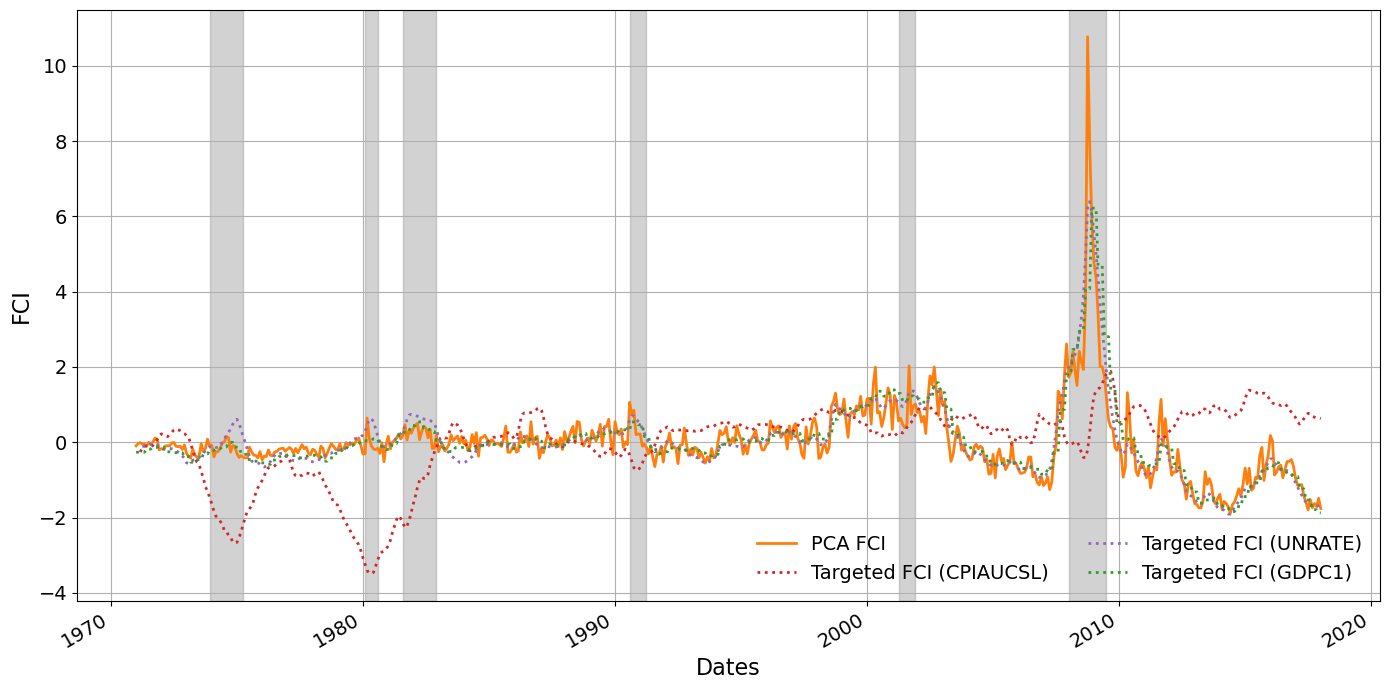}
  \end{subfigure}

  \vspace{1em} 

  \caption{Targeted Financial Conditions}
  \label{FCI_all}
  \vspace{0.5mm}
  \begin{minipage}{0.95\textwidth}
    \footnotesize
    \textbf{Notes:} The targeted FCI refers to the common component estimated with a targeted Dynamic Factor Model. The figure contrasts targeted FCIs with different univariate target inputs and signal extraction method.The red, green, and purple lines target unemployment rate, GDP growth, and CPI inflation, respectively. The solid orange line exhibits an FCI computed with PCA and therefore no target.
  \end{minipage}

\end{figure}

Next, we fit our targeted Dynamic Factor Model to this data to examine the host of potential FCIs one can construct. Figure \ref{FCI_all} plots different Targeted FCIs, which differ with respect to the variables they target. The main result is that the target choice matters: the GDP and unemployment targeted FCIs are the closest to the PCA counterpart and the most volatile, reacting strongly around growth slowdowns, while the inflation-targeted FCI is more muted and its dynamics are more pronounced in later stages of recessions rather than at mature stages of the business cycle.

An interesting result emerges from the inspection of the loadings of each financial variables onto the alternative FCIs we consider. Figure \ref{loadings} describes the most influential variables per FCI. We can observe that credit-related variables such as the treasury bill spread, debt-to-GDP ratio, or the mortgage spread load more heavily on the inflation targeted FCI. In contrast, market volatility (VIX) and equity returns (S\&P500) weigh more heavily in the GDP and unemployment targeted indices. It is apparent that Financial Conditions Indices are sensitive to their targets and their dynamics vary accordingly.

\subsection{Macroeconomic Forecasting}

Partial Least Squares may be valuable when the key motivation for dimensionality reduction is prediction. To demonstrate the practical relevance of PTFA, we conduct a simple macroeconomic forecasting exercise using the Federal Reserve Economic Data -- Monthly Database \citep[FRED-MD; for details, see,][]{mccracken_fred-md_2016}. The data includes $p = 126$ variables that track macroeconomic developments in the United States at a monthly frequency. In this section, we discuss the main results and compare PTFA with popular alternatives in the literature.

\begin{table}[!htbp]
\centering
\begin{tabular}{lcccccc}
\toprule
\textbf{Horizon} & \textbf{PLS} & \textbf{PTFA} & \textbf{PTFA-SV} & \textbf{PTFA-DFM} & \textbf{PPCA} & \textbf{PCA} \\
\midrule
\multicolumn{7}{c}{\textbf{Industrial Production}} \\
1     & 0.8536 & 0.7228 & 0.8441 & 0.7172 & 0.7397 & 0.7397 \\
6     & 1.3816 & 0.9106 & 1.0268 & 0.9090 & 0.9172 & 0.9172 \\
12    & 1.4917 & 0.9296 & 1.0585 & 0.9177 & 0.9459 & 0.9459 \\
\midrule
\multicolumn{7}{c}{\textbf{CPI Inflation}} \\
1     & 1.1560 & 1.0380 & 1.0246 & 1.0373 & 1.0503 & 1.0503 \\
6     & 1.8003 & 1.0173 & 1.1452 & 1.0162 & 1.0271 & 1.0271 \\
12    & 1.8119 & 1.0095 & 1.1076 & 1.0159 & 1.0225 & 1.0225 \\
\midrule
\multicolumn{7}{c}{\textbf{Unemployment Rate}} \\
1     & 1.0654 & 0.9182 & 0.9319 & 0.9094 & 0.9227 & 0.9227 \\
6     & 1.3985 & 1.0392 & 1.0996 & 1.0342 & 1.0638 & 1.0638 \\
12    & 1.6456 & 1.0215 & 1.0852 & 1.0132 & 1.0448 & 1.0448 \\
\bottomrule
\end{tabular}
\caption{Out-of-Sample Forecast Performance Across Models}
\label{Tab:Forecast_Performance}
\caption*{\small \textbf{Notes:} Mean Squared Forecast Errors (MSFE) are calculated out-of-sample using a rolling window of 180 monthly observations for forecast horizons of 1, 6, and 12 months, with the sample spanning 1961M7--2023M3. PTFA, PTFA-SV and PTFA-DFM, correspond to probabilistic targeted factor models with static, stochastic volatility, dynamic factor extensions, respectively. PPCA and PCA (Factors) correspond to probabilistic and standard principal component factor models. All targets and predictors are standardized prior to estimation.}
\end{table}

Table \ref{Tab:Forecast_Performance} shows out-of-sample (OOS) Mean Squared Forecast Error (MSFE) of prediction for each target variable using $k = 7$ factors extracted with either PLS, PTFA, PTFA-DFM, PTFA-SV, PPCA or PCA. The targets in our exercise are Industrial Production, CPI inflation and the Unemployment Rate (such that $q=1$ for each panel). The number of factors $k$ is chosen to be consistent with the FRED-MD factors calculated according to \cite{mccracken_fred-md_2016}, with code made available by the authors. The key message from Table \ref{Tab:Forecast_Performance} is that the PTFA outperforms both PLS and PCA, adding value to forecasts across forecast horizons and macroeconomic variables considered. PTFA delivers systematic OOS gains by extracting factors that maximize predictive covariance with targets, down-weighting directions in predictors that inflate variance but do not aid prediction. This mechanism mirrors our Monte Carlo exercise, discussed in previous section, where improvements intensify as the noise in targets rise. We note that allowing for stochastic volatility (PTFA-SV) does not materially improve macro forecasts over the baseline PTFA except for the case of CPI inflation for $h=1$, suggesting the gains stem primarily from targeted factor extraction rather than time-varying error dynamics. We also note that the Dynamic Factor extension of PTFA performs quite well, outperforming competitors for many combinations of forecast horizons $h$ and target variables.

\subsection{Predicting the Equity Premium}

Attempts to predict stock returns or the equity premium are in no short supply in the Finance literature. \cite{welch_comprehensive_2007} and \cite{goyal_comprehensive_2024}, provide a review and comprehensive assessment of the performance of 46 different variables that have been suggested by the academic literature to be good predictors of the equity premium. Following this large body of empirical work, our financial application studies the predictability of U.S. aggregate stock returns, using the \cite{goyal_comprehensive_2024} dataset.

The goal of this exercise is to predict the equity risk premia, using $p = 26$ signals that are available at a monthly frequency. Thus, in this case $\mathbf{Y}$ is the equity premium and $\mathbf{X}$ are the various predictors, lagged by one period, following standard practice. In this setting, it is less clear how many factors $k$ should be considered. Therefore, we estimate PTFA and all competing models for different values of $k$ and report the forecasting performance results in Table \ref{Tab:Forecast_Performance_New}. As for the forecasting horizon, we only consider 1 month ahead forecasts given the nature of the problem of forecasting stock returns, since information is priced-in quite fast.

\begin{table}[!htbp]
\centering
\begin{tabular}{lccc}
\toprule
\textbf{Model} & $\bm{k=1}$ & $\bm{k=2}$ & $\bm{k=3}$ \\
\midrule
\textbf{PLS}          & 0.7941 & 0.8624 & 0.9253 \\
\textbf{PTFA}         & 0.7329 & 0.7508 & 0.7476 \\
\textbf{PTFA-SV}      & 0.7691 & 0.8746 & 0.9150 \\
\textbf{PTFA-DFM}     & 0.7319 & 0.7425 & 0.7411 \\
\textbf{PPCA}         & 0.7562 & 0.7570 & 0.7569 \\
\textbf{PCA} & 0.7528 & 0.7492 & 0.7553 \\
\bottomrule
\end{tabular}
\caption{Out-of-Sample Forecast Performance Across Factor Models}
\label{Tab:Forecast_Performance_New}
\caption*{\small \textbf{Notes:} Mean Squared Forecast Error (MSFE) statistics are calculated out-of-sample using a rolling window of 120 monthly observations on the sample 1926M1--2023M12. Results are shown for different numbers of factors $k = 1, 2, 3$ across probabilistic targeted factor models (PTFM, PTFM-SV, PTFM-DFM), partial least squares (PLS), and principal component-based models (PPCA and PCA). Both the equity premium and predictor variables \citep[26 monthly signals from][]{goyal_comprehensive_2024} are standardized prior to estimation.}
\end{table}

The main message from Table \ref{Tab:Forecast_Performance_New} is that PTFA and extensions add value as compared to PLS and PCA in predicting the equity risk premia. Similar to our application with macroeconomic data, our model with stochastic volatility (PTFA-SV) does not seem to outperform PTFA, that is relatively more parsimonious.  We observe that MSFE associated to PTFA forecasts are quite competitive, regardless of how many factors one chooses. PTFA-DFM delivers uniformly lower MSFEs than PCA and PLS across different choices of $k$ with the tightest gains for $k=1$. As $k$ increases, OOS MSFE rise for all models, reflecting higher variance from estimating weakly identifiable directions in high-noise, short-sample settings. The one-factor PTFA solution concentrates on the most stable predictor–target covariance direction and avoids the overfitting that afflicts multi-component PLS, a clear advantage of PTFA.

%% file: Sections/Conclusion.tex
We introduce a probabilistic framework for targeted factor extraction called PTFA and derive a fast Expectation-Maximization (EM) algorithm to estimate the model. PTFA is flexible and naturally handles parameter uncertainty, noise, and missing data in estimation.  Through simulation exercises and three real-world applications in macroeconomic forecasting, equity premium prediction, and the construction of financial conditions indices, we demonstrate the superior performance of PTFA, especially in noisy and incomplete data environments. Along the way, we provide additional contributions to mixed-frequency data, stochastic volatility, time-series persistence in the latent factors, and give further theoretical insight to the probabilistic PLS solutions.

Our probabilistic foundation also opens many avenues for future research, including interesting methodological extensions using probabilistic (fully Bayesian) or variational inference. By providing an open-source implementation of the method, our hope is that practitioners of time-series forecasting and researchers alike will continue to expand and improve upon the technique.


%% file: Sections/Appendix_MLE.tex
Our PTFA is formulated to provide a probabilistic foundation to PLS. In the body of the paper we propose an iterative procedure to estimate the loadings and data variances associated to our model. In this section we show that a maximum likelihood estimator (MLE) of these quantities reproduces the standard PLS solution. This is akin to the results by \cite{tipping_probabilistic_1999} who propose their probabilistic version of PCA and show the same axes of maximal variance are obtained from a MLE of their model. Therefore, our setup also allows us to recover the frequentist properties of estimators for factor loadings and data variances. Related derivations from the point of view of canonical correlation analysis (CCA) can be found in \cite{bach2005probabilistic}.

Recall from \eqref{Eq:Conditional_Likelihood} that the distribution of one realization of the data conditional on the factors is Gaussian with mean vector $\bm{\mu} \coloneqq [\bm{f}\T \mathbf{P}\T, \bm{f}\T \mathbf{Q}\T]\T$ and covariance matrix $\bm{\Sigma} \coloneqq \diag(\sigma^2_x \mathbf{I}_p, \sigma^2_y \mathbf{I}_q)$. Recall we have also stacked the data and loadings into $\mathbf{Z} = [\mathbf{X}, \mathbf{Y}]$ and $\mathbf{L} = [\mathbf{P}\T, \mathbf{Q}\T]\T$, respectively. Collecting all disturbances $\bm{e}_t \coloneqq [\bm{e}_{x, t}\T, \bm{e}_{y, t}\T]$ and stacking into $\mathbf{E} \coloneqq [\bm{e}_1\T, \ldots, \bm{e}_n\T]$, we can express our model equations \eqref{Eq:Feature_Equation} and \eqref{Eq:Target_Equation} succinctly as
\begin{align}
    \label{Eq:Stacked_Feature_Target}
    \mathbf{Z} = \mathbf{F}\mathbf{L}\T + \mathbf{E} \, .
\end{align}
Integrating out the factors according to their $\N(\bm{0}_k, \mathbf{V}_F)$ prior distribution results in the marginal log-likelihood of the data as a function of the loadings, denoted as $\ell(\theta)$. Write $\mathbf{S} \coloneqq n^{-1}  \mathbf{Z}\T \mathbf{Z}$ for the sample data covariance (the columns of $\mathbf{X}$ and $\mathbf{Y}$ are standardized independently) and $\mathbf{C} \coloneqq \mathbf{L} \mathbf{V}_F \mathbf{L}\T + \bm{\Sigma}$ for the model variance. The log-likelihood can then be expressed as:
\begin{align}
    \label{Eq:Marginal_Likelihood}
    \ell(\theta) & = -\frac{T}{2} \left[ d \log(2\pi) + \log |\mathbf{C}| + \Tr(\mathbf{C}^{-1} \mathbf{S} ) \right] \, .
\end{align}
The MLE of $\theta$, denoted as $\widehat{\theta} = (\widehat{\mathbf{P}}, \widehat{\mathbf{Q}}, \widehat{\sigma}_x^2, \widehat{\sigma}_y^2)$, is therefore given as the solution to
\begin{equation}
    \label{Eq:Maximum_Likelihood_Estimator}
    \widehat{\theta} \coloneqq \arg \max_\theta \ell(\theta) \, .
\end{equation}
Note that the assumption of uncorrelated Gaussian errors is made for convenience as it results in a particularly simple likelihood structure. If we assume the factor decomposition to be correctly specified ---i.e., that there exists an $\mathbf{L} \in \mathbb{R}^{d \times k}$ such that $\E[\mathbf{Z} \mid \mathbf{F}] = \mathbf{F}\mathbf{L}\T$) --- then, under regularity conditions, our estimator will remain consistent even if the data is subject to other kind of more complex error processes \citep[due to standard quasi-maximum likelihood results; see for example][]{gong_pseudo_1981, gourieroux_pseudo_1984}.

While we use the maximum likelihood moniker following earlier work by \citet{tipping_probabilistic_1999}, $\widehat{\theta}$ is technically a \emph{maximum a-posteriori} (MAP) estimator as it can depend on the assumed factor prior variance $\mathbf{V}_F$. For the remainder of the derivation and to economize on notation we set $\mathbf{V}_F = \mathbf{I}_k$, as it also results in the canonical PLS formulation \citep[for more details, see][]{frank_statistical_1993, hastie_elements_2001}.

The parameters $\theta$ enter the likelihood only through $\mathbf{C}$, with a gradient equal to
\begin{equation}
    \label{Eq:Variance_Gradient}
    \frac{\partial \ell(\theta)}{\partial \mathbf{C}} = -\frac{n}{2} \left( \mathbf{C}^{-1} - \mathbf{C}^{-1} \mathbf{S} \mathbf{C}^{-1} \right) \, .
\end{equation}
Writing $\widehat{\mathbf{L}} = [\widehat{\mathbf{P}}\T, \widehat{\mathbf{Q}}\T]\T$, $\widehat{\bm{\Sigma}} \coloneqq \diag(\widehat{\sigma}^2_x \mathbf{I}_p, \widehat{\sigma}^2_y \mathbf{I}_q)$, and $\widehat{\mathbf{C}} \coloneqq \widehat{\mathbf{L}} \widehat{\mathbf{L}}\T + \widehat{\bm{\Sigma}}$, this means that the maximum likelihood solutions for the loadings $\mathbf{L}$ are characterized by
\begin{equation}
    \label{Eq:FOC_MLE_Stacked}
    \bm{0}_{d \times k} = \left[ \frac{\partial \ell(\widehat{\theta})}{\partial \mathbf{C}} \right]\T \frac{\partial \ell(\widehat{\theta})}{\partial \mathbf{L}} \implies (\mathbf{S} \widehat{\mathbf{C}}^{-1} - \mathbf{I}_d) \begin{bmatrix} \widehat{\mathbf{P}} \\ \widehat{\mathbf{Q}} \end{bmatrix} = \bm{0}_{d \times k} \, .
\end{equation}
The first-order condition \eqref{Eq:FOC_MLE_Stacked} exhibits the same three classes of solutions as explored by \cite{tipping_probabilistic_1999}, plus an additional interesting special case. First, the trivial solution sets $\widehat{\mathbf{L}} = \bm{0}_{d \times k}$, which represents a minimum of the log-likelihood $\ell(\theta)$. Second, we obtain a solution if we assume our implied model variance to equal the data variance, such that $\widehat{\mathbf{C}} = \mathbf{S}$. Letting $\mathbf{S}_X$, $\mathbf{S}_Y$ and $\mathbf{S}_{XY}$ denote the blocks of $\mathbf{S}$ partitioned according to features and targets, we can then identify the components $\widehat{\mathbf{P}}$ and $\widehat{\mathbf{Q}}$ from the set of equations given by
\begin{align}
    \label{Eq:Loadings_FOC_TrueVariance}
    \begin{aligned}
        \widehat{\mathbf{P}} \widehat{\mathbf{P}}\T & = \mathbf{S}_X - \widehat{\sigma}_x^2 \mathbf{I}_p \, , \\
    \widehat{\mathbf{Q}} \widehat{\mathbf{Q}}\T & = \mathbf{S}_Y - \widehat{\sigma}_y^2 \mathbf{I}_q \, ,\\
    \widehat{\mathbf{P}} \widehat{\mathbf{Q}}\T & = \mathbf{S}_{XY} \, .
    \end{aligned}
\end{align}
Note that equations \eqref{Eq:Loadings_FOC_TrueVariance} require the last $p - k$ eigenvalues of $\mathbf{S}_X$ to be equal to each other, and similarly for the last $q - k$ eigenvalues of $\mathbf{S}_Y$. In this case, $\widehat{\mathbf{P}}$ is constructed from the eigenvectors of $\mathbf{S}_X$, $\widehat{\mathbf{Q}}$ from the eigenvectors of $\mathbf{S}_Y$, and the components are rotated to ensure $\widehat{\mathbf{P}} \widehat{\mathbf{Q}}\T = \mathbf{S}_{XY}$.

An explicit construction for the solutions in this case can be provided as follows. Let $\mathbf{X}_k \coloneqq \mathbf{U}_X \mathbf{D}_X \mathbf{V}_X\T$ and $\mathbf{Y}_k \coloneqq \mathbf{U}_Y \mathbf{D}_Y \mathbf{V}_Y\T$ be the best rank $k$ approximations to the data matrices $\mathbf{X}$ and $\mathbf{Y}$, respectively. Therefore, $\mathbf{U}_X$ and $\mathbf{U}_Y$ are $T \times k$ matrices with orthogonal columns containing the left singular vectors, $\mathbf{V}_X$ and $\mathbf{V}_Y$ are $k \times k$ orthogonal matrices containing the right singular vectors, and $\mathbf{D}_X$ and $\mathbf{D}_Y$ are $k \times k$ diagonal matrices holding the largest $k$ singular values for features $\mathbf{X}$ and targets $\mathbf{Y}$, respectively. One can then check the following solutions satisfy $\eqref{Eq:Loadings_FOC_TrueVariance}$:
\begin{align}
    \label{Eq:Loadings_Solutions_TrueVariance}
    \begin{aligned}
        \widehat{\mathbf{P}} & = \mathbf{V}_X \left(\frac{1}{T} \mathbf{D}_X^2 - \widehat{\sigma}_x^2 \mathbf{I}_k \right)^{1/2} \mathbf{V}_P\T \, , \\
        \widehat{\mathbf{Q}} & = \mathbf{V}_Y \left(\frac{1}{T} \mathbf{D}_Y^2 - \widehat{\sigma}_y^2 \mathbf{I}_k \right)^{1/2} \mathbf{V}_Q\T \, , \\
        \mathbf{V}_P & = \mathbf{V}_Q \left(\mathbf{D}_Y^2 - T \cdot \widehat{\sigma}_y^2 \mathbf{I}_k \right)^{-1/2} \mathbf{D}_Y \mathbf{U}_Y\T \mathbf{U}_X \mathbf{D}_X \left(\mathbf{D}_X^2 - T \cdot \widehat{\sigma}_x^2 \mathbf{I}_k \right)^{-1/2}
    \end{aligned}
\end{align}
This solution leaves $\mathbf{V}_Q$ as an arbitrary $k \times k$ orthogonal matrix, and sets $\widehat{\sigma}_x^2$ equal to the smallest eigenvalue of $\mathbf{S}_X$ and $\widehat{\sigma}_y^2$ equal to the smallest eigenvalue of $\mathbf{S}_Y$ (recall the trailing eigenvalues after the $k$-th one are assumed equal for both $\mathbf{S}_X$ and $\mathbf{S}_X$). Note that this solution corresponds to extracting factors for $\mathbf{X}$ and $\mathbf{Y}$ independently using PCA, and then rotating the principal axes of the feature loadings according to the weighted covariance $\mathbf{S}_{XY}$ between features and targets.

The third class of solutions represents the most interesting case likely to be encountered in practice, where \eqref{Eq:FOC_MLE_Stacked} is satisfied but $\widehat{\mathbf{C}} \neq \mathbf{S}$ so that our covariance model is misspecified. That is, we recognize that the isotropic Gaussian error terms in \eqref{Eq:Feature_Equation}--\eqref{Eq:Target_Equation} are not to be taken as true modeling choices for the features nor targets, and are rather used as a working assumption to obtain a framework for probabilistic targeted factor recovery.

Define the marginal covariance matrices implied from the model as $\widehat{\mathbf{C}}_X \coloneqq \widehat{\mathbf{P}} \widehat{\mathbf{P}}\T + \widehat{\sigma}_x^2 \mathbf{I}_p$ and $\widehat{\mathbf{C}}_Y \coloneqq \widehat{\mathbf{Q}} \widehat{\mathbf{Q}}\T + \widehat{\sigma}_y^2 \mathbf{I}_q$. Additionally, define the conditional covariance of $\mathbf{Y}$ given $\mathbf{X}$ implied from the model as $\widehat{\mathbf{C}}_{Y \mid X} \coloneqq \widehat{\mathbf{C}}_Y - \widehat{\mathbf{Q}} \widehat{\mathbf{P}}\T \widehat{\mathbf{C}}_X^{-1} \widehat{\mathbf{P}} \widehat{\mathbf{Q}}\T$. Using these definitions, the system of equations \eqref{Eq:FOC_MLE_Stacked} can be expressed as
\begin{align}
    \label{Eq:FOC_MLE_Unstacked_PQ}
    \begin{aligned}
        (\mathbf{S}_X \widehat{\mathbf{C}}_X^{-1} - \mathbf{I}_p) \widehat{\mathbf{P}} + \frac{1}{T} \mathbf{X}\T (\mathbf{Y} - \mathbf{X} \widehat{\mathbf{C}}_X^{-1} \widehat{\mathbf{P}} \widehat{\mathbf{Q}}\T) \widehat{\mathbf{C}}_{Y \mid X}^{-1} \widehat{\mathbf{Q}} (\mathbf{I}_k - \widehat{\mathbf{P}}\T \widehat{\mathbf{C}}_X^{-1} \widehat{\mathbf{P}}) & = \bm{0}_{p \times k} \\
        (\mathbf{S}_{XY}\T \widehat{\mathbf{C}}_X^{-1} \widehat{\mathbf{P}} - \widehat{\mathbf{Q}}) + \frac{1}{T} \mathbf{Y}\T (\mathbf{Y} - \mathbf{X} \widehat{\mathbf{C}}_X^{-1} \widehat{\mathbf{P}} \widehat{\mathbf{Q}}\T) \widehat{\mathbf{C}}_{Y \mid X}^{-1} \widehat{\mathbf{Q}} (\mathbf{I}_k - \widehat{\mathbf{P}}\T \widehat{\mathbf{C}}_X^{-1} \widehat{\mathbf{P}}) & =  \bm{0}_{q \times k}
    \end{aligned}
\end{align}
If the loadings $\widehat{\mathbf{P}}$ and $\widehat{\mathbf{Q}}$ are solutions to \eqref{Eq:FOC_MLE_Stacked}, and recalling $\widehat{\mathbf{C}} \neq \mathbf{S}$ holds, \eqref{Eq:FOC_MLE_Unstacked_PQ} showcases these loadings are equivalently determined as the solution to the simpler system
\begin{align}
    \label{Eq:Loadings_Solutions_General_P} \widehat{\mathbf{P}} & = \mathbf{S}_X \widehat{\mathbf{C}}_X^{-1} \widehat{\mathbf{P}} \\
    \label{Eq:Loadings_Solutions_General_Q} \widehat{\mathbf{Q}} & = \mathbf{S}_{XY}\T \widehat{\mathbf{C}}_X^{-1} \widehat{\mathbf{P}} \\
    \label{Eq:Loadings_Solutions_General_PLS} \mathbf{Y} & = \mathbf{X} \widehat{\mathbf{C}}_X^{-1} \widehat{\mathbf{P}} \widehat{\mathbf{Q}}\T
\end{align}
Equations \eqref{Eq:Loadings_Solutions_General_P} and \eqref{Eq:Loadings_Solutions_General_Q} jointly imply that the columns of $\widehat{\mathbf{P}}$ take into account information from the eigenvectors of the feature variance $\mathbf{S}_X$ and covariance matrix $\mathbf{S}_{XY}$. Finally, equation \eqref{Eq:Loadings_Solutions_General_PLS} provides the explicit decomposition satisfied by the estimated loadings $\widehat{\mathbf{P}}$ and $\widehat{\mathbf{Q}}$. This is akin to the prediction produced by the canonical PLS setting with multiple targets. These results showcase that our PTFA framework indeed provides a probabilistic foundation for standard PLS as it reproduces the spirit of the non-random solution from the NIPALS algorithm.

Finally, we note one more special case of interest that does not have an analogue to the PPCA framework in \cite{tipping_probabilistic_1999}. Specifically, this occurs when one is interested in targeted recovery of factors for a set of response variables with a \emph{smaller} number of variables than the suspected number of components, such that $q \leq k$. In this case, the column space of $\mathbf{Y}$ can be spanned by linear combinations of the $k$ scores without loss of generality (i.e., as long as they are orthogonal and non-zero). We can then directly assume $\widehat{\mathbf{C}}_Y = \mathbf{S}_Y$. In this case, loadings $\widehat{\mathbf{Q}}$ can be assumed orthogonal and can be recovered using PCA. However, we should still allow for $\widehat{\mathbf{C}}_X \neq \mathbf{S}_X$. In this case, the solution is a hybrid solution between \eqref{Eq:Loadings_Solutions_TrueVariance} and \eqref{Eq:Loadings_Solutions_General_P}--\eqref{Eq:Loadings_Solutions_General_PLS}.

\newpage

%% file: Sections/Appendix_Implementation.tex
We provide a fast and simple  expectation-maximization (EM) solution to efficiently learn the parameters of our PTFA formulation, collected into $\theta = (\mathbf{P}, \mathbf{Q}, \sigma_x^2, \sigma_y^2)$. We assume we have access to a random sample $\{\bm{x}_t, \bm{y}_t\}_{t=1}^{T}$, collected into the matrices $\mathbf{X} \in \mathbb{R}^{T \times p}$ and $\mathbf{Y} \in \mathbb{R}^{T \times q}$. Similarly, we assume factors $\{\bm{f}_t\}_{t=1}^{T}$ are collected into a matrix $\mathbf{F} \in \mathbb{R}^{T \times k}$.

The independence across rows of factor components in matrix $\mathbf{F}$ is translated into a prior $\bm{f}_t \sim p(\bm{f})$ independently across observations $t = 1, \ldots, T$. Letting $\vecv(\mathbf{F})$ be the column-vectorized version of $\mathbf{F}$ and recalling $p(\bm{f}) = \N_k(\bm{f} \mid \bm{0}_k, \mathbf{V}_F)$, we can obtain the posterior of the stacked scores as
\begin{align}
    \label{Eq:Posterior_Factors}
    \vecv(\mathbf{F}) \mid \mathbf{X}, \mathbf{Y}; \theta \sim \N_{Tk} (\vecv(\mathbf{M}), \bm{\Omega} \otimes \mathbf{I}_T) \, .
\end{align}
The posterior mean matrix can be expressed succinctly as
\begin{align}
    \label{Eq:Posterior_Mean}
    \mathbf{M} = \mathbf{Z} \bm{\Sigma}^{-1} \mathbf{L} \bm{\Omega} \, ,
\end{align}
where we stack the data into $\mathbf{Z} = [\mathbf{X}, \mathbf{Y}]$ and all loadings into $\mathbf{L} = [\mathbf{P}\T, \mathbf{Q}\T]\T$. Under standard statistical loss functions (such as quadratic, absolute, or zero-one losses), decision theory arguments guarantee that $\mathbf{M}$ is the optimal prediction of the scores $\mathbf{F}$ in this model \citep[see, e.g.,][pp. 29--31]{greenberg_introduction_2012}.

The first step in deriving an EM algorithm to learn the parameters and latent vectors in the PTFA model is to derive the complete data log-likelihood $\log p(\mathbf{X},\mathbf{Y},\mathbf{F} \mid \theta)$. The expectation (E) step finds the observed-data likelihood by integrating out the factors under their posterior distribution:
\begin{align}
    Q(\theta) = \E_{\mathbf{F} \mid \mathbf{X}, \mathbf{Y}; \theta} \left[ \log p(\mathbf{X} \mid \mathbf{F} ; \mathbf{P}, \sigma_x^2) + \log p(\mathbf{Y} \mid \mathbf{F} ; \mathbf{Q}, \sigma_y^2) + \log p(\mathbf{F}) \right] \, . \label{Eq:Observed_Likelihood}
\end{align}
The maximization (M) step consists in optimizing $Q(\theta)$ (given an initial value) with a view of deriving updating rules for $\theta$. We use $\mathbf{M} \coloneqq \E_{\mathbf{F} \mid \mathbf{X},\mathbf{Y}; \theta}[\mathbf{F}]$ and $\Cov_{\mathbf{F} \mid \mathbf{X},\mathbf{Y}; \theta}(\mathbf{F}|\mathbf{X},\mathbf{Y}) = \bm{\Omega} \otimes \mathbf{I}_T$ to obtain $\mathbf{V} \coloneqq \E_{\mathbf{F} \mid \mathbf{X},\mathbf{Y}; \theta}[\mathbf{F}\T \mathbf{F}] = T \cdot \bm{\Omega} + \mathbf{M}\T \mathbf{M}$. The result of maximizing \eqref{Eq:Observed_Likelihood} is a stacked update rule for the loadings (see Appendix \ref{Sec:Appendix_EM_Derivation} for a thorough derivation of these equations):
\begin{align}
    \label{Eq:Loading_Updates}
    \mathbf{L} = \mathbf{Z}\T \mathbf{M} \mathbf{V}^{-1}\, .
\end{align}
Similar simple update rules can be found for the variance parameters $\sigma_x^2$ and $\sigma_y^2$ as
\begin{align}
    \label{Eq:Sigma2_EM_Estimates}
    \sigma_x^2 = \frac{1}{T p} \left[ \| \mathbf{X} \|_F^2 - \Tr(\mathbf{P}\T \mathbf{P} \mathbf{V}) \right] \quad \text{and} \quad \sigma_y^2 = \frac{1}{T q} \left[ \| \mathbf{Y} \|_F^2 - \Tr(\mathbf{Q}\T \mathbf{Q} \mathbf{V}) \right] \, .
\end{align}
Our algorithm achieves performance gains compared to PLS without incurring in large computational complexity by efficiently computing all updates using simple matrix operations. A full implementation of our EM algorithm that computes these steps along with their formal derivation can be found in Appendices \ref{Sec:Appendix_EM_Derivation} and \ref{Sec:Appendix_Algorithms}. This EM estimator shares the asymptotic properties of the MLE (up to a logarithmic term), and factors themselves are consistently estimated as the number of predictors $p$ diverges with the sample size \citep{barigozzi2024}.

Previous research has shown that the number of factors $k$ in standard PLS can influence forecasting performance, leading to overfitting when a large $k$ is used \citep{fuentes_sparse_2015, ahn_forecasting_2022, giglio_test_2025}. There are well-established criteria for choosing $k$ as it plays a role in both forecasting performance and structural properties of the time series \citep{velicer_construct_2000, bai_determining_2002}. As the criteria used to determine this number depends on the model structure, the asymptotic properties of the constructed factors, the sample size and dimensionality of the predictors, deriving such a criteria would require a complete asymptotic treatment of the properties of PTFA, which is out of the scope of the current paper. We abstract away from this decision by considering other data-driven selection rules such as exploring forecasting performance across a range of $k$ values or using cross-validation. As determining the number of factors is likely to play a larger role in structural analysis than in forecasting, this problem is less relevant in the three applications we present in Section \ref{Sec:Application}.

%% file: Sections/Appendix_EM_Derivation.tex
We let $\widetilde{\theta}$ represent an initial or current fixed value of the parameters. The expectation (E) step requires us to obtain the observed-data likelihood by integrating out the factors according to their \emph{posterior} distribution. This results in the objective function:
\begin{align*}
    Q(\theta \mid \widetilde{\theta}) = \E_{\mathbf{F} \mid \mathbf{X}, \mathbf{Y}; \widetilde{\theta}} \left[\log p(\mathbf{X}, \mathbf{Y}, \mathbf{F} \mid \theta)\right]
\end{align*}
Our framework allows us to produce a closed-form solution for the E-step. The maximization (M) step, then consists in optimizing $Q(\cdot)$ with respect to the variables and parameters with a view of deriving updating rules. These updating rules are derived using the following steps:
\begin{enumerate}
    \item To update $\mathbf{P}$, we maximize the expected log-likelihood term involving $\mathbf{X}$    
    \begin{align*}
        Q_X(\mathbf{P}) \coloneqq \E_{\mathbf{F} \mid \mathbf{X},\mathbf{Y}; \widetilde{\theta}} \left[ -\frac{1}{2\sigma_x^2} \|\mathbf{X}-\mathbf{F}\mathbf{P}\T\|_F^2 \right]
    \end{align*}	
    expanding the norm and taking expectations we get 	
    \begin{align*}
        Q_X(\mathbf{P}) & = -\frac{1}{2\sigma_x^2} \E_{\mathbf{F} \mid \mathbf{X},\mathbf{Y}; \widetilde{\theta}} \left[\Tr\left((\mathbf{X} - \mathbf{F} \mathbf{P}\T)\T (\mathbf{X} - \mathbf{F} \mathbf{P}\T)\right)\right]
    \end{align*}
    This simplifies to:
    \begin{align*}
        Q_X(\mathbf{P}) & = -\frac{1}{2\sigma_x^2} \left[\|\mathbf{X}\|_F^2 - 2 \Tr(\mathbf{X}\T \mathbf{M} \mathbf{P}\T) + \Tr(\mathbf{P} \mathbf{V} \mathbf{P}\T)\right]
    \end{align*}
    Maximizing this quadratic form in $\mathbf{P}$ gives the first-order condition $\mathbf{X}\T \mathbf{M} = \mathbf{P} \mathbf{V}$, which can be solved to yield:
    \begin{align}
        \label{Eq:P_EM_Update}
        \mathbf{P} = \mathbf{X}\T \mathbf{M} \mathbf{V}^{-1}
    \end{align}
    \item To update $\mathbf{Q}$, we maximize the expected log-likelihood term involving $\mathbf{Y}$:
    \begin{align*}
        Q_Y(\mathbf{Q}) \coloneqq \E_{\mathbf{F} \mid \mathbf{X},\mathbf{Y}; \widetilde{\theta}} \left[ -\frac{1}{2\sigma_y^2} \|\mathbf{Y} - \mathbf{F} \mathbf{Q}\T\|_F^2 \right]
    \end{align*}
    Expanding and simplifying, we obtain an expression similar to before:
    \begin{align*}
        Q_Y(\mathbf{Q}) = -\frac{1}{2\sigma_y^2} \left[ \|\mathbf{Y}\|_F^2 - 2 \Tr(\mathbf{Y}\T \mathbf{M} \mathbf{Q}\T) + \Tr(\mathbf{Q} \mathbf{V} \mathbf{Q}\T) \right]
    \end{align*}
    Maximizing this quadratic form in $\mathbf{Q}$ gives the first-order condition $\mathbf{Y}\T \mathbf{M} = \mathbf{Q} \mathbf{V}$, which can be solved to yield:
    \begin{align}
        \label{Eq:Q_EM_Update}
        \mathbf{Q} = \mathbf{Y}\T \mathbf{M} \mathbf{V}^{-1}
    \end{align}    
    \item To update $\sigma^2_x$, We only need the term involving $\sigma_x^2$:
    \begin{align*}
        Q(\sigma_x^2) & = -\frac{T p}{2}\log(\sigma_x^2) - \frac{1}{2\sigma_x^2} \E_{\mathbf{F}|\mathbf{X},\mathbf{Y}; \widetilde{\theta}}\left[\|\mathbf{X} - \mathbf{F} \mathbf{P}\T\|_F^2\right] \\
        & = -\frac{T p}{2}\log(\sigma_x^2) - \frac{1}{2\sigma_x^2} \left[\|\mathbf{X}\|_F^2 - 2 \Tr(\mathbf{X}\T \mathbf{M} \mathbf{P}\T) + \Tr(\mathbf{P} \mathbf{V} \mathbf{P}\T)\right]
    \end{align*}
    To maximize $Q(\sigma_x^2)$ with respect to $\sigma_x^2$, take the derivative and set it to zero:
    \begin{align*}
        \frac{\partial Q(\sigma_x^2)}{\partial \sigma_x^2} = -\frac{T p}{2\sigma_x^2} + \frac{1}{2\sigma_x^4} \left[\|\mathbf{X}\|_F^2 - 2 \Tr(\mathbf{X}\T \mathbf{M} \mathbf{P}\T) + \Tr(\mathbf{P} \mathbf{V} \mathbf{P}\T)\right] = 0
    \end{align*}
    Solving for $\sigma_x^2$:
    \begin{align*}
        \sigma_x^2 = \frac{1}{T p} \left[\|\mathbf{X}\|_F^2 - 2 \Tr(\mathbf{X}\T \mathbf{M} \mathbf{P}\T) + \Tr(\mathbf{P} \mathbf{V} \mathbf{P}\T)\right]
    \end{align*}
    Using the first-order condition satisfied by $\mathbf{P}$ and combining the terms in the previous expression, our estimate can be succinctly computed as:
    \begin{align*}
        \sigma_x^2 = \frac{1}{T p} \left[ \| \mathbf{X} \|_F^2 - \Tr(\mathbf{P}\T \mathbf{P} \mathbf{V}) \right]
    \end{align*}
    \item To update $\sigma^2_y$, similar calculations can be performed as in the last step to find:
    \begin{align*}
        \sigma_y^2 = \frac{1}{T q} \left[ \| \mathbf{Y} \|_F^2 - \Tr(\mathbf{Q}\T \mathbf{Q} \mathbf{V}) \right]
    \end{align*}
\end{enumerate}
\newpage

%% file: Sections/Appendix_Algorithms.tex
\begin{algorithm}
    \caption{EM Algorithm for Probabilistic Targeted Factor Extraction}
    \label{Algo:EM}
    \begin{algorithmic}[1]
        \Require Predictor matrix $\mathbf{X} \in \mathbb{R}^{T \times p}$, target matrix $\mathbf{Y} \in \mathbb{R}^{T \times q}$, number of components $k$, starting values $(\mathbf{P}_0, \mathbf{Q}_0, \sigma_{x, 0}^2, \sigma_{y, 0}^2)$, prior variance $\mathbf{V}_F$, tolerance $\epsilon$, iteration limit $S$
        \State \textbf{Center and scale the Data:}
        \[
            \mathbf{X} \gets ( \mathbf{X} - \bm{1}_T \bm{m}_x\T) \diag(\bm{s}_x)^{-1}, \quad \mathbf{Y} \gets ( \mathbf{Y} - \bm{1}_T \bm{m}_y\T) \diag(\bm{s}_y)^{-1}
        \]
        where for the columns of $\mathbf{X}$ and $\mathbf{Y}$, $\bm{m}_x \coloneqq (1/T) \sum_{t=1}^T \bm{x}_t$ and $\bm{m}_y \coloneqq (1/T) \sum_{t=1}^T \bm{y}_t$ are the vector of means, whereas $\bm{s}_x$ and $\bm{s}_y$ are the vectors of standard deviations, respectively.
        \Repeat
        \State \textbf{E-step: Expectation}
        \State Collect all initial parameters into $\bm{\theta}_0 \gets (\mathbf{P}_0, \mathbf{Q}_0, \sigma_{x, 0}^2, \sigma_{y, 0}^2)$
        \State Compute Posterior Covariance ($\bm{\Omega}$):
        \[
            \bm{\Omega} \gets \left( \mathbf{V}_F^{-1} + \frac{1}{\sigma_{x, 0}^2} \mathbf{P}\T_0 \mathbf{P}_0 + \frac{1}{\sigma_{y, 0}^2} \mathbf{Q}\T_0 \mathbf{Q}_0 \right)^{-1}
        \]
        \State Compute Posterior Mean ($\mathbf{M}$):
        \[
            \mathbf{M} \gets \left(\frac{1}{\sigma_{x, 0}^2} \mathbf{X}\mathbf{P}_0 + \frac{1}{\sigma_{y, 0}^2} \mathbf{Y}\mathbf{Q}_0 \right) \bm{\Omega}
        \]
        \State \textbf{M-step: Maximization}
        \State $\mathbf{V} \gets T \cdot \bm{\Omega} + \mathbf{M}\T \mathbf{M}$
        \State Update $\mathbf{P}$ and $\mathbf{Q}$ jointly as:
        \[
            \begin{bmatrix} \mathbf{P}_1 \\ \mathbf{Q}_1 \end{bmatrix} \gets \begin{bmatrix} \mathbf{X}\T \\ \mathbf{Y}\T \end{bmatrix} \mathbf{M} \mathbf{V}^{-1}
        \]
        \State Update $\sigma_x^2$:
        \[
            \sigma_{x, 1}^2 \gets \frac{1}{T p} \left[\|\mathbf{X}\|_F^2 - \Tr(\mathbf{P}\T_1 \mathbf{P}_1 \mathbf{V}) \right]
        \]
        \State Update $\sigma_y^2$:
        \[
            \sigma_{y, 1}^2 \gets \frac{1}{T q} \left[\|\mathbf{Y}\|_F^2 - \Tr(\mathbf{Q}\T_1 \mathbf{Q}_1 \mathbf{V}) \right]
        \]
        \State Collect updated parameters as $\bm{\theta}_1 \gets (\mathbf{P}_1, \mathbf{Q}_1, \sigma^2_{x, 1}, \sigma^2_{y, 1})$
        \Until{convergence $\|\bm{\theta}_1 - \bm{\theta}_0\| < \epsilon$ or $S$ iterations are reached}
        \State \Return Loading matrices $\mathbf{P} \in \mathbb{R}^{p \times k}$ and $\mathbf{Q} \in \mathbb{R}^{q \times k}$, as well as noise variances $\sigma_x^2$ and $\sigma_y^2$ from final estimate $\bm{\theta}_1$
    \end{algorithmic}
\end{algorithm}

\newpage
    
\begin{algorithm}
    \caption{EM Algorithm for Probabilistic Targeted Factor Extraction with Missing-at-Random Data}
    \label{Algo:EM_Missing}
    \begin{algorithmic}[1]
        \Require Predictor matrix $\mathbf{X} \in \mathbb{R}^{T \times p}$, target matrix $\mathbf{Y} \in \mathbb{R}^{T \times q}$, number of components $k$, starting values $(\mathbf{P}_0, \mathbf{Q}_0, \sigma_{x, 0}^2, \sigma_{y, 0}^2)$, prior variance $\mathbf{V}_F$, tolerance $\epsilon$, iteration limit $S$
        \State Missing value indices: let $\tau_{t, j}^{(X)} = 1$ if entry $i, j$ of matrix $\mathbf{X}$ is missing, 0 otherwise. Define $\tau_{t, j}^{(Y)}$ similarly for $\mathbf{Y}$
        \State \textbf{Initial imputation step:} Replace $\mathbf{X}_{t, j} \gets 0$ and $\mathbf{Y}_{t, j} \gets 0$ if $\tau_{t, j}^{(X)} = 1$ and $\tau_{t, j}^{(Y)} = 1$, respectively
        \State \textbf{Center and scale the Data:}
        \[
            \mathbf{X} \gets ( \mathbf{X} - \bm{1}_T \bm{m}_x\T) \diag(\bm{s}_x)^{-1}, \quad \mathbf{Y} \gets ( \mathbf{Y} - \bm{1}_T \bm{m}_y\T) \diag(\bm{s}_y)^{-1}
        \]
        where for the columns of $\mathbf{X}$ and $\mathbf{Y}$, $\bm{m}_x \coloneqq (1/T) \sum_{t=1}^T \bm{x}_t$ and $\bm{m}_y \coloneqq (1/T) \sum_{t=1}^T \bm{y}_t$ are the vector of means, whereas $\bm{s}_x$ and $\bm{s}_y$ are the vectors of standard deviations, respectively.
        \Repeat
        \State \textbf{E-step: Expectation}
        \State Collect all initial parameters into $\bm{\theta}_0 \gets (\mathbf{P}_0, \mathbf{Q}_0, \sigma_{x, 0}^2, \sigma_{y, 0}^2)$
        \State Compute Posterior Covariance ($\bm{\Omega}$):
        \[
            \bm{\Omega} \gets \left( \mathbf{V}_F^{-1} + \frac{1}{\sigma_{x, 0}^2} \mathbf{P}\T_0 \mathbf{P}_0 + \frac{1}{\sigma_{y, 0}^2} \mathbf{Q}\T_0 \mathbf{Q}_0 \right)^{-1}
        \]
        \State Compute Posterior Mean ($\mathbf{M}$):
        \[
            \mathbf{M} \gets \left(\frac{1}{\sigma_{x, 0}^2} \mathbf{X}\mathbf{P}_0 + \frac{1}{\sigma_{y, 0}^2} \mathbf{Y}\mathbf{Q}_0 \right) \bm{\Omega}
        \]      
        \State \textbf{M-step: Maximization}
        \State $\mathbf{V} \gets T \cdot \bm{\Omega} + \mathbf{M}\T \mathbf{M}$
        \State Update the missing value entries with the latest EM fit:
        \begin{align*}
            \mathbf{X}_{t, j} & \gets \sum_{c = 1}^{k} \mathbf{M}_{i c} \mathbf{P}_{j c} \quad \text{ if } \quad \tau_{t, j}^{(X)} = 1 \\
            \mathbf{Y}_{t, j} & \gets \sum_{c = 1}^{k} \mathbf{M}_{i c} \mathbf{Q}_{j c} \quad \text{ if } \quad \tau_{t, j}^{(Y)} = 1
        \end{align*}
        \algstore{missing_break}
    \end{algorithmic}
\end{algorithm}

\newpage

\begin{algorithm}[t]
    \begin{algorithmic}
        \algrestore{missing_break}
        \State Update $\mathbf{P}$ and $\mathbf{Q}$ jointly as:
        \[
            \begin{bmatrix} \mathbf{P}_1 \\ \mathbf{Q}_1 \end{bmatrix} \gets \begin{bmatrix} \mathbf{X}\T \\ \mathbf{Y}\T \end{bmatrix} \mathbf{M} \mathbf{V}^{-1}
        \]
        \State Update $\sigma_x^2$:
        \[
            \sigma_{x, 1}^2 \gets \frac{1}{T p} \left[\|\mathbf{X}\|_F^2 - \Tr(\mathbf{P}\T_1 \mathbf{P}_1 \mathbf{V}) \right]
        \]
        \State Update $\sigma_y^2$:
        \[
            \sigma_{y, 1}^2 \gets \frac{1}{T q} \left[\|\mathbf{Y}\|_F^2 - \Tr(\mathbf{Q}\T_1 \mathbf{Q}_1 \mathbf{V}) \right]
        \]
        \State Collect updated parameters as $\bm{\theta}_1 \gets (\mathbf{P}_1, \mathbf{Q}_1, \sigma^2_{x, 1}, \sigma^2_{y, 1})$
        \Until{convergence $\|\bm{\theta}_1 - \bm{\theta}_0\| < \epsilon$ or $S$ iterations are reached}
        \State \Return Loading matrices $\mathbf{P} \in \mathbb{R}^{p \times k}$ and $\mathbf{Q} \in \mathbb{R}^{q \times k}$, as well as noise variances $\sigma_x^2$ and $\sigma_y^2$ from final estimate $\bm{\theta}_1$
    \end{algorithmic}
\end{algorithm}

\newpage

\begin{algorithm}[hb]
    \caption{EM Algorithm for Probabilistic PLS with Mixed-Frequency Data}
    \label{Algo:EM_MixedFrequency}
    \begin{algorithmic}[1]
        \Require High-frequency predictor matrix $\widetilde{\mathbf{X}} \in \mathbb{R}^{(TL) \times p}$, low-frequency target matrix $\mathbf{Y} \in \mathbb{R}^{T \times q}$, low-to-high-frequency period $L$, number of components $k$, starting values $(\mathbf{P}_0, \mathbf{Q}_0, \sigma_{x, 0}^2, \sigma_{y, 0}^2)$, prior variance $\mathbf{V}_F$, tolerance $\epsilon$, iteration limit $S$
        \State Reshape $\widetilde{\mathbf{X}}$ into a $T \times (pL)$ matrix $\mathbf{X}$:
        \[
            \widetilde{\mathbf{X}} = \begin{bmatrix}
                \bm{x}_1\T \\ \vdots \\ \bm{x}_L\T \\ \vdots \\ \bm{x}_{(T-1)L+1}\T \\ \vdots \\ \bm{x}_{TL}\T
            \end{bmatrix} \mapsto \begin{bmatrix}
                \bm{x}_1\T & \cdots & \bm{x}_L\T \\
                \vdots & \ddots & \vdots \\
                \bm{x}_{(T-1)L+1}\T & \cdots & \bm{x}_{TL}\T
            \end{bmatrix} = \mathbf{X}
        \]
        \State \textbf{Center and scale the Data:}
        \[
            \mathbf{X} \gets ( \mathbf{X} - \bm{1}_T \bm{m}_x\T) \diag(\bm{s}_x)^{-1}, \quad \mathbf{Y} \gets ( \mathbf{Y} - \bm{1}_T \bm{m}_y\T) \diag(\bm{s}_y)^{-1}
        \]
        where for the columns of $\mathbf{X}$ and $\mathbf{Y}$, $\bm{m}_x \coloneqq (1/T) \sum_{t=1}^T \bm{x}_t$ and $\bm{m}_y \coloneqq (1/T) \sum_{t=1}^T \bm{y}_t$ are the vector of means, whereas $\bm{s}_x$ and $\bm{s}_y$ are the vectors of standard deviations, respectively.
        \Repeat
        \State \textbf{E-step: Expectation}
        \State Collect all initial parameters into $\bm{\theta}_0 \gets (\mathbf{P}_0, \mathbf{Q}_0, \sigma_{x, 0}^2, \sigma_{y, 0}^2)$
        \State Compute Posterior Covariance ($\bm{\Omega}$):
        \[
            \bm{\Omega} \gets \left[ \mathbf{I}_L \otimes \left( \frac{1}{\sigma_{x, 0}^2} \mathbf{P}_0\T \mathbf{P}_0 + \mathbf{V}_F^{-1} \right) + \frac{1}{L \cdot \sigma_{y, 0}^2} \bm{1}_{L \times L} \otimes \left( \mathbf{Q}_0\T \mathbf{Q}_0 \right) \right]^{-1}
        \]
        \State Compute Posterior Mean ($\mathbf{M}$):
        \[
            \mathbf{M} \gets \begin{bmatrix} \frac{1}{\sigma_{x, 0}^2} \mathbf{X}^{(1)} \mathbf{P}_0 + \frac{1}{\sigma_{y, 0}^2} \mathbf{Y} \mathbf{Q}_0 & \cdots & \frac{1}{\sigma_{x, 0}^2} \mathbf{X}^{(L)} \mathbf{P}_0 + \frac{1}{\sigma_{y, 0}^2} \mathbf{Y} \mathbf{Q}_0 \end{bmatrix} \bm{\Omega}
        \]      
        \algstore{mixedfrequency_break}
    \end{algorithmic}
\end{algorithm}

\newpage

\begin{algorithm}[t]
    \begin{algorithmic}
        \algrestore{mixedfrequency_break}
        \State \textbf{M-step: Maximization}
        \State $\mathbf{V} = \begin{bmatrix}
            \mathbf{V}_{1, 1} & \cdots & \mathbf{V}_{1, L} \\
            \vdots & \ddots & \vdots \\
            \mathbf{V}_{L, 1} & \cdots & \mathbf{V}_{L, L}
        \end{bmatrix} \gets T \cdot \bm{\Omega} + \mathbf{M}\T \mathbf{M}$
        \State Update $\mathbf{P}$ as:
        \[
            \mathbf{P}_1 \gets \left( \sum_{\ell = 1}^{L} \mathbf{X}^{(\ell)\top} \mathbf{M}^{(\ell)} \right) \left( \sum_{\ell = 1}^{L} \mathbf{V}_{\ell, \ell} \right)^{-1}
        \]
        \State Update $\mathbf{Q}$ as:
        \[
            \mathbf{Q}_1 \gets L \cdot \left( \mathbf{Y}\T \sum_{\ell = 1}^{L} \mathbf{M}^{(\ell)} \right) \left( \sum_{r = 1}^{L} \sum_{\ell = 1}^{L} \mathbf{V}_{\ell, r} \right)^{-1}
        \]
        \State Update $\sigma_x^2$:
        \[
            \sigma_{x, 1}^2 \gets \frac{1}{T L p} \left\{\|\mathbf{X}\|_F^2 - \Tr \left[\mathbf{P}\T_1 \mathbf{P}_1 \left( \sum_{\ell = 1}^{L} \mathbf{V}_{\ell, \ell} \right) \right] \right\}
        \]
        \State Update $\sigma_y^2$:
        \[
            \sigma_{y, 1}^2 \gets \frac{L}{T q} \Tr \left\{\mathbf{Y}\T \left[\mathbf{Y} - \frac{1}{L} \left( \sum_{\ell = 1}^{L} \mathbf{M}^{(\ell)} \right) \mathbf{Q}_1\T \right] \right\}
        \]
        \State Collect updated parameters as $\bm{\theta}_1 \gets (\mathbf{P}_1, \mathbf{Q}_1, \sigma^2_{x, 1}, \sigma^2_{y, 1})$
        \Until{convergence $\|\bm{\theta}_1 - \bm{\theta}_0\| < \epsilon$ or $S$ iterations are reached}
        \State \Return Loading matrices $\mathbf{P} \in \mathbb{R}^{p \times k}$ and $\mathbf{Q} \in \mathbb{R}^{q \times k}$, as well as noise variances $\sigma_x^2$ and $\sigma_y^2$ from final estimate $\bm{\theta}_1$
    \end{algorithmic}
\end{algorithm}

\newpage

\begin{algorithm}[hb]
    \caption{EM Algorithm for Probabilistic PLS with Exponential Weighted Moving Average (EWMA) Stochastic Volatility}
    \label{Algo:EM_StochasticVolatility}
    \begin{algorithmic}[1]
        \Require Predictor matrix $\mathbf{X} \in \mathbb{R}^{T \times p}$, target matrix $\mathbf{Y} \in \mathbb{R}^{T \times q}$, number of components $k$, starting values $(\mathbf{P}_0, \mathbf{Q}_0, \Bar{\sigma}_x^2, \Bar{\sigma}_y^2)$, prior variance $\mathbf{V}_F$, EWMA smoothing parameters $(\lambda_x, \lambda_y)$, tolerance $\epsilon$, iteration limit $S$
        \State \textbf{Center and scale the Data:}
        \[
            \mathbf{X} \gets ( \mathbf{X} - \bm{1}_T \bm{m}_x\T) \diag(\bm{s}_x)^{-1}, \quad \mathbf{Y} \gets ( \mathbf{Y} - \bm{1}_T \bm{m}_y\T) \diag(\bm{s}_y)^{-1}
        \]
        where for the columns of $\mathbf{X}$ and $\mathbf{Y}$, $\bm{m}_x \coloneqq (1/T) \sum_{t=1}^T \bm{x}_t$ and $\bm{m}_y \coloneqq (1/T) \sum_{t=1}^T \bm{y}_t$ are the vector of means, whereas $\bm{s}_x$ and $\bm{s}_y$ are the vectors of standard deviations, respectively.
        \State Start the $T$-dimensional stochastic volatility vectors $\bm{\sigma}_{x, 0}^2$ and $\bm{\sigma}_{y, 0}^2$ as constant:
        \[
            \sigma_{x, 0}^2(t) \gets \Bar{\sigma}_x^2 \quad \text{and} \quad \sigma_{y, 0}^2(t) \gets \Bar{\sigma}_y^2 \quad \text{for all} \quad t = 1, \ldots, T
        \]
        \Repeat
        \State \textbf{E-step: Expectation}
        \State Collect initial parameters into $\bm{\theta}_0 \gets (\mathbf{P}_0, \mathbf{Q}_0, \bm{\sigma}_{x, 0}^2, \bm{\sigma}_{y, 0}^2)$
        \For{$t = 1, \ldots, T$}
            \State Compute and store posterior covariance per period ($\bm{\Omega}_t$):
            \[
                \bm{\Omega}_t \gets \left( \mathbf{V}_F^{-1} + \frac{1}{\sigma_{x, 0}^2(t)} \mathbf{P}_0\T \mathbf{P}_0 + \frac{1}{\sigma_{y, 0}^2(t)} \mathbf{Q}_0\T \mathbf{Q}_0 \right)^{-1}
            \]
            \State Compute posterior mean per period ($\bm{m}_t$):
            \[
                \bm{m}_t \gets \bm{\Omega}_t \left(\frac{1}{\sigma_{x, 0}^2(t)} \mathbf{P}_0\T \bm{x}_t + \frac{1}{\sigma_{y, 0}^2(t)} \mathbf{Q}_0\T \bm{y}_t \right)
            \]
        \EndFor
        \State Stack posterior means into $T \times k$ matrix:
        \[
            \mathbf{M} \gets \begin{bmatrix}
                \bm{m}_1\T \\ \vdots \\ \bm{m}_T\T
            \end{bmatrix}
        \]
        \State \textbf{M-step: Maximization}
        \State $\mathbf{V} \gets \sum_{t=1}^T \bm{\Omega}_t + \mathbf{M}\T \mathbf{M}$
        \algstore{sv_break}
    \end{algorithmic}
\end{algorithm}

\newpage

\begin{algorithm}[t]
    \begin{algorithmic}
        \algrestore{sv_break}
        \State Update $\mathbf{P}$ and $\mathbf{Q}$ jointly as:
        \[
            \begin{bmatrix} \mathbf{P}_1 \\ \mathbf{Q}_1 \end{bmatrix} \gets \begin{bmatrix} \mathbf{X}\T \\ \mathbf{Y}\T \end{bmatrix} \mathbf{M} \mathbf{V}^{-1}
        \]

        \State Compute Residuals:
        \[
            \widehat{\mathbf{E}}_{x} = \begin{bmatrix}
                \widehat{\bm{e}}_{x, 1}\T \\ \vdots \\ \widehat{\bm{e}}_{x, T}\T
            \end{bmatrix} \gets \mathbf{X} - \mathbf{M} \mathbf{P}_1\T \quad \text{and} \quad \widehat{\mathbf{E}}_{y} = \begin{bmatrix}
                \widehat{\bm{e}}_{y, 1}\T \\ \vdots \\ \widehat{\bm{e}}_{y, T}\T
            \end{bmatrix} \gets \mathbf{Y} - \mathbf{M} \mathbf{Q}_1\T
        \]
        \State Update first-period stochastic volatility estimates:
        \begin{align*}
            \sigma_{x, 1}^2(1) & \gets \frac{1}{p} \left[ \|\widehat{\bm{e}}_{x, 1}\|_2^2 + \Tr (\mathbf{P}_1\T \mathbf{P}_1 \bm{\Omega}_1) \right] \\
            \sigma_{y, 1}^2(1) & \gets \frac{1}{q} \left[ \|\widehat{\bm{e}}_{y, 1}\|_2^2 + \Tr (\mathbf{Q}_1\T \mathbf{Q}_1 \bm{\Omega}_1) \right]
        \end{align*}
        \State Update remaining stochastic volatility estimates in $\bm{\sigma}_{x, 1}^2$ and $\bm{\sigma}_{y, 1}^2$ using EWMA:
        \For{$t = 2, \ldots, T$}
            \begin{align*}
                \sigma_{x, 1}^2(t) & \gets \lambda_x \cdot \sigma_{x, 1}^2(t-1) + (1 - \lambda_x) \cdot \frac{1}{p} \left[ \|\widehat{\bm{e}}_{x, t}\|_2^2 + \Tr (\mathbf{P}_1\T \mathbf{P}_1 \bm{\Omega}_t) \right] \\
                \sigma_{y, 1}^2(t) & \gets \lambda_y \cdot \sigma_{y, 1}^2(t-1) + (1 - \lambda_y) \cdot \frac{1}{q} \left[ \|\widehat{\bm{e}}_{y, t}\|_2^2 + \Tr (\mathbf{Q}_1\T \mathbf{Q}_1 \bm{\Omega}_t) \right]
            \end{align*}
        \EndFor
        \State Collect updated parameters into $\bm{\theta}_1 \gets (\mathbf{P}_1, \mathbf{Q}_1, \bm{\sigma}_{x, 1}^2, \bm{\sigma}_{y, 1}^2)$
        \Until{convergence $\|\bm{\theta}_1 - \bm{\theta}_0\| < \epsilon$ or $S$ iterations are reached}
        \State \Return Loading matrices $\mathbf{P} \in \mathbb{R}^{p \times k}$ and $\mathbf{Q} \in \mathbb{R}^{q \times k}$, as well as time-varying noise variances $\bm{\sigma}_x^2$ and $\bm{\sigma}_y^2$ from final estimates $\bm{\theta}_1$
    \end{algorithmic}
\end{algorithm}

\begin{algorithm}
    \caption{EM Algorithm for Probabilistic Targeted Factor Extraction with Factor Dynamics}
    \label{Algo:EM_FactorDynamics}
    \begin{algorithmic}[1]
        \Require Predictor matrix $\mathbf{X} \in \mathbb{R}^{T \times p}$, target matrix $\mathbf{Y} \in \mathbb{R}^{T \times q}$, number of components $k$, starting values $(\mathbf{P}_0, \mathbf{Q}_0, \sigma_{x, 0}^2, \sigma_{y, 0}^2, \mathbf{A}_0, \bm{\Sigma}_{v, 0}, \bm{f}_0)$, tolerance $\epsilon$, iteration limit $S$
        \State \textbf{Center and scale the Data:}
        \[
            \mathbf{X} \gets ( \mathbf{X} - \bm{1}_T \bm{m}_x\T) \diag(\bm{s}_x)^{-1}, \quad \mathbf{Y} \gets ( \mathbf{Y} - \bm{1}_T \bm{m}_y\T) \diag(\bm{s}_y)^{-1}
        \]
        where for the columns of $\mathbf{X}$ and $\mathbf{Y}$, $\bm{m}_x \coloneqq (1/T) \sum_{t=1}^T \bm{x}_t$ and $\bm{m}_y \coloneqq (1/T) \sum_{t=1}^T \bm{y}_t$ are the vector of means, whereas $\bm{s}_x$ and $\bm{s}_y$ are the vectors of standard deviations, respectively.
        \Repeat
        \State \textbf{E-step: Expectation}
        \State Collect all initial parameters into $\bm{\theta}_0 \gets (\mathbf{P}_0, \mathbf{Q}_0, \sigma_{x, 0}^2, \sigma_{y, 0}^2, \mathbf{A}_0, \bm{\Sigma}_{v, 0}, \bm{f}_0)$
        \State Store matrix $\mathbf{H}_A$ as a block-banded matrix:
        \[
            \mathbf{H}_A \gets \begin{bmatrix}
                \mathbf{I}_k &  & & & \\
                -\mathbf{A}_0 & \mathbf{I}_k & & & \\
                \vdots & \vdots & \ddots & & \\
                \mathbf{0}_{k \times k} & \mathbf{0}_{k \times k} & \cdots & \mathbf{I}_k & \\
                \mathbf{0}_{k \times k} & \mathbf{0}_{k \times k} & \cdots & -\mathbf{A}_0 & \mathbf{I}_k
            \end{bmatrix}
        \]
        \State Pre-compute vectorized prior mean $\vecv(\mathbf{M}_0\T) \coloneqq \mathbf{H}_A \bm{\mu}_0$:
        \[
            \vecv(\mathbf{M}_0\T) \gets \begin{bmatrix} \mathbf{A}_0 \bm{f}_0 \\ \bm{0}_k \\ \vdots \\ \bm{0}_k \end{bmatrix} = \mathbf{H}_A \bm{\mu}_0
        \]
        \State Compute posterior precision ($\bm{\Omega}^{-1}$) and store as banded matrix:
        \[
            \bm{\Omega}^{-1} \gets \mathbf{H}_A\T \left(\mathbf{I}_T \otimes \bm{\Sigma}_{v, 0}^{-1} \right) \mathbf{H}_A + I_T \otimes \left( \frac{1}{\sigma_{x, 0}^2} \mathbf{P}\T_0 \mathbf{P}_0 + \frac{1}{\sigma_{y, 0}^2} \mathbf{Q}\T_0 \mathbf{Q}_0 \right)
        \]
        \State Compute vectorized posterior mean ($\mathbf{M}$) using a banded matrix system solver:
        \begin{multline*}
            \vecv(\mathbf{M}\T) \gets \bm{\Omega} \Bigg\{ \mathbf{H}_A\T \left(\mathbf{I}_T \otimes \bm{\Sigma}_{v, 0}^{-1} \right) \vecv(\mathbf{M}_0\T) \\ {} + \left[I_T \otimes \left(\frac{1}{\sigma_{x, 0}^2} \mathbf{X}\mathbf{P}_0 + \frac{1}{\sigma_{y, 0}^2} \mathbf{Y}\mathbf{Q}_0 \right) \right] \vecv(\mathbf{Z}\T) \Bigg\}
        \end{multline*}
        \algstore{dfm_break}
    \end{algorithmic}
\end{algorithm}

\newpage

\begin{algorithm}[t]
    \begin{algorithmic}
        \algrestore{dfm_break}
        \State \textbf{M-step: Maximization}
        \State Compute only the following quantities from the banded Cholesky decomposition of the posterior covariance matrix:
        \begin{align*}
            \mathbf{V}_0 & \gets \sum_{t=1}^T \bm{\Omega}_{t, t} + \bm{m}_t \bm{m}_t\T \\
            \mathbf{V}_1 & \gets \sum_{t=2}^T \bm{\Omega}_{t-1, t-1} + \bm{m}_{t-1} \bm{m}_{t-1}\T \\
            \mathbf{V}_{10} & \gets \sum_{t=2}^T \bm{\Omega}_{t-1, t} + \bm{m}_{t-1} \bm{m}_t\T \\
            \mathbf{V}_{20} & \gets \sum_{t=3}^T \bm{\Omega}_{t-2, t} + \bm{m}_{t-2} \bm{m}_t\T
        \end{align*}
        \State Update $\mathbf{P}$ and $\mathbf{Q}$ jointly as:
        \[
            \begin{bmatrix} \mathbf{P}_1 \\ \mathbf{Q}_1 \end{bmatrix} \gets \begin{bmatrix} \mathbf{X}\T \\ \mathbf{Y}\T \end{bmatrix} \mathbf{M} \mathbf{V}_0^{-1}
        \]
        \State Update $\sigma_x^2$:
        \[
            \sigma_{x, 1}^2 \gets \frac{1}{T p} \left[\|\mathbf{X}\|_F^2 - \Tr(\mathbf{P}\T_1 \mathbf{P}_1 \mathbf{V}_0) \right]
        \]
        \State Update $\sigma_y^2$:
        \[
            \sigma_{y, 1}^2 \gets \frac{1}{T q} \left[\|\mathbf{Y}\|_F^2 - \Tr(\mathbf{Q}\T_1 \mathbf{Q}_1 \mathbf{V}_0) \right]
        \]
        \State Update $\mathbf{A}$:
        \[
            \mathbf{A}_1 \gets \mathbf{V}_{20}^{-1} \mathbf{V}_{10}
        \]
        \State Update $\bm{\Sigma}_v$:
        \[
            \bm{\Sigma}_{v, 1} \gets \diag \left[ \mathbf{V}_0 + (\bm{m}_1 - \mathbf{A}_1 \bm{f}_0)(\bm{m}_1 - \mathbf{A}_1 \bm{f}_0)\T + \mathbf{A}_1\T \mathbf{V}_1 \mathbf{A}_1 - 2 \, \mathbf{V}_{10} \mathbf{A}_1 \right]^{-1} \, .
        \]
        \State Update $\bm{f}_0$:
        \[
            \bm{f}_0 \gets \left( \mathbf{A}_1\T \bm{\Sigma}_{v, 1} \mathbf{A}_1 \right)^{-1} \mathbf{A}_1\T \bm{\Sigma}_{v, 1} \bm{m}_1
        \]
        \State Collect updated parameters as $\bm{\theta}_1 \gets (\mathbf{P}_1, \mathbf{Q}_1, \sigma^2_{x, 1}, \sigma^2_{y, 1}, \mathbf{A}_1, \bm{\Sigma}_{v, 1}, \bm{f}_0)$
        \Until{convergence $\|\bm{\theta}_1 - \bm{\theta}_0\| < \epsilon$ or $S$ iterations are reached}
        \State \Return Final estimates from $\bm{\theta}_1$
    \end{algorithmic}
\end{algorithm}

%% file: Sections/Appendix_Results.tex
\subsection{EM convergence}

\begin{figure}[H]
    \centering
    \includegraphics[width = 0.7\textwidth]{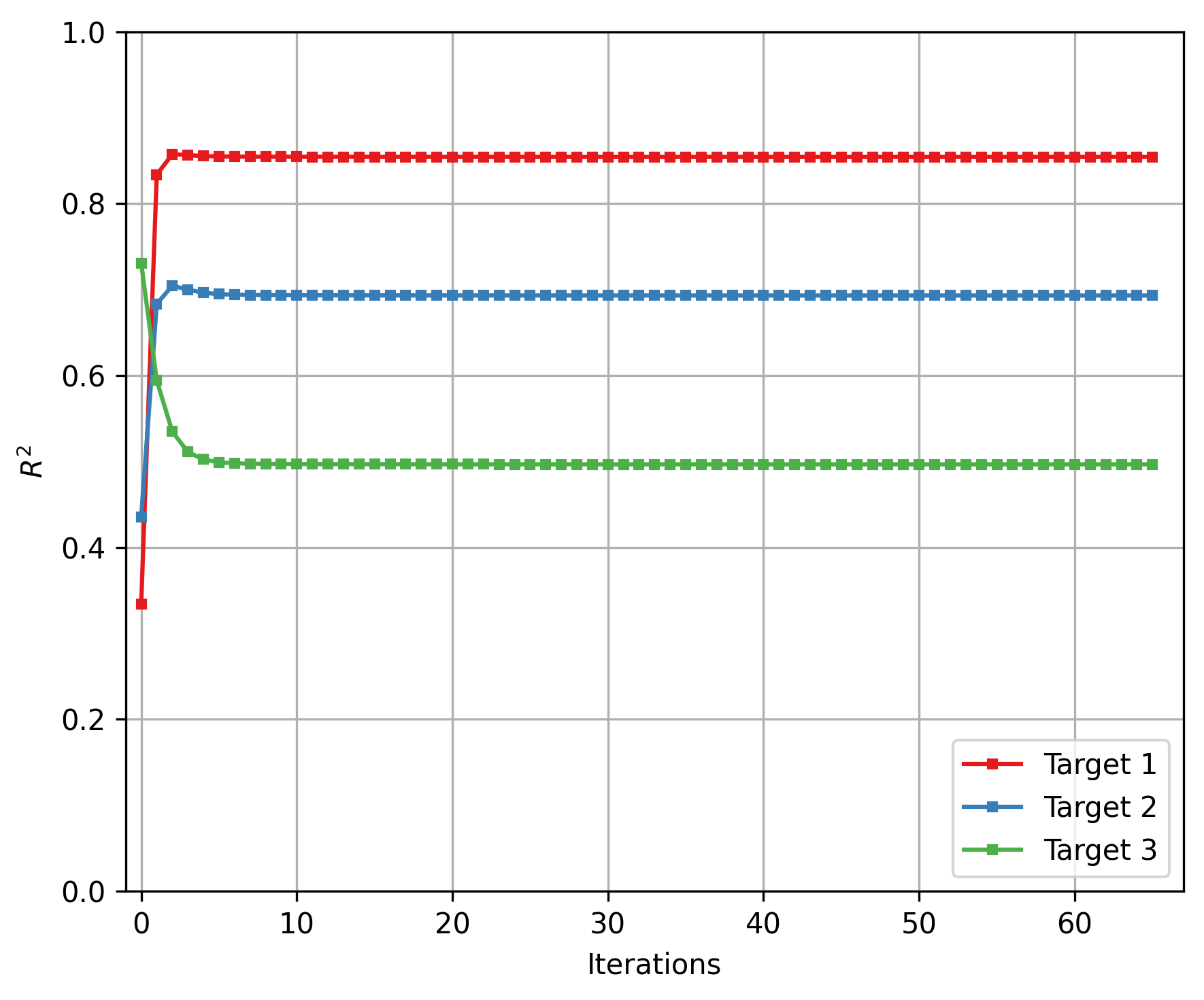}
    \caption{Path of $R^2$ of fit on a single realization of simulated data with independent Gaussian errors \eqref{DGP:Simple}}
    \label{fig:rsquared_em_iterations_simple}
\end{figure}

\begin{figure}
    \centering
    \includegraphics[width = 0.7\textwidth]{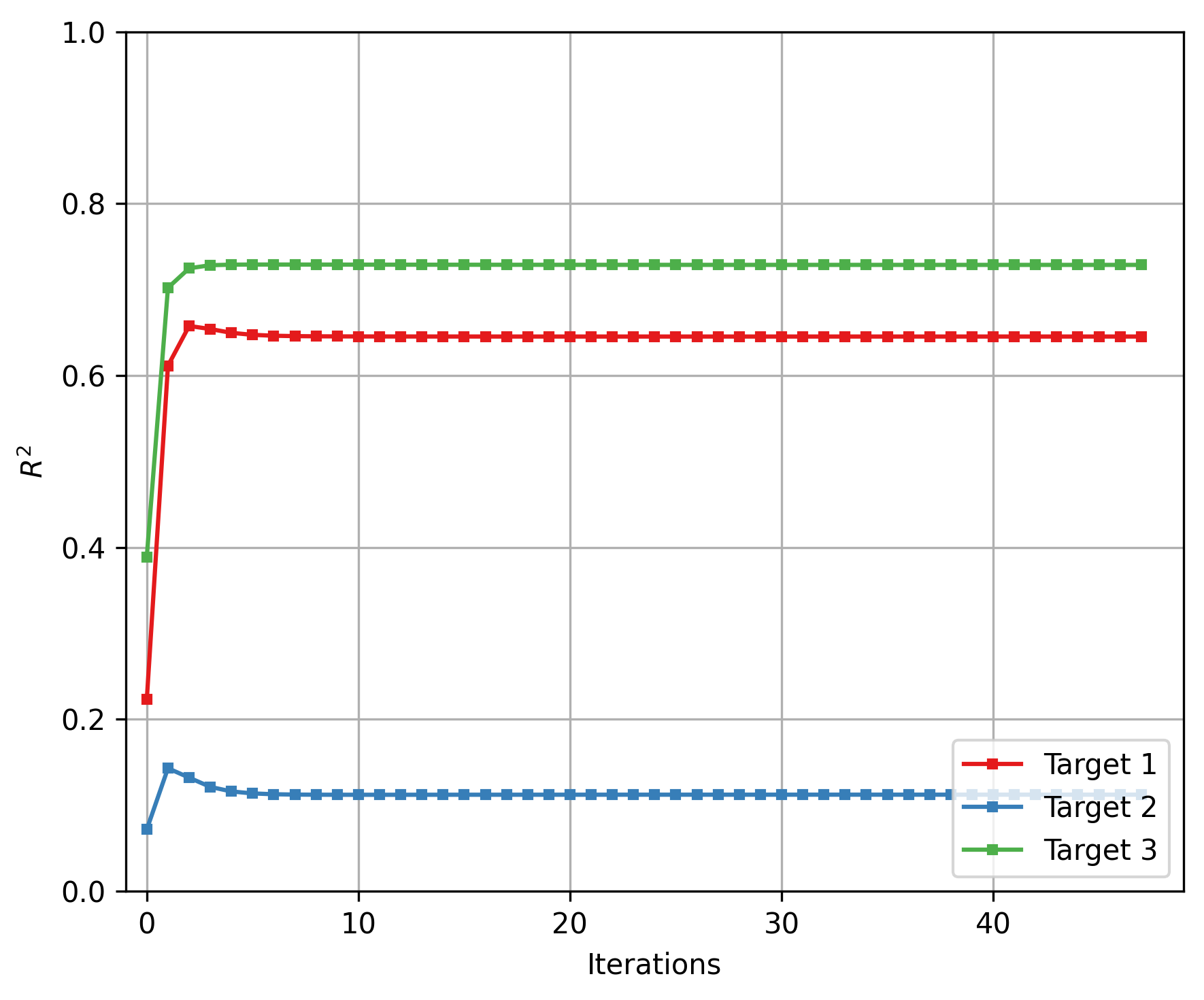}
    \caption{Path of $R^2$ of fit on a single realization of simulated data with correlated Gaussian errors \eqref{DGP:System}}
    \label{fig:rsquared_em_iterations_system}
\end{figure}

\begin{figure}
    \centering
    \includegraphics[width = 0.7\textwidth]{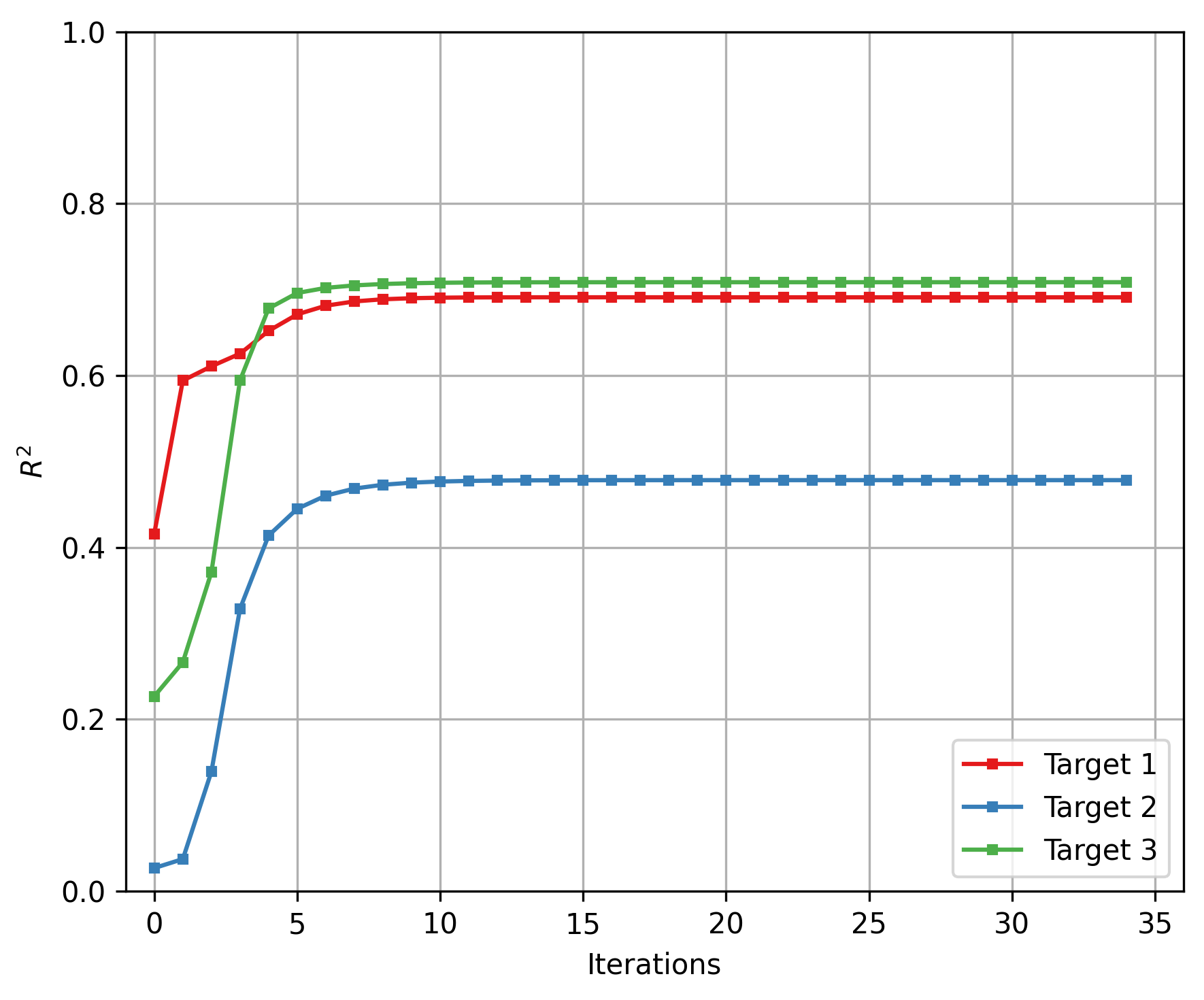}
    \caption{Path of $R^2$ of fit on a single realization of simulated data with heavy-tailed non-Gaussian errors \eqref{DGP:NonGaussian}}
    \label{fig:rsquared_em_iterations_nongaussian}
\end{figure}

\subsection{Missing data scenario}

An additional advantage of the probabilistic formulation of PLS is that it also allows us to directly deal with missing observations. Missing observations are the staple in most real-world scenarios as these can arise from distinct data release schedules, different collection techniques, data corruption, etc. To compare our PTFA method, we introduce an additional modification allowing us to deal with missing data.

This is a standard approach when using EM-type algorithms, as the expectation step with missing data can be easily computed by a simple imputation step based on the current EM estimate. The full procedure is presented in Algorithm \ref{Algo:EM_Missing}. In our provided package, we allow all of our methods and extensions to deal with missing data using this same idea.

To test the performance of our proposed method under missing data, we generate observations for $\mathbf{X}$ and $\mathbf{Y}$ as in the previous examples, additionally introducing a given percentage of missing-at-random observations for both. That is, we choose corruption levels $\rho_x$\% and $\rho_y$\% to set that fraction of elements randomly as missing. We compare two potential solutions to the missing data issue:
\begin{enumerate}
    \item Create an imputed version of the data ($\widetilde{\mathbf{X}}$ and $\widetilde{\mathbf{Y}}$) that sets the missing values in $\mathbf{X}$ and $\mathbf{Y}$ to a fixed value, such as 0 or the sample average. We can then apply both PLS or our PTFA $\widetilde{\mathbf{X}}$ and $\widetilde{\mathbf{Y}}$.
    \item On the other hand, we directly implement our PTFA with an imputation step on the inner loop of the EM algorithm based on the current predicted value for $\mathbf{X}$ and $\mathbf{Y}$ (as proposed in Eq. \ref{Eq:Missing_Data_Imputation}).
\end{enumerate}

\begin{figure}[H]
    \centering
    \begin{subfigure}{0.49 \textwidth}
        \includegraphics[width = \textwidth]{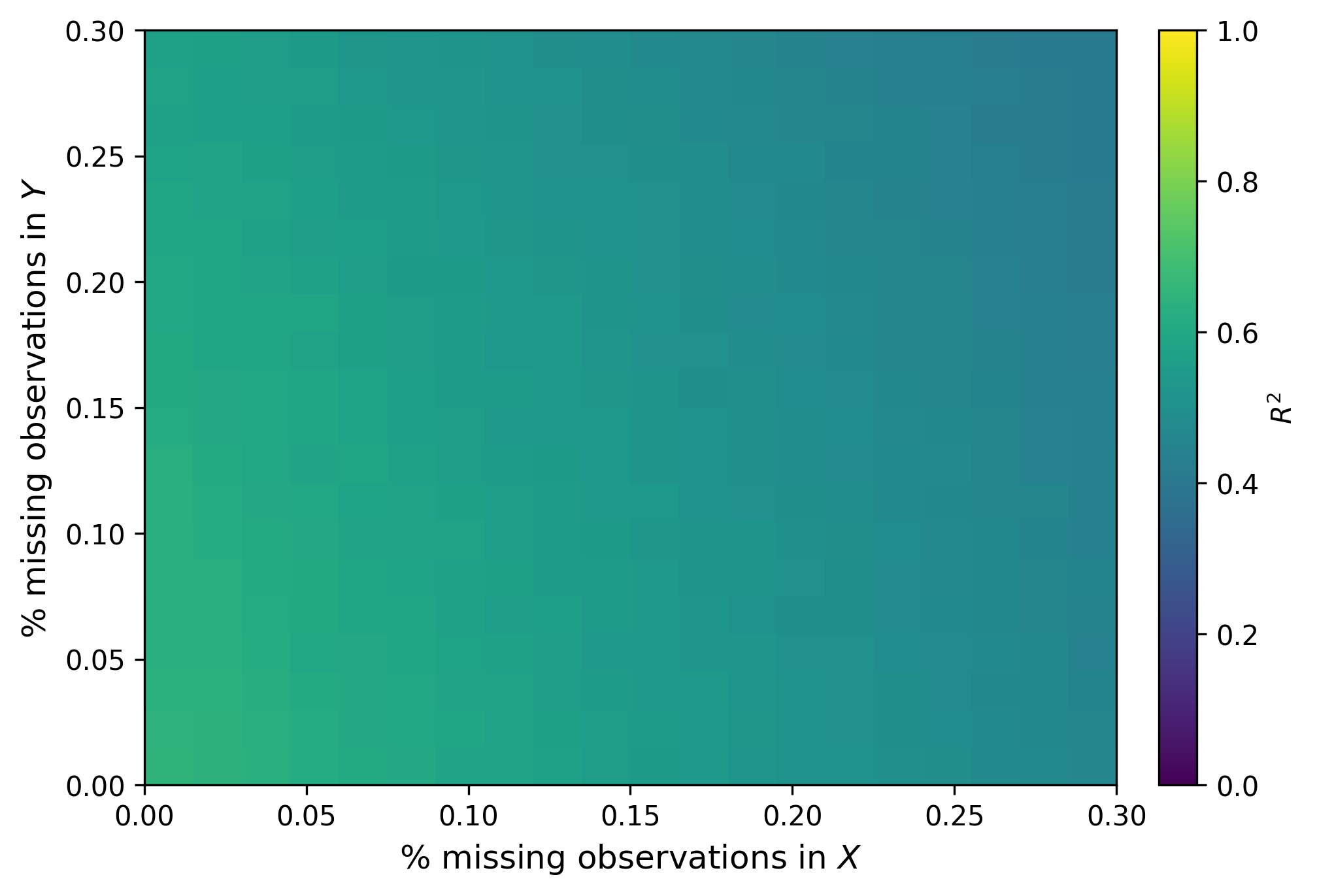}
        \caption{\footnotesize{Partial Least Squares (PLS)}}
    \end{subfigure}
    \begin{subfigure}{0.49 \textwidth}
        \includegraphics[width = \textwidth]{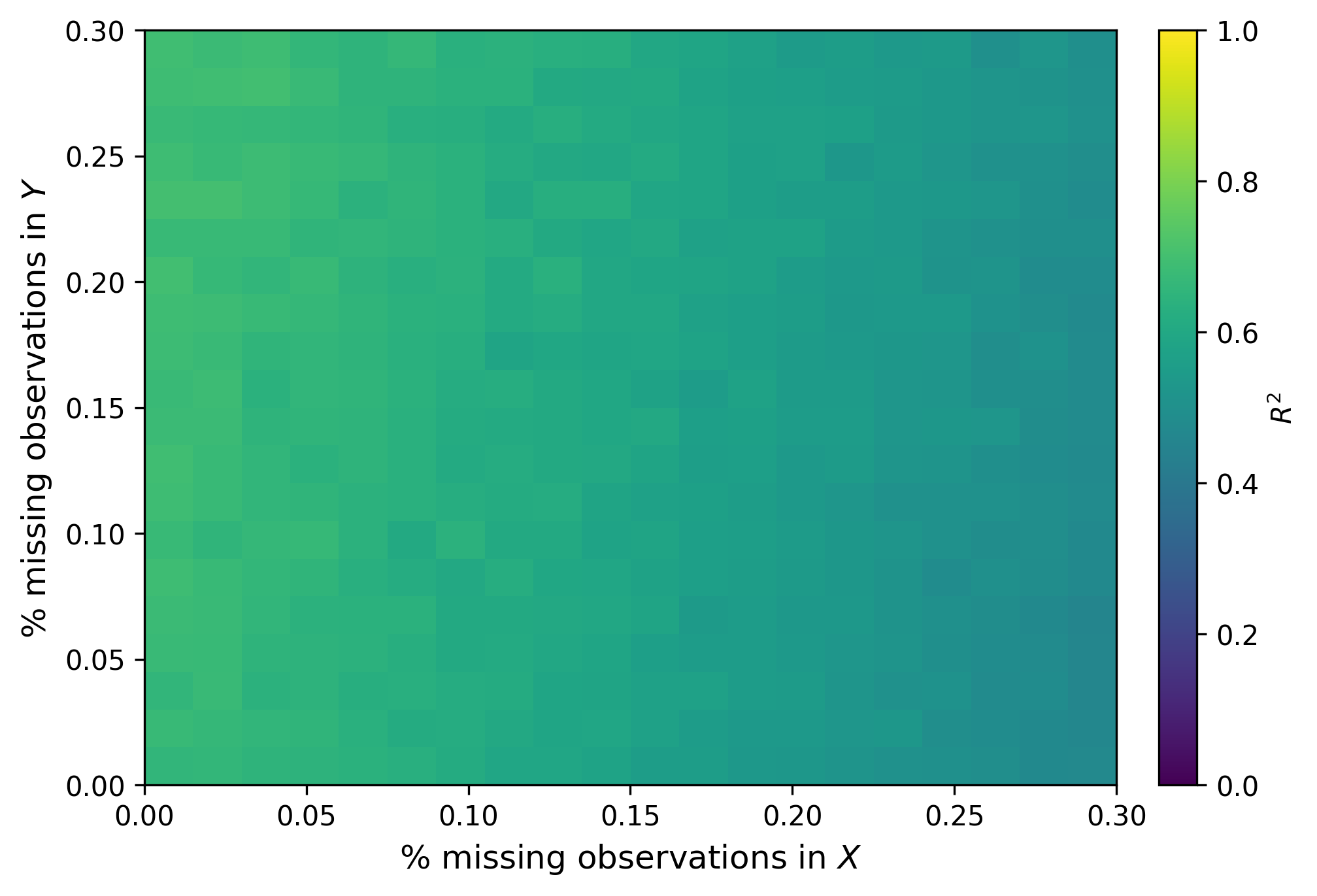}
        \caption{\footnotesize{Probabilistic Targeted Factor Analysis (PTFA)}}
    \end{subfigure}
    \caption{Comparison of PLS vs PTFA based on \% of missing-at-random observations in features ($X$) and targets ($Y$)}
    \label{fig:missing_comparison}
\end{figure}

Figure \ref{fig:missing_comparison} presents median $R^2$ statistics across 100 replications for this exercise, when we vary the level of missing-at-random observations in both features ($\rho_x$) and targets ($\rho_y$). The $R^2$ in this scenario is computed with respect to the true, infeasible values for the targets $\mathbf{Y}$ that have no missing observations. The first and second panels represent a direct comparison of applying either PLS or PTFA on the imputed data ($\widetilde{\mathbf{X}}$ and $\widetilde{\mathbf{Y}}$), while the third represents the results from including the missing values into our EM algorithm.

Note that PLS never achieves a large value for the $R^2$ with imputation, which is not the case for our PTFA method. However, by implementing the imputation step in the inner loop, our method is able to deal with large amounts of missing observations in the features, only breaking down with extreme amounts of missing observations close to 50\%.

\subsection{Unsupservised methods performance}

\begin{figure}[H]
    \centering
    \begin{subfigure}{0.49 \textwidth}
        \includegraphics[width = \textwidth]{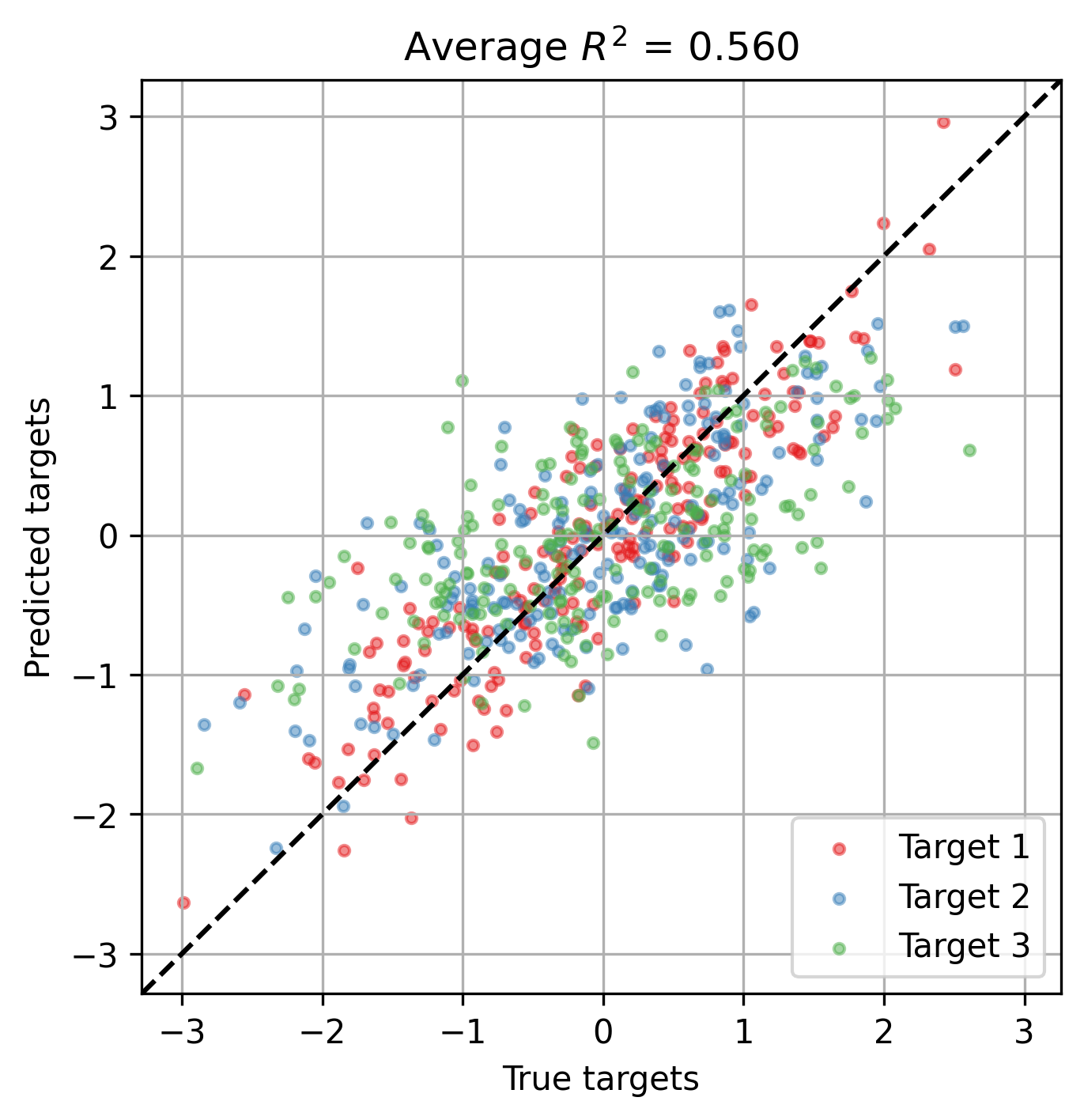}
        \caption{Principal Component Analysis (PCA)}
    \end{subfigure}
    \begin{subfigure}{0.49 \textwidth}
        \includegraphics[width = \textwidth]{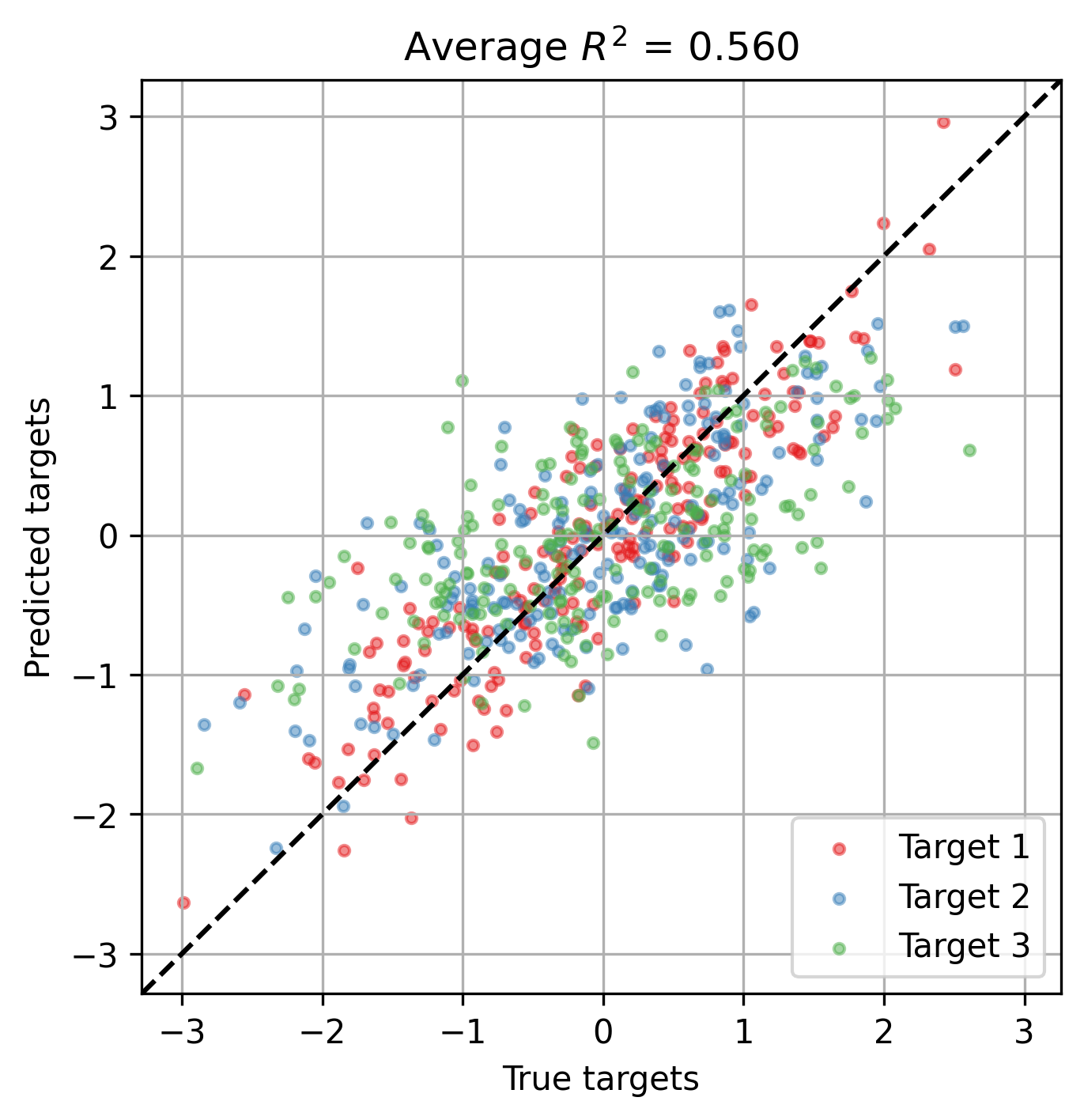}
        \caption{Probabilistic PCA (PPCA)}
    \end{subfigure}
    \caption{Comparison of PCA and PPCA on a single realization of simulated data with independent Gaussian errors \eqref{DGP:Simple}}
    \label{fig:single_fit_unsupervised_simple}
\end{figure}

\begin{figure}[H]
    \centering
    \begin{subfigure}{0.49 \textwidth}
        \includegraphics[width = \textwidth]{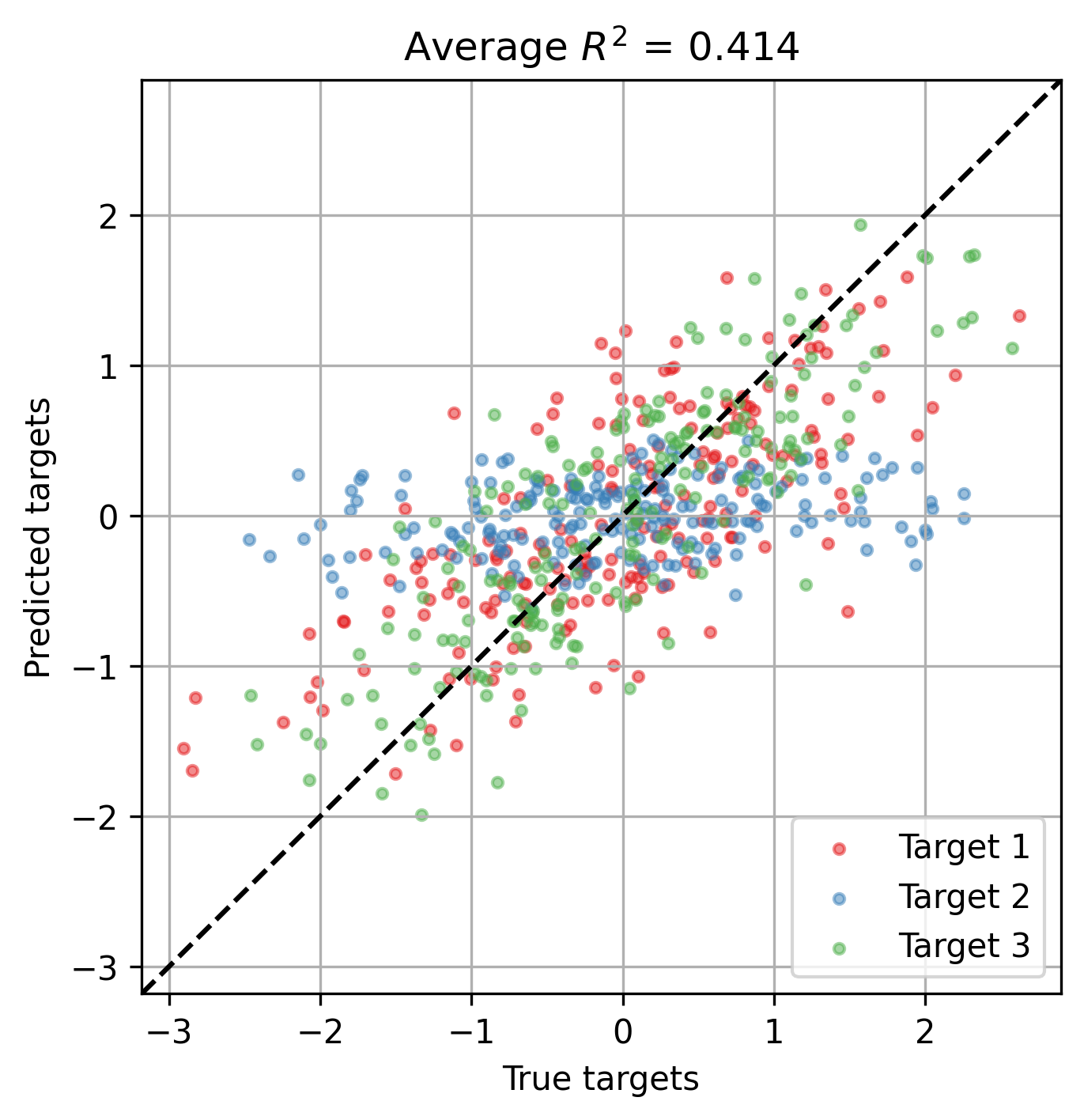}
        \caption{Principal Component Analysis (PCA)}
    \end{subfigure}
    \begin{subfigure}{0.49 \textwidth}
        \includegraphics[width = \textwidth]{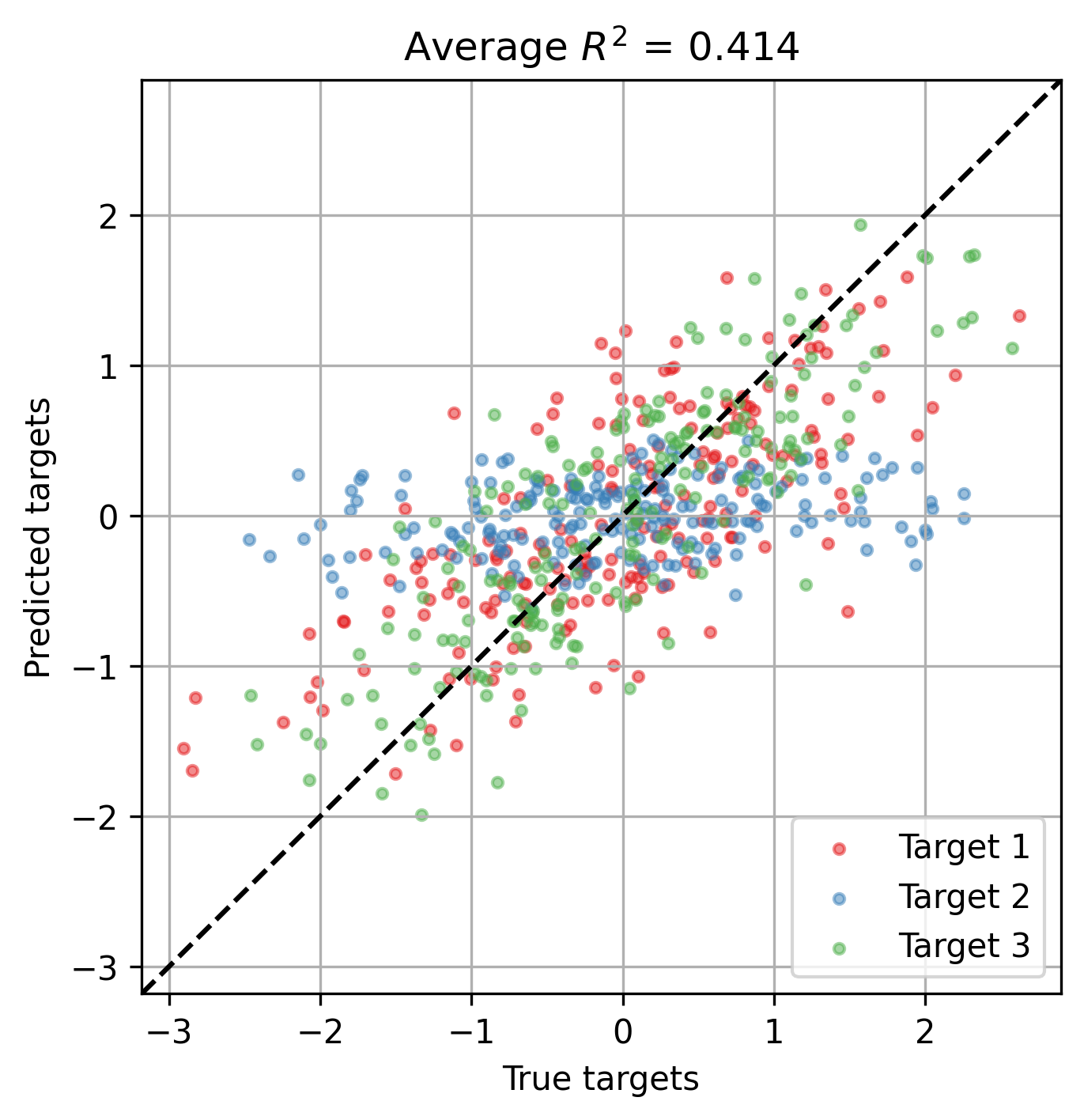}
        \caption{Probabilistic PCA (PPCA)}
    \end{subfigure}
    \caption{Comparison of PCA and PPCA on a single realization of simulated data with correlated Gaussian errors \eqref{DGP:System}}
    \label{fig:single_fit_unsupervised_system}
\end{figure}

\begin{figure}[H]
    \centering
    \begin{subfigure}{0.49 \textwidth}
        \includegraphics[width = \textwidth]{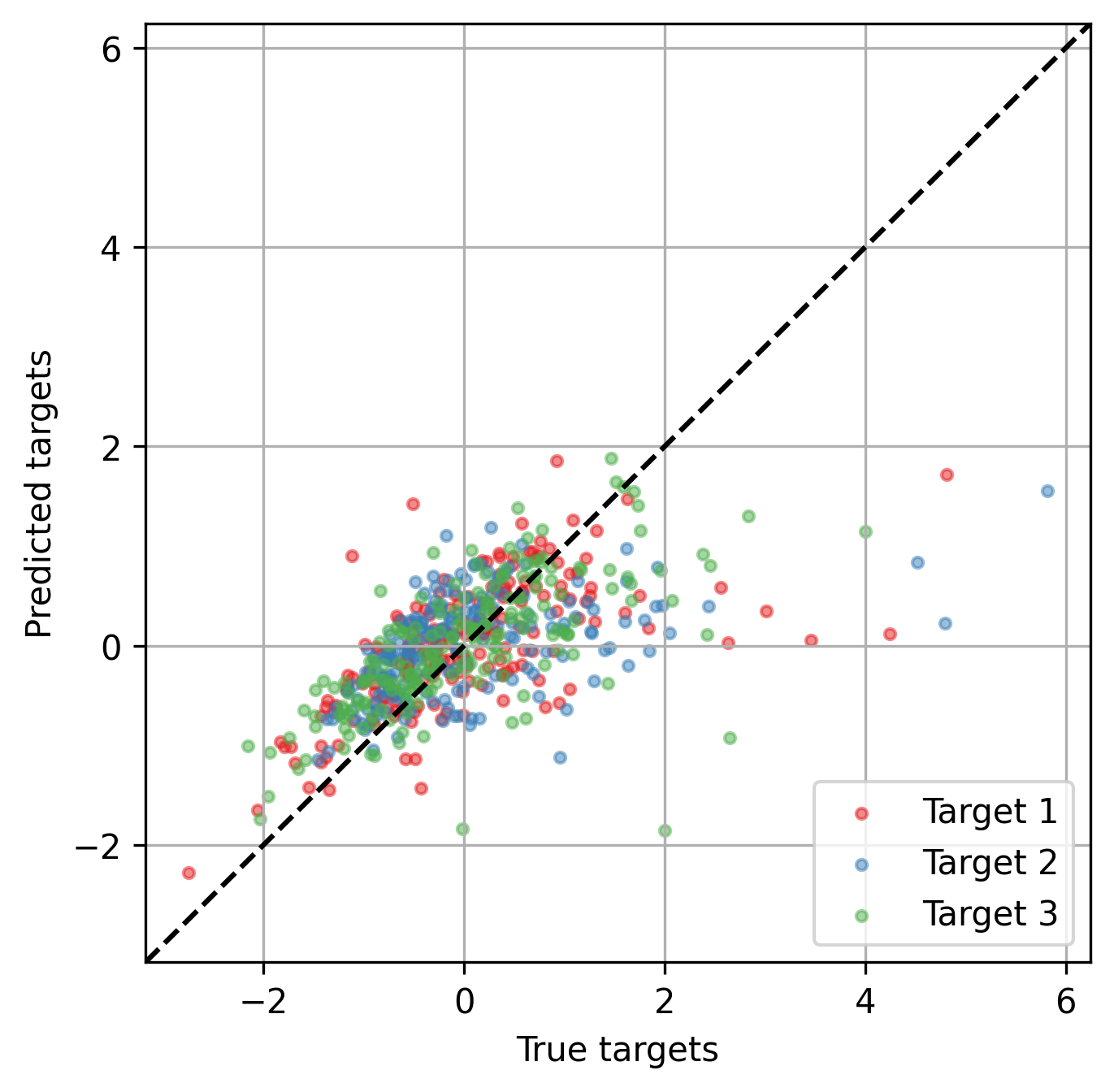}
        \caption{Principal Component Analysis (PCA)}
    \end{subfigure}
    \begin{subfigure}{0.49 \textwidth}
        \includegraphics[width = \textwidth]{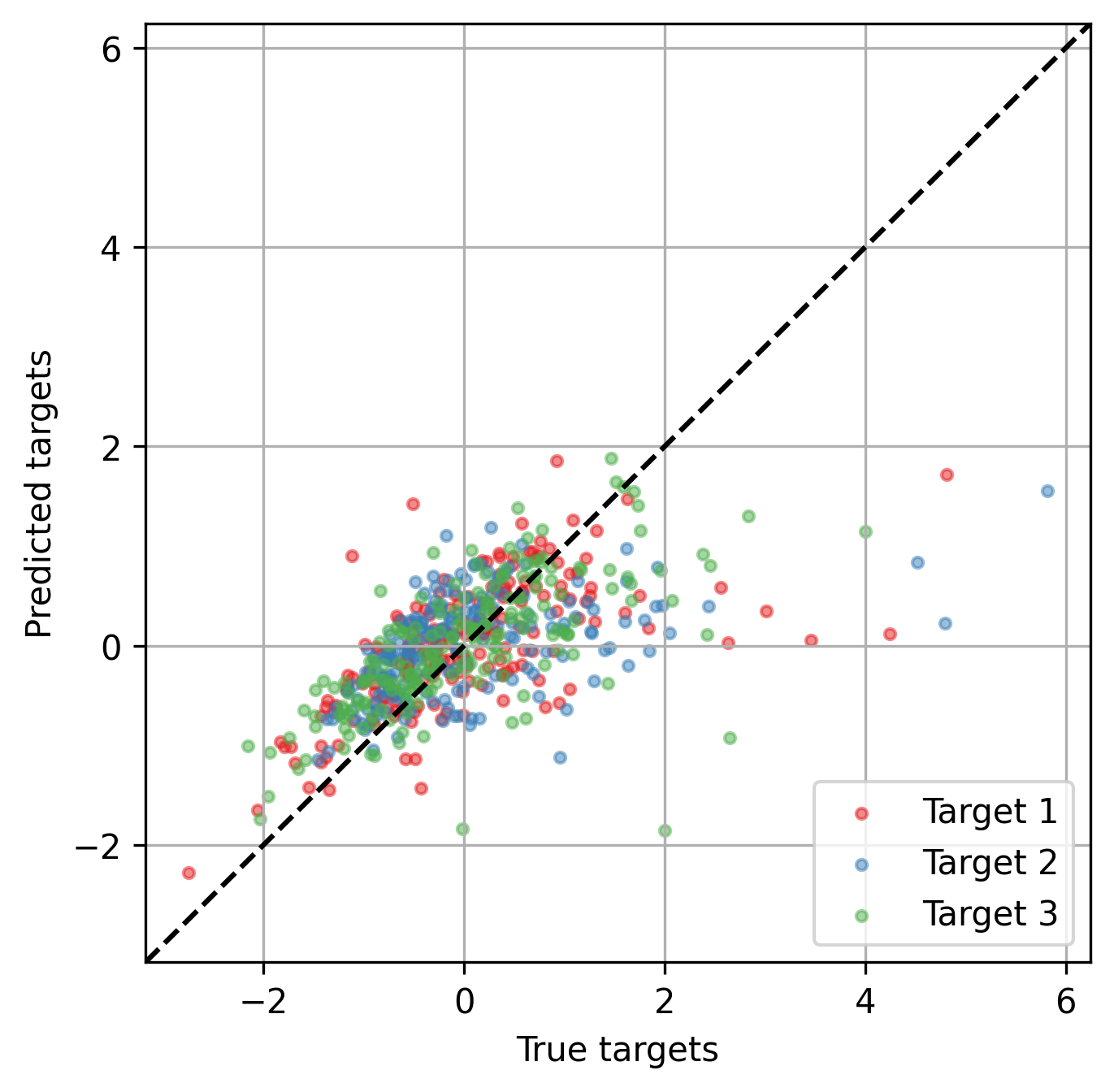}
        \caption{Probabilistic PCA (PPCA)}
    \end{subfigure}
    \caption{Comparison of PCA and PPCA on a single realization of simulated data with correlated Gaussian errors \eqref{DGP:NonGaussian}}
    \label{fig:single_fit_unsupervised_nongaussian}
\end{figure}

\begin{figure}[H]
    \centering
    \begin{subfigure}[c]{0.49\textwidth}
        \centering
        \includegraphics[width = \textwidth]{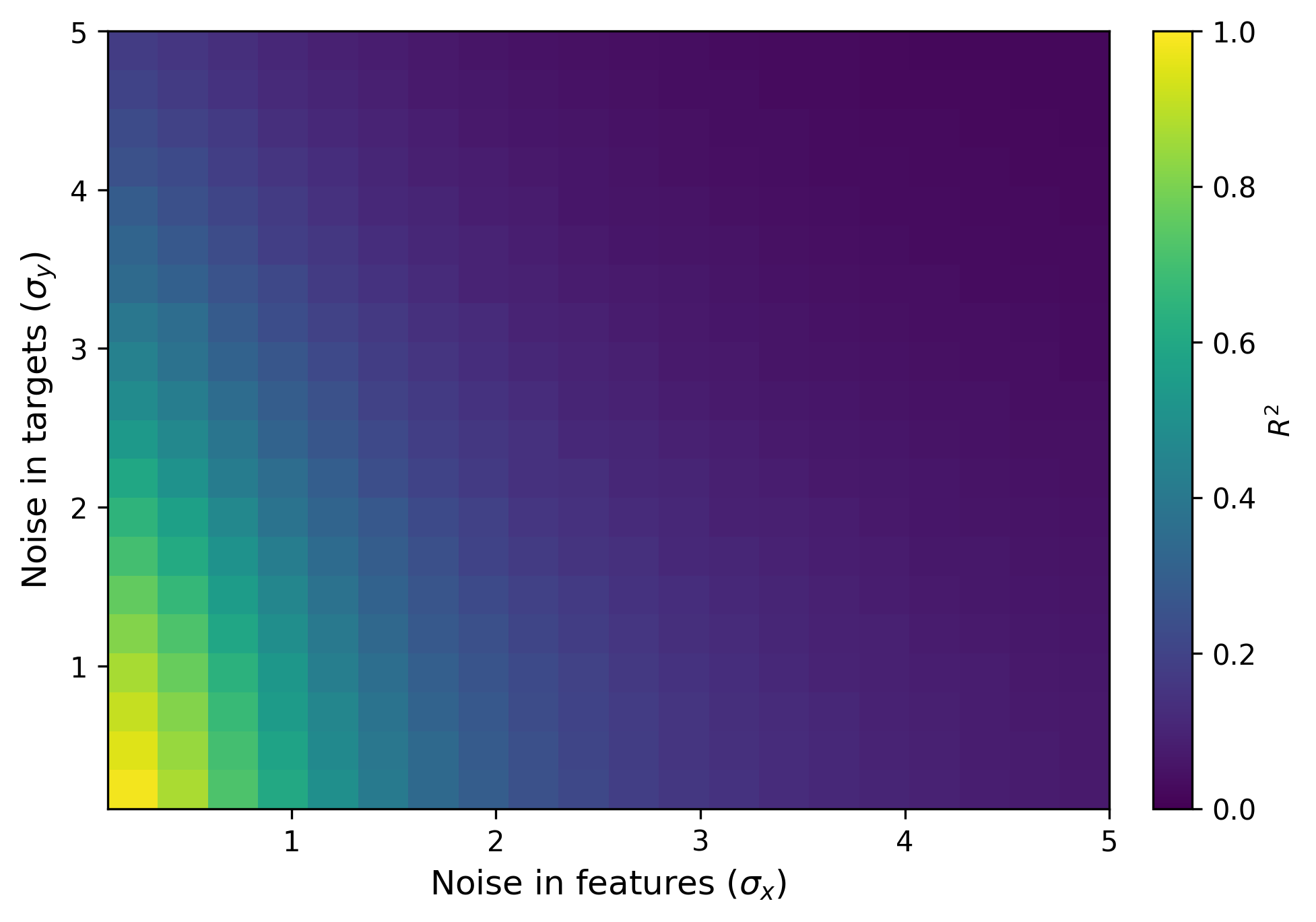}
        \caption{Principal Component Analysis (PCA)}
    \end{subfigure}
    \hfill
    \begin{subfigure}[c]{0.49\textwidth}
        \centering
        \includegraphics[width = \textwidth]{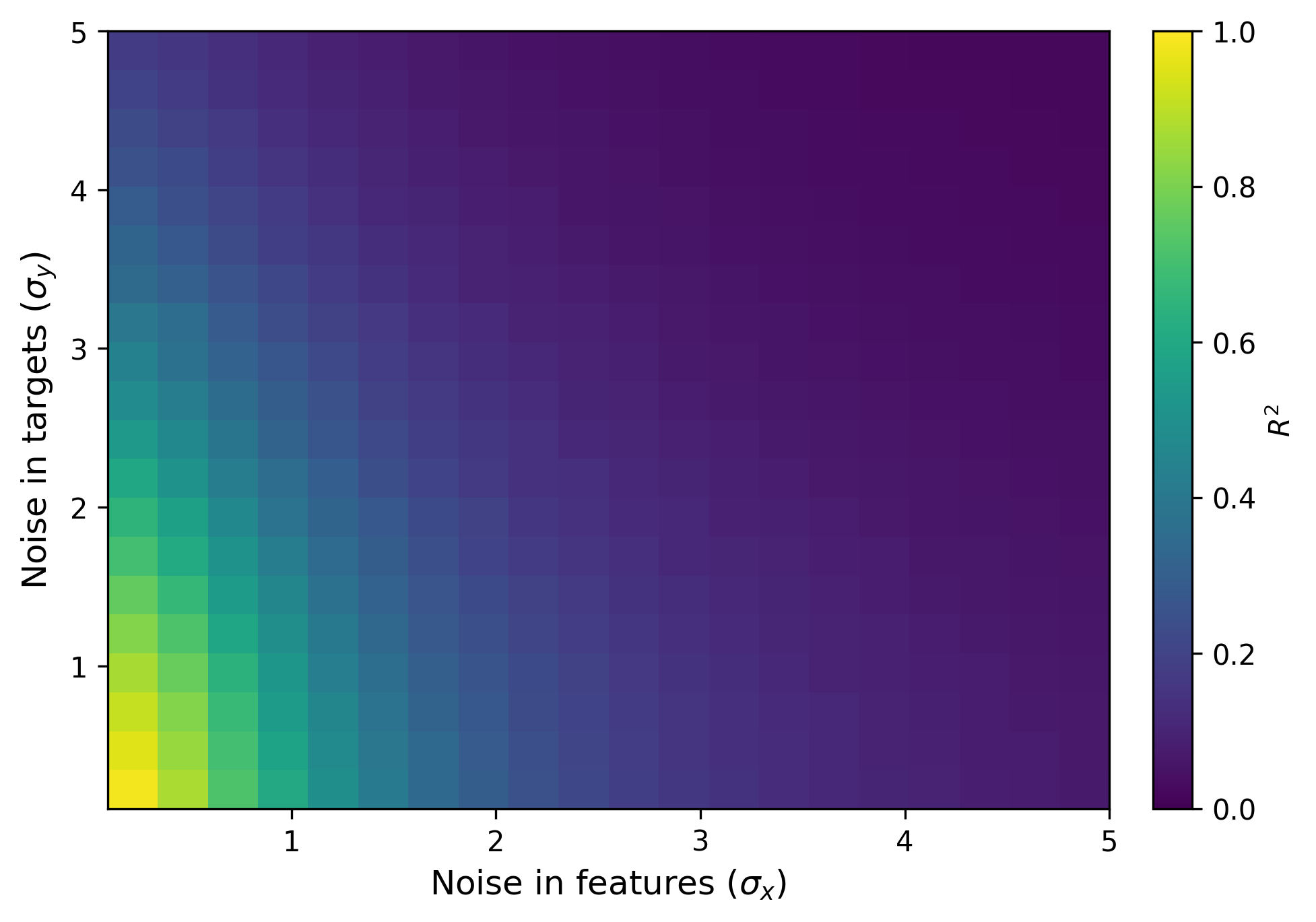}
        \caption{Probabilistic PCA (PPCA)}
    \end{subfigure}
    \caption{Comparison of the median $R^2$ statistics for PCA (a) and PPCA (b) across 1000 replications from \ref{DGP:Simple}, varying noise in features ($\sigma_x$) and targets ($\sigma_y$)}
    \label{fig:noise_supervised}
\end{figure}

\begin{figure}[H]
    \centering
    \begin{subfigure}{0.49 \textwidth}
        \includegraphics[width = \textwidth]{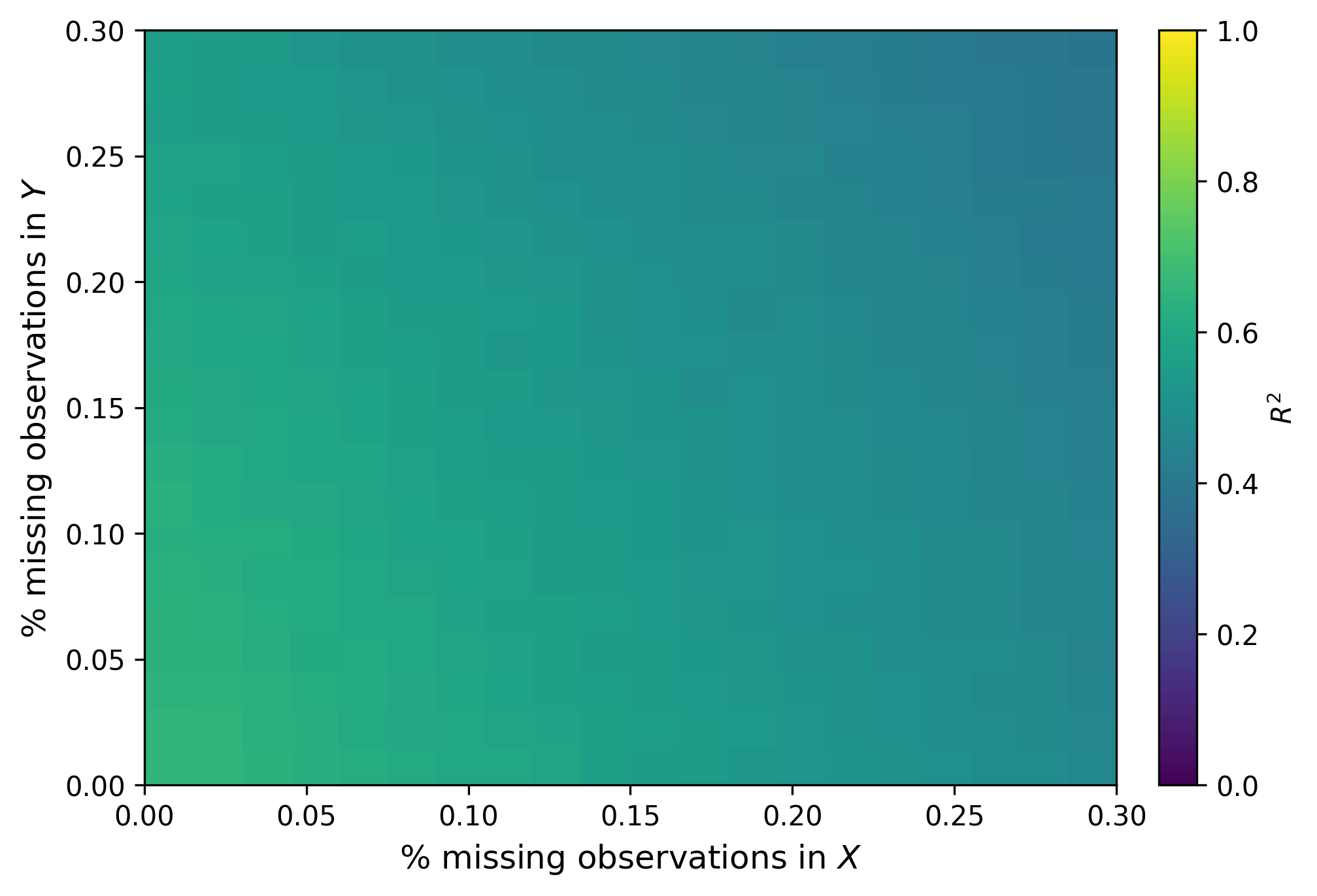}
        \caption{Principal Component Analysis (PCA)}
    \end{subfigure}
    \begin{subfigure}{0.49 \textwidth}
        \includegraphics[width = \textwidth]{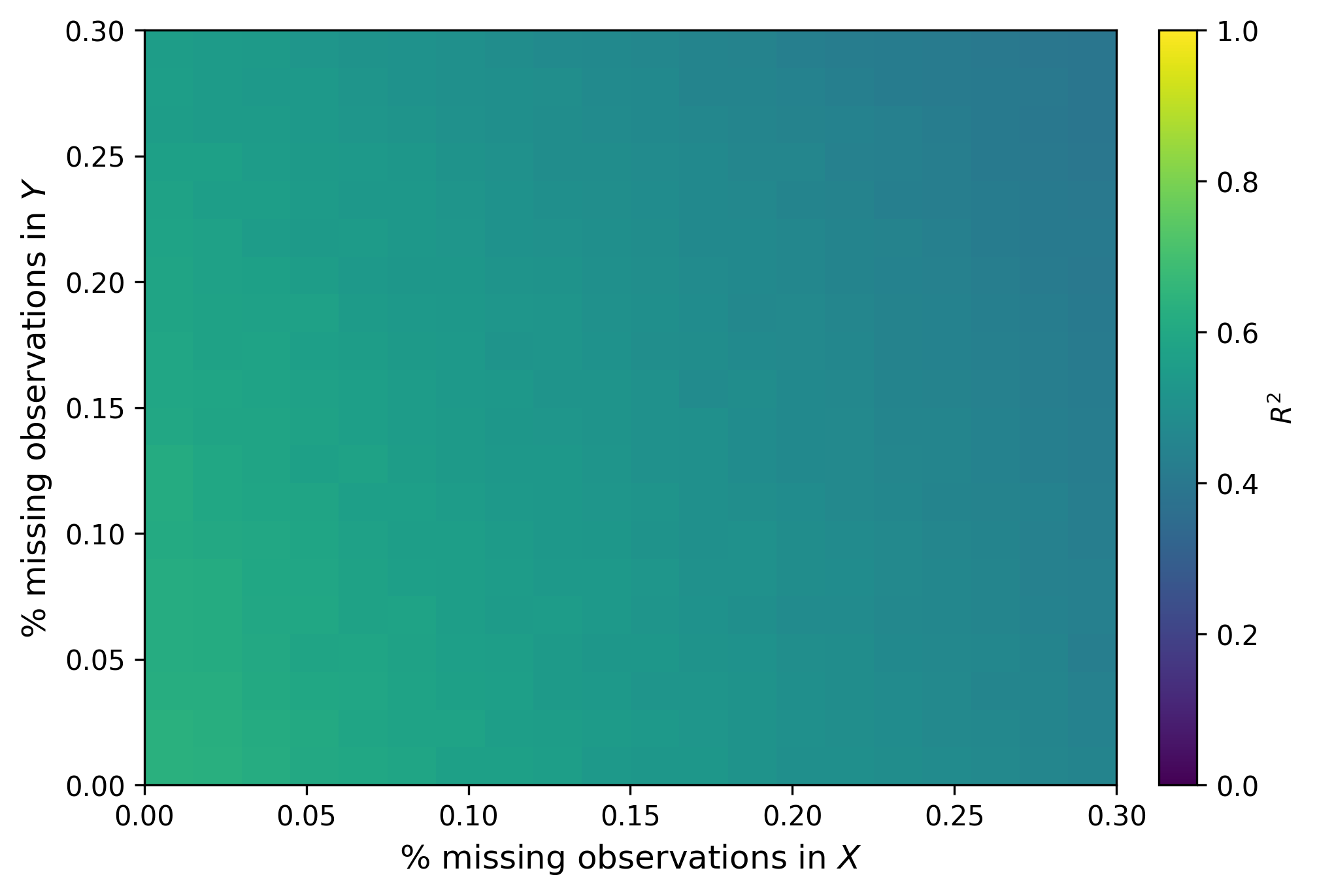}
        \caption{Probabilistic PCA (PPCA)}
    \end{subfigure}
    \caption{Comparison of PLS vs PTFA based on \% of missing-at-random observations in features ($X$) and targets ($Y$)}
    \label{fig:missing_supervised}
\end{figure}

\subsection{Applications}

\begin{table}[H]
\centering
\resizebox{\textwidth}{!}{
\begin{tabular}{lllccl}
\toprule
\textbf{\#} & \textbf{Mnemonic} & \textbf{Description} & \textbf{Sample Start} & \textbf{t-code} & \textbf{Source} \\
\midrule
1  & GDPC1           & Real Gross Domestic Product SA. Annual Rate                            & 1971-01-01 & 5 & St. Louis FRED \\
2  & CPIAUCSL        & Consumer Price Index for All Urban Consumers                           & 1971-01-01 & 5 & St. Louis FRED \\
3  & UNRATE          & Unemployment Rate, Percent, SA.                                        & 1971-01-01 & 5 & St. Louis FRED \\
4  & DFF             & Effective Federal Funds Rate                                           & 1971-01-01 & 1 & St. Louis FRED \\
5  & TOTALSL         & Total Consumer Credit Owned and Securitized \% GDP                     & 1971-01-01 & 1 & St. Louis FRED \\
6  & BPLR            & Bank Prime Loan Rate / Libor spread                                    & 1971-01-01 & 1 & St. Louis FRED \\
7  & JPMNEER         & JPMorgan Broad Nominal Effective Exchange Rate (2010=100)              & 1971-01-01 & 5 & Bloomberg      \\
8  & LDR             & All Commercial Banks Loan to Deposit Ratio                             & 1973-01-01 & 1 & Haver Analytics \\
9  & 2/3TBS          & 2yr/3m Treasury bill spread                                            & 1976-06-01 & 1 & St. Louis FRED \\
10 & MORTGAGE30US    & Mortgage rate / 10yr Treasury Bill spread                              & 1971-04-02 & 1 & St. Louis FRED \\
11 & T10Y2Y          & 10-Year Minus 2-Year Treasury Constant Maturity yield, Percent         & 1976-06-01 & 1 & St. Louis FRED \\
12 & BAMLH0A0HYM2EY  & ICE BofAML US High Yield Master II Effective Yield, Percent            & 1996-12-31 & 1 & Bloomberg      \\
\multirow{2}{*}{13} & \multirow{2}{*}{MOVE Index}      & Yield curve weighted index of normalized implied volatility & \multirow{2}{*}{1988-04-04} & \multirow{2}{*}{1} & \multirow{2}{*}{Bloomberg}      \\
& & on 1-month Treasury options & & & \\
14 & CRY Index       & Thomson Reuters/CoreCommodity CRB Commodity Index                     & 1994-01-03 & 1 & Bloomberg      \\
15 & VXOVIX          & Cboe S\&P 100/500 Volatility Index                                     & 1990-01-02 & 1 & St. Louis FRED \\
16 & BASPTDSP        & Ted Spread                                                            & 2001-01-02 & 1 & St. Louis FRED \\
17 & WILL5000PRFC    & Wilshire 5000 Full Cap Price Index                                     & 1971-01-01 & 5 & St. Louis FRED \\
\multirow{2}{*}{18} & \multirow{2}{*}{CPFF}            & 3-Month Commercial Paper Minus Federal Funds Rate, & \multirow{2}{*}{1997-01-02} & \multirow{2}{*}{1} & \multirow{2}{*}{St. Louis FRED} \\
& & Percent, Daily, Not Seasonally Adjusted & & & \\
19 & SP500           & S\&P 500 price index                                                   & 1971-01-01 & 5 & St. Louis FRED \\
\bottomrule
\end{tabular}
}
\medskip{}
\caption{Financial variables that proxy financial conditions}
\label{FCI_data}
\hfill{}%
\parbox[c]{6.5in}{%
\footnotesize{\textbf{Notes}: Mnemonic refers to the statistical reference with which the time series can be fetched from the source. Sample Start date refers to the first observation for a specific time-series in our sample. t-code refers to transformation applied to each variable. 1: levels; 5: log-differences.}
}

\end{table}

\begin{figure}[H]
  \centering
  \vspace{0.5em} 
    \centering
    \includegraphics[width=\linewidth]{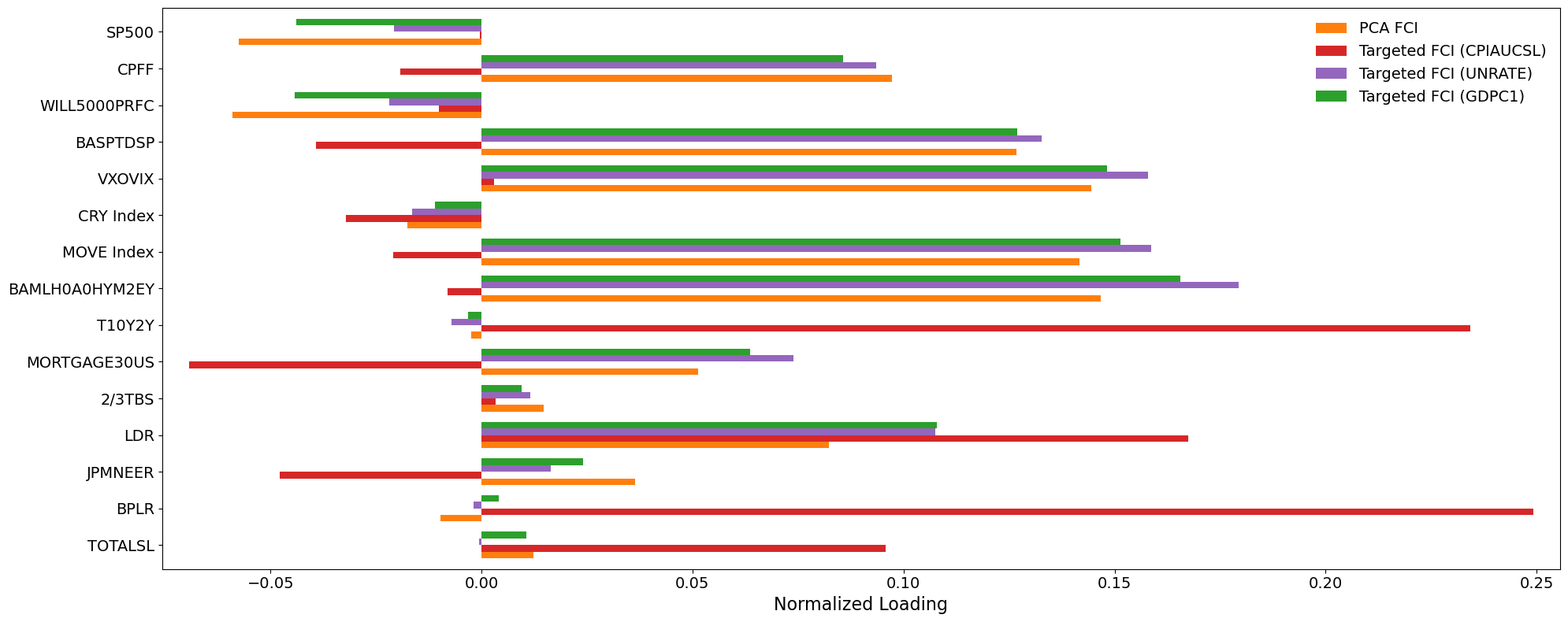}
        \caption{Comparison of FCI loadings across alternative specifications}
    \label{loadings}
      \vspace{0.5em} 
  \begin{minipage}{0.95\textwidth}
    \footnotesize
    \textbf{Notes:} The chart describes the loadings of each underlying financial variable to the respective FCI which targets: (i) all macroeconomic variables, (ii) GDP growth, (iii) CPI inflation and (iv) Unemployment rate. The PCA FCI loadings estimates are also included for comparison. All loadings are standardized for comparison purposes. 
  \end{minipage}

\end{figure}

%% file: PTFA.bib
@article{zheng_probabilistic_2016,
	title = {Probabilistic learning of partial least squares regression model: {Theory} and industrial applications},
	volume = {158},
	issn = {0169-7439},
	shorttitle = {Probabilistic learning of partial least squares regression model},
	url = {https://www.sciencedirect.com/science/article/pii/S0169743916302672},
	doi = {10.1016/j.chemolab.2016.08.014},
	urldate = {2024-11-20},
	journal = {Chemometrics and Intelligent Laboratory Systems},
	author = {Zheng, Junhua and Song, Zhihuan and Ge, Zhiqiang},
	month = nov,
	year = {2016},
	pages = {80--90},
}

@article{Petrella2017,
  author  = {Antol{\'\i}n-D{\'\i}az, Juan and Drechsel, Thomas and Petrella, Ivan},
  title   = {Tracking the Slowdown in Long-Run GDP Growth},
  journal = {The Review of Economics and Statistics},
  year    = {2017},
  volume  = {99},
  number  = {2},
  pages   = {343--356},
  doi     = {10.1162/REST_a_00646}
}

@article{vidaurre_bayesian_2013,
	title = {Bayesian {Sparse} {Partial} {Least} {Squares}},
	volume = {25},
	issn = {0899-7667},
	url = {https://ieeexplore.ieee.org/abstract/document/6797396},
	doi = {10.1162/NECO_a_00524},
	number = {12},
	urldate = {2024-11-20},
	journal = {Neural Computation},
	author = {Vidaurre, Diego and van Gerven, Marcel A. J. and Bielza, Concha and Larrañaga, Pedro and Heskes, Tom},
	month = dec,
	year = {2013},
	note = {Conference Name: Neural Computation},
	pages = {3318--3339},
}

@article{xie_fault_2019,
	title = {Fault monitoring based on locally weighted probabilistic kernel partial least square for nonlinear time-varying processes},
	volume = {33},
	copyright = {© 2019 John Wiley \& Sons, Ltd.},
	issn = {1099-128X},
	url = {https://onlinelibrary.wiley.com/doi/abs/10.1002/cem.3196},
	doi = {10.1002/cem.3196},
	language = {en},
	number = {12},
	urldate = {2024-11-20},
	journal = {Journal of Chemometrics},
	author = {Xie, Ying},
	year = {2019},
	pages = {e3196},
}

@article{el_bouhaddani_statistical_2022,
	title = {Statistical {Integration} of {Heterogeneous} {Omics} {Data}: {Probabilistic} {Two}-{Way} {Partial} {Least} {Squares} ({PO2PLS})},
	volume = {71},
	issn = {0035-9254},
	shorttitle = {Statistical {Integration} of {Heterogeneous} {Omics} {Data}},
	url = {https://doi.org/10.1111/rssc.12583},
	doi = {10.1111/rssc.12583},
	number = {5},
	urldate = {2024-11-20},
	journal = {Journal of the Royal Statistical Society Series C: Applied Statistics},
	author = {el Bouhaddani, Said and Uh, Hae-Won and Jongbloed, Geurt and Houwing-Duistermaat, Jeanine},
	month = nov,
	year = {2022},
	pages = {1451--1470},
}

@article{yang_robust_2021,
	title = {Robust {Mixture} {Probabilistic} {Partial} {Least} {Squares} {Model} for {Soft} {Sensing} {With} {Multivariate} {Laplace} {Distribution}},
	volume = {70},
	issn = {1557-9662},
	url = {https://ieeexplore.ieee.org/abstract/document/9141330},
	doi = {10.1109/TIM.2020.3009354},
	urldate = {2024-11-20},
	journal = {IEEE Transactions on Instrumentation and Measurement},
	author = {Yang, Xianqiang and Liu, Xinpeng and Xu, Chao},
	year = {2021},
	note = {Conference Name: IEEE Transactions on Instrumentation and Measurement},
	pages = {1--9},
}

@article{li_process_2018,
	title = {Process {Modeling} and {Monitoring} {With} {Incomplete} {Data} {Based} on {Robust} {Probabilistic} {Partial} {Least} {Square} {Method}},
	volume = {6},
	issn = {2169-3536},
	url = {https://ieeexplore.ieee.org/abstract/document/8304745},
	doi = {10.1109/ACCESS.2018.2810079},
	urldate = {2024-11-20},
	journal = {IEEE Access},
	author = {Li, Qinghua and Pan, Feng and Zhao, Zhonggai and Yu, Junzhi},
	year = {2018},
	note = {Conference Name: IEEE Access},
	pages = {10160--10168},
}

@article{gustafsson_probabilistic_2001,
	title = {A {Probabilistic} {Derivation} of the {Partial} {Least}-{Squares} {Algorithm}},
	volume = {41},
	issn = {0095-2338},
	url = {https://doi.org/10.1021/ci0003909},
	doi = {10.1021/ci0003909},
	number = {2},
	urldate = {2024-11-20},
	journal = {Journal of Chemical Information and Computer Sciences},
	author = {Gustafsson, Mats G.},
	month = mar,
	year = {2001},
	pages = {288--294},
}

@article{zheng_semisupervised_2018,
	title = {Semisupervised learning for probabilistic partial least squares regression model and soft sensor application},
	volume = {64},
	issn = {0959-1524},
	url = {https://www.sciencedirect.com/science/article/pii/S0959152418300209},
	doi = {10.1016/j.jprocont.2018.01.008},
	urldate = {2024-11-20},
	journal = {Journal of Process Control},
	author = {Zheng, Junhua and Song, Zhihuan},
	month = apr,
	year = {2018},
	pages = {123--131},
}

@article{el_bouhaddani_probabilistic_2018,
	title = {Probabilistic partial least squares model: {Identifiability}, estimation and application},
	volume = {167},
	issn = {0047-259X},
	shorttitle = {Probabilistic partial least squares model},
	url = {https://www.sciencedirect.com/science/article/pii/S0047259X17302762},
	doi = {10.1016/j.jmva.2018.05.009},
	urldate = {2024-11-20},
	journal = {Journal of Multivariate Analysis},
	author = {el Bouhaddani, Said and Uh, Hae-Won and Hayward, Caroline and Jongbloed, Geurt and Houwing-Duistermaat, Jeanine},
	month = sep,
	year = {2018},
	pages = {331--346},
}

@inproceedings{li_probabilistic_2011,
	title = {Probabilistic {Partial} {Least} {Square} {Regression}: {A} {Robust} {Model} for {Quantitative} {Analysis} of {Raman} {Spectroscopy} {Data}},
	shorttitle = {Probabilistic {Partial} {Least} {Square} {Regression}},
	url = {https://ieeexplore.ieee.org/document/6120496},
	doi = {10.1109/BIBM.2011.94},
	urldate = {2024-11-20},
	booktitle = {2011 {IEEE} {International} {Conference} on {Bioinformatics} and {Biomedicine}},
	author = {Li, Shuo and Gao, Jean and Nyagilo, James O. and Dave, Digant P.},
	month = nov,
	year = {2011},
	pages = {526--531},
}

@inproceedings{durrande_banded_2019,
	series = {Proceedings of {Machine} {Learning} {Research}},
	title = {Banded {Matrix} {Operators} for {Gaussian} {Markov} {Models} in the {Automatic} {Differentiation} {Era}},
	volume = {89},
	url = {https://proceedings.mlr.press/v89/durrande19a.html},
	booktitle = {Proceedings of the {Twenty}-{Second} {International} {Conference} on {Artificial} {Intelligence} and {Statistics}},
	publisher = {PMLR},
	author = {Durrande, Nicolas and Adam, Vincent and Bordeaux, Lucas and Eleftheriadis, Stefanos and Hensman, James},
	editor = {Chaudhuri, Kamalika and Sugiyama, Masashi},
	year = {2019},
	pages = {2780--2789},
}

@article{primiceri_time_2005,
	title = {Time varying structural vector autoregressions and monetary policy},
	volume = {72},
	number = {3},
	journal = {Review of Economic Studies},
	author = {Primiceri, Giorgio E.},
	year = {2005},
	pages = {821--852},
}

@book{chan_bayesian_2019,
	edition = {Second},
	series = {Econometric {Exercises}},
	title = {Bayesian {Econometric} {Methods}},
	publisher = {Cambridge University Press},
	author = {Chan, Joshua and Koop, Gary and Poirier, Dale J. and Tobias, Justin L.},
	year = {2019},
}

@article{forni_generalized_2000,
	title = {The {Generalized} {Dynamic}-{Factor} {Model}: {Identification} and {Estimation}},
	volume = {82},
	issn = {0034-6535},
	url = {https://doi.org/10.1162/003465300559037},
	doi = {10.1162/003465300559037},
	number = {4},
	journal = {Review of Economics and Statistics},
	author = {Forni, Mario and Hallin, Marc and Lippi, Marco and Reichlin, Lucrezia},
	month = nov,
	year = {2000},
	pages = {540--554},
}

@incollection{doz_dynamic_2020,
    title = {{Dynamic Factor Models}},
    author = {Doz, Catherine and Fuleky, Peter},
    url = {https://shs.hal.science/halshs-02491811},
    booktitle = {{Macroeconomic Forecasting in the Era of Big Data: Theory and Practice}},
    editor = {Peter Fuleky},
    publisher = {{Springer}},
    pages = {27-64},
    year = {2020},
    month = Nov,
    doi = {10.1007/978-3-030-31150-6\_2},
    hal_id = {halshs-02491811},
    hal_version = {v1},
}

@article{foroni_mixed-frequency_2014,
	title = {Mixed-frequency {Structural} {Models}: {Identification}, {Estimation}, and {Policy} {Analysis}},
	volume = {29},
	url = {https://onlinelibrary.wiley.com/doi/abs/10.1002/jae.2396},
	doi = {https://doi.org/10.1002/jae.2396},
	number = {7},
	journal = {Journal of Applied Econometrics},
	author = {Foroni, Claudia and Marcellino, Massimiliano},
	year = {2014},
	pages = {1118--1144},
}

@incollection{stock_dynamic_2011,
	title = {Dynamic {Factor} {Models}},
	isbn = {978-0-19-539864-9},
	url = {https://doi.org/10.1093/oxfordhb/9780195398649.013.0003},
	booktitle = {The {Oxford} {Handbook} of {Economic} {Forecasting}},
	publisher = {Oxford University Press},
	author = {Stock, James H. and Watson, Mark W.},
	month = jul,
	year = {2011},
	doi = {10.1093/oxfordhb/9780195398649.013.0003},
    pages = {35--60}
}

@article{giannone_nowcasting_2008,
	title = {Nowcasting: {The} real-time informational content of macroeconomic data},
	volume = {55},
	issn = {0304-3932},
	url = {https://www.sciencedirect.com/science/article/pii/S0304393208000652},
	doi = {https://doi.org/10.1016/j.jmoneco.2008.05.010},
	number = {4},
	journal = {Journal of Monetary Economics},
	author = {Giannone, Domenico and Reichlin, Lucrezia and Small, David},
	year = {2008},
	pages = {665--676},
}

@article{frank_statistical_1993,
	title = {A statistical view of some chemometrics regression tools},
	volume = {35},
	number = {2},
	journal = {Technometrics},
	author = {Frank, Ildiko E. and Friedman, Jerome H.},
	year = {1993},
	pages = {109--135},
}

@article{welch_comprehensive_2007,
	title = {A {Comprehensive} {Look} at {The} {Empirical} {Performance} of {Equity} {Premium} {Prediction}},
	volume = {21},
	issn = {0893-9454},
	url = {https://doi.org/10.1093/rfs/hhm014},
	doi = {10.1093/rfs/hhm014},
	number = {4},
	journal = {Review of Financial Studies},
	author = {Welch, Ivo and Goyal, Amit},
	year = {2007},
	pages = {1455--1508},
}

@article{goyal_comprehensive_2024,
	title = {A {Comprehensive} 2022 {Look} at the {Empirical} {Performance} of {Equity} {Premium} {Prediction}},
	issn = {0893-9454},
	url = {https://doi.org/10.1093/rfs/hhae044},
	doi = {10.1093/rfs/hhae044},
	journal = {Review of Financial Studies},
	author = {Goyal, Amit and Welch, Ivo and Zafirov, Athanasse},
	year = {2024},
	pages = {hhae044},
}

@article{gong_pseudo_1981,
	title = {Pseudo {Maximum} {Likelihood} {Estimation}: {Theory} and {Applications}},
	volume = {9},
	url = {http://www.jstor.org/stable/2240854},
	number = {4},
	journal = {The Annals of Statistics},
	author = {Gong, Gail and Samaniego, Francisco J.},
	year = {1981},
	pages = {861--869},
}

@article{gourieroux_pseudo_1984,
	title = {Pseudo {Maximum} {Likelihood} {Methods}: {Theory}},
	volume = {52},
	url = {http://www.jstor.org/stable/1913471},
	number = {3},
	journal = {Econometrica},
	author = {Gourieroux, C. and Monfort, A. and Trognon, A.},
	year = {1984},
	pages = {681--700},
}

@book{chen_spectral_2021,
	address = {Hanover, MA, USA},
	title = {Spectral {Methods} for {Data} {Science}: {A} {Statistical} {Perspective}},
	volume = {14},
	url = {https://doi.org/10.1561/2200000079},
	issue = {5},
	publisher = {Now Publishers Inc.},
	author = {Chen, Yuxin and Chi, Yuejie and Fan, Jianqing and Ma, Cong},
	month = oct,
	year = {2021},
	doi = {10.1561/2200000079},
	note = {Foundations and Trends® in Machine Learning},
}

@book{greenberg_introduction_2012,
	title = {Introduction to {Bayesian} {Econometrics}},
	isbn = {978-1-139-78933-2},
	url = {https://books.google.co.uk/books?id=etMhAwAAQBAJ},
	publisher = {Cambridge University Press},
	author = {Greenberg, E.},
	year = {2012},
}

@article{butler_peculiar_2002,
	title = {The {Peculiar} {Shrinkage} {Properties} of {Partial} {Least} {Squares} {Regression}},
	volume = {62},
	issn = {1369-7412},
	url = {https://doi.org/10.1111/1467-9868.00252},
	doi = {10.1111/1467-9868.00252},
	number = {3},
	journal = {Journal of the Royal Statistical Society Series B: Statistical Methodology},
	author = {Butler, Neil A. and Denham, Michael C.},
	year = {2002},
	pages = {585--593},
}

@article{kelly_market_2013,
	title = {Market {Expectations} in the {Cross}-{Section} of {Present} {Values}},
	volume = {68},
	url = {https://onlinelibrary.wiley.com/doi/abs/10.1111/jofi.12060},
	doi = {https://doi.org/10.1111/jofi.12060},
	number = {5},
	journal = {The Journal of Finance},
	author = {Kelly, Bryan and Pruitt, Seth},
	year = {2013},
	pages = {1721--1756},
}

@article{giglio_systemic_2016,
	title = {Systemic risk and the macroeconomy: {An} empirical evaluation},
	volume = {119},
	issn = {0304-405X},
	url = {https://www.sciencedirect.com/science/article/pii/S0304405X16000143},
	doi = {https://doi.org/10.1016/j.jfineco.2016.01.010},
	number = {3},
	journal = {Journal of Financial Economics},
	author = {Giglio, Stefano and Kelly, Bryan and Pruitt, Seth},
	year = {2016},
	pages = {457--471},
}

@article{groen_revisiting_2016,
	title = {Revisiting useful approaches to data-rich macroeconomic forecasting},
	volume = {100},
	issn = {0167-9473},
	url = {https://www.sciencedirect.com/science/article/pii/S0167947315002972},
	doi = {https://doi.org/10.1016/j.csda.2015.11.014},
	journal = {Computational Statistics \& Data Analysis},
	author = {Groen, Jan J. J. and Kapetanios, George},
	year = {2016},
	pages = {221--239},
}

@book{hastie_elements_2001,
	address = {New York, NY, USA},
	series = {Springer {Series} in {Statistics}},
	title = {The {Elements} of {Statistical} {Learning}},
	publisher = {Springer New York Inc.},
	author = {Hastie, Trevor and Tibshirani, Robert and Friedman, Jerome},
	year = {2001},
}

@book{little_statistical_2019,
	title = {Statistical analysis with missing data},
	volume = {793},
	publisher = {John Wiley \& Sons},
	author = {Little, Roderick J. A. and Rubin, Donald B.},
	year = {2019},
}

@article{wold_soft_1975,
	title = {Soft {Modelling} by {Latent} {Variables}: {The} {Non}-{Linear} {Iterative} {Partial} {Least} {Squares} ({NIPALS}) {Approach}},
	volume = {12},
	doi = {10.1017/S0021900200047604},
	number = {S1},
	journal = {Journal of Applied Probability},
	author = {Wold, Herman},
	year = {1975},
	pages = {117--142},
}

@article{mccracken_fred-md_2016,
	title = {{FRED}-{MD}: {A} {Monthly} {Database} for {Macroeconomic} {Research}},
	volume = {34},
	doi = {10.1080/07350015.2015.1086655},
	number = {4},
	journal = {Journal of Business \& Economic Statistics},
	author = {McCracken, Michael W. and Ng, Serena},
	year = {2016},
	pages = {574--589},
}

@article{tipping_probabilistic_1999,
	title = {Probabilistic principal component analysis},
	volume = {61},
	number = {3},
	journal = {Journal of the Royal Statistical Society Series B: Statistical Methodology},
	author = {Tipping, Michael E. and Bishop, Christopher M.},
	year = {1999},
	pages = {611--622},
}

@article{kelly_three-pass_2015,
	title = {The three-pass regression filter: {A} new approach to forecasting using many predictors},
	volume = {186},
	number = {2},
	journal = {Journal of Econometrics},
	author = {Kelly, Bryan and Pruitt, Seth},
	year = {2015},
	pages = {294--316},
}

@article{fuentes_sparse_2015,
	title = {Sparse {Partial} {Least} {Squares} in {Time} {Series} for {Macroeconomic} {Forecasting}},
	volume = {30},
	copyright = {Copyright © 2014 John Wiley \& Sons, Ltd.},
	issn = {1099-1255},
	url = {https://onlinelibrary.wiley.com/doi/abs/10.1002/jae.2384},
	doi = {10.1002/jae.2384},
	language = {en},
	number = {4},
	urldate = {2025-08-15},
	journal = {Journal of Applied Econometrics},
	author = {Fuentes, Julieta and Poncela, Pilar and Rodríguez, Julio},
	year = {2015},
	pages = {576--595},
}

@article{bai_determining_2002,
	title = {Determining the {Number} of {Factors} in {Approximate} {Factor} {Models}},
	volume = {70},
	copyright = {The Econometric Society 2001},
	issn = {1468-0262},
	url = {https://onlinelibrary.wiley.com/doi/abs/10.1111/1468-0262.00273},
	doi = {10.1111/1468-0262.00273},
	language = {en},
	number = {1},
	urldate = {2025-08-15},
	journal = {Econometrica},
	author = {Bai, Jushan and Ng, Serena},
	year = {2002},
	pages = {191--221},
}

@article{giglio_test_2025,
	title = {Test {Assets} and {Weak} {Factors}},
	volume = {80},
	copyright = {© 2024 the American Finance Association.},
	issn = {1540-6261},
	url = {https://onlinelibrary.wiley.com/doi/abs/10.1111/jofi.13415},
	doi = {10.1111/jofi.13415},
	language = {en},
	number = {1},
	urldate = {2025-08-15},
	journal = {The Journal of Finance},
	author = {Giglio, Stefano and Xiu, Dacheng and Zhang, Dake},
	year = {2025},
	pages = {259--319},
}

@misc{ahn_forecasting_2022,
	address = {Rochester, NY},
	type = {{SSRN} {Scholarly} {Paper}},
	title = {Forecasting with {Partial} {Least} {Squares} {Using} {Many} {Predictors}},
	url = {https://papers.ssrn.com/abstract=4248450},
	doi = {10.2139/ssrn.4248450},
	language = {en},
	urldate = {2025-08-15},
	publisher = {Social Science Research Network},
	author = {Ahn, Seung C. and Bae, Juhee},
	month = feb,
	year = {2022},
}

@article{doz_quasimaximum_2012,
	title = {A {Quasi}–{Maximum} {Likelihood} {Approach} for {Large}, {Approximate} {Dynamic} {Factor} {Models}},
	volume = {94},
	issn = {0034-6535},
	url = {https://doi.org/10.1162/REST_a_00225},
	doi = {10.1162/REST_a_00225},
	number = {4},
	urldate = {2025-08-15},
	journal = {The Review of Economics and Statistics},
	author = {Doz, Catherine and Giannone, Domenico and Reichlin, Lucrezia},
	month = nov,
	year = {2012},
	pages = {1014--1024},
}

@article{koop_new_2014,
	title = {A new index of financial conditions},
	volume = {71},
	issn = {00142921},
	url = {https://linkinghub.elsevier.com/retrieve/pii/S0014292114001068},
	doi = {10.1016/j.euroecorev.2014.07.002},
	language = {en},
	urldate = {2025-08-15},
	journal = {European Economic Review},
	author = {Koop, Gary and Korobilis, Dimitris},
	month = oct,
	year = {2014},
	pages = {101--116},
}

@article{brave_monitoring_2011,
	title = {Monitoring financial stability: {A} financial conditions index approach},
	volume = {35},
	number = {1},
	journal = {Economic Perspectives},
	author = {Brave, Scott A. and Butters, R. Andrew},
	year = {2011},
	pages = {22},
}

@techreport{hatzius_financial_2010,
	address = {Cambridge, MA},
	title = {Financial {Conditions} {Indexes}: {A} {Fresh} {Look} after the {Financial} {Crisis}},
	shorttitle = {Financial {Conditions} {Indexes}},
	url = {http://www.nber.org/papers/w16150.pdf},
	language = {en},
	number = {w16150},
	urldate = {2025-08-15},
	institution = {National Bureau of Economic Research},
	author = {Hatzius, Jan and Hooper, Peter and Mishkin, Frederic and Schoenholtz, Kermit and Watson, Mark},
	month = jul,
	year = {2010},
	doi = {10.3386/w16150},
	pages = {w16150},
}

@article{JuradoLudvigsonNg2015,
  author  = {Jurado, Kyle and Ludvigson, Sydney C. and Ng, Serena},
  title   = {Measuring Uncertainty},
  journal = {American Economic Review},
  year    = {2015},
  volume  = {105},
  number  = {3},
  pages   = {1177--1216},
  doi     = {10.1257/aer.20131193}
}

@article{quintana_time_1988,
	title = {Time series analysis of compositional data},
	volume = {3},
	journal = {Bayesian Statistics},
	author = {Quintana, Jose M and West, Mike},
	year = {1988},
	pages = {747--756},
}

@article{Lombardi2025,
	title = {Financial conditions and
the macroeconomy:
a two-factor view},
	volume = {1272},
	journal = {BIS Working Papers},
	author = {Lombardi, M. and Manea, C. and
Schrimpf, S.},
	year = {2025},
	pages = {747--756},
}

@incollection{Stock2016,
title = {{Chapter 8 - Dynamic Factor Models, Factor-Augmented Vector Autoregressions, and Structural Vector Autoregressions in Macroeconomics}},
editor = {John B. Taylor and Harald Uhlig},
booktitle = {Handbook of Macroeconomics},
publisher = {Elsevier},
volume = {2},
pages = {415-525},
year = {2016},
issn = {1574-0048},
doi = {https://doi.org/10.1016/bs.hesmac.2016.04.002},
url = {https://www.sciencedirect.com/science/article/pii/S1574004816300027},
author = {J.H. Stock and M.W. Watson},
}

@misc{barigozzi2024,
      title={{Quasi Maximum Likelihood Estimation and Inference of Large Approximate Dynamic Factor Models via the EM algorithm}}, 
      author={Matteo Barigozzi and Matteo Luciani},
      year={2024},
      eprint={1910.03821},
      archivePrefix={arXiv},
      primaryClass={math.ST},
      url={https://arxiv.org/abs/1910.03821}, 
}

@techreport{welch_introduction_1995,
	address = {USA},
	type = {Technical {Report}},
	title = {An {Introduction} to the {Kalman} {Filter}},
	institution = {University of North Carolina at Chapel Hill},
	author = {Welch, Greg and Bishop, Gary},
	month = oct,
	year = {1995},
}

@article{adrian_vulnerable_2019,
	title = {Vulnerable {Growth}},
	volume = {109},
	issn = {0002-8282},
	url = {https://www.aeaweb.org/articles?id=10.1257/aer.20161923},
	doi = {10.1257/aer.20161923},
	language = {en},
	number = {4},
	urldate = {2025-08-15},
	journal = {American Economic Review},
	author = {Adrian, Tobias and Boyarchenko, Nina and Giannone, Domenico},
	month = apr,
	year = {2019},
	pages = {1263--1289},
}

@incollection{velicer_construct_2000,
	address = {Boston, MA},
	title = {Construct {Explication} through {Factor} or {Component} {Analysis}: {A} {Review} and {Evaluation} of {Alternative} {Procedures} for {Determining} the {Number} of {Factors} or {Components}},
	isbn = {978-1-4615-4397-8},
	shorttitle = {Construct {Explication} through {Factor} or {Component} {Analysis}},
	url = {https://doi.org/10.1007/978-1-4615-4397-8_3},
	language = {en},
	urldate = {2025-08-15},
	booktitle = {Problems and {Solutions} in {Human} {Assessment}: {Honoring} {Douglas} {N}. {Jackson} at {Seventy}},
	publisher = {Springer US},
	author = {Velicer, Wayne F. and Eaton, Cheryl A. and Fava, Joseph L.},
	editor = {Goffin, Richard D. and Helmes, Edward},
	year = {2000},
	doi = {10.1007/978-1-4615-4397-8_3},
	pages = {41--71},
}

@article{doz_two-step_2011,
	series = {Annals {Issue} on {Forecasting}},
	title = {A two-step estimator for large approximate dynamic factor models based on {Kalman} filtering},
	volume = {164},
	issn = {0304-4076},
	url = {https://www.sciencedirect.com/science/article/pii/S030440761100039X},
	doi = {10.1016/j.jeconom.2011.02.012},
	number = {1},
	urldate = {2025-08-15},
	journal = {Journal of Econometrics},
	author = {Doz, Catherine and Giannone, Domenico and Reichlin, Lucrezia},
	month = sep,
	year = {2011},
	pages = {188--205},
}

@article{jungbacker_koopman_2015,
  title     = {Likelihood-Based Dynamic Factor Analysis for Measurement and Forecasting},
  author    = {Jungbacker, Borus and Koopman, Siem Jan},
  journal   = {Econometrics Journal},
  volume    = {18},
  number    = {2},
  pages     = {C1--C21},
  year      = {2015},
  doi       = {10.1111/ectj.12043}
}

@article{coroneo2016,
  title     = {Unspanned Macro Factors in the Yield Curve},
  author    = {Coroneo, Laura and Giannone, Domenico and Modugno, Michele},
  journal   = {Journal of Business \& Economic Statistics},
  volume    = {34},
  number    = {3},
  pages     = {472--485},
  year      = {2016},
  doi       = {10.1080/07350015.2015.1041770}
}

@techreport{bach2005probabilistic,
  title={A probabilistic interpretation of canonical correlation analysis},
  author={Bach, Francis R and Jordan, Michael I},
  year={2005},
  institution={Department of Statistics, University of California-Berkeley}
}

@article{klami2013bayesian,
  title={Bayesian canonical correlation analysis},
  author={Klami, Arto and Virtanen, Seppo and Kaski, Samuel},
  journal={The Journal of Machine Learning Research},
  volume={14},
  number={1},
  pages={965--1003},
  year={2013},
  publisher={JMLR. org}
}
